\documentclass[12pt]{article}
\usepackage{epsfig}
\usepackage{amsfonts}
\usepackage{amsmath}
\usepackage{amssymb}

\usepackage{color} 
  \newcommand{\beq}{\begin{equation}}
  \newcommand{\eeq}{\end{equation}} 
  \def\nuc#1#2{\relax\ifmmode{}^{#1}{\protect\text{#2}}\else${}^{#1}$#2\fi}
  \def\itnuc#1#2{\setbox\@tempboxa=\hbox{\scriptsize\it #1}
    \def\@tempa{{}^{\box\@tempboxa}\!\protect\text{\it #2}}\relax
    \ifmmode \@tempa \else $\@tempa$\fi}
  
\newcommand{\simge}{\hspace*{0.2em}\raisebox{0.5ex}{$>$}
     \hspace{-0.8em}\raisebox{-0.3em}{$\sim$}\hspace*{0.2em}}
\newcommand{\simle}{\hspace*{0.2em}\raisebox{0.5ex}{$<$}
     \hspace{-0.8em}\raisebox{-0.3em}{$\sim$}\hspace*{0.2em}}

\newcommand{\dslash}[1]{#1 \llap{/\kern-0.5pt}}
\newcommand{\Dslash}[1]{#1 \llap{/\kern+1.2pt}}
\newcommand{\DDslash}[1]{#1 \llap{/\kern+2.3pt}}
\newcommand{\dslashh}[1]{#1 \llap{/\kern+1pt}}

\def\bdm{\begin{displaymath}}
\def\edm{\end{displaymath}}

\begin{document}

\begin{titlepage}

\vspace*{0.75cm}

\begin{center}
{\Large\bf Les Houches Lectures on}
\\
\vspace{0.25cm}
{\Large\bf Effective Field Theories for}
\\
\vspace{0.25cm}
{\Large\bf Nuclear and (some) Atomic Physics}\\

\vspace{1cm}

{\large \bf U. van Kolck}

\vspace{0.3cm}
{
{\it Institut de Physique Nucl\'eaire, CNRS/IN2P3,
\\
Universit\'e Paris-Sud, Universit\'e Paris-Saclay,
\\
91406 Orsay, France}
\\
and
\\
{\it Department of Physics, University of Arizona,
\\
Tucson, AZ 85721, USA}
}

\vspace{0.75cm}
\today

\vspace{0.75cm}
{\it Dedicated to the memory of Professor C\'ecile DeWitt-Morette
\footnote{Professor DeWitt-Morette was a towering figure in 
mathematical physics who founded the Les Houches School.
I had the privilege of her guidance while studying stochastic
systems in my early graduate-student days at the University of Texas.
Like other great theorists, she did not look at physics
from the perspective of a specific field, but strove
to build a consistent view of nature.  
Besides an accomplished scientist, she was 
passionate about science, kind and supportive.
Near the end of my Ph.D. she recommended me to come to Les Houches,
but it took me a whole quarter of a century to get here. 
It is a sad twist of fate that, when writing this dedication,
I found out I had to add ``the memory of'' to it.}}

\end{center}

\vspace{0.4cm}

\begin{abstract}
These lectures are a pedagogical --- not comprehensive --- introduction
to the applications of effective field theory in the context of
nuclear and atomic physics. 
A common feature of these applications is the interplay
between nonperturbative physics (needed at leading order
to produce nonrelativistic bound states and resonances)
and controlled perturbative corrections (crucial for predictive power).
The essential ideas are illustrated
with the simplest nuclear EFT, Pionless EFT,
which contains only contact interactions
and, with minor changes, can be adapted to certain atomic systems.
This EFT exploits the two-body unitarity limit, where
renormalization leads to discrete scale invariance in systems of
three and more bodies.
Remarkably complex structures then
arise from very simple leading-order interactions.
Some of the challenges and rewards of including 
long-range forces --- pion exchange in Chiral EFT for nuclear systems 
or Van der Waals forces between atoms --- are briefly described.
\end{abstract}

\end{titlepage}

\section{Introduction}
\label{intro}

For us humans, who unlike frogs are born basically smaller
versions of ourselves at the time we have babies,
one of the most natural transformations of the physical
world is that of scale, where all distances change by a common
numerical factor. 
Despite its familiarity, scale invariance is usually not directly useful
in physics
--- we do not particularly resemble stars or molecules.
As we embark on an exploration of a physical problem, the first task is 
always to 
identify its relevant scales
\footnote{And, for convenience, simple units.
Since I am interested in problems that are rooted in quantum mechanics
and relativity, I use throughout these lectures
natural units where Planck's constant
and the speed of light are $\hbar=c=1$. Then distance and time have
the same units as inverse momentum, energy and mass. 
(To convert to more conventional units, use $\hbar c\simeq 200$ MeV fm.)}.

The existence of characteristic scales means that our resolution
when probing a system is important. 
Tracking changes in resolution is the task of the renormalization
group (RG). 
Once a resolution is chosen, we have at least two momentum scales
to contend with,
that of the physics of interest (call it $M_{\rm lo}$)
and that of the physics at much shorter distances 
(call it $M_{\rm hi}\gg M_{\rm lo}$).
Effective field theory (EFT) is the formalism to 
exploit this separation of scales 
and expand observable quantities
in powers of $M_{\rm lo}/M_{\rm hi}$ without making assumptions
about the short-range dynamics, other than its symmetries 
\cite{Manohar:2018aog}.
Schematically, the $T$ matrix for momentum $Q\sim M_{\rm lo}$ 
(from which the $S$ matrix and other observables can be obtained)
can be written as
\begin{equation}
T(Q)=
\sum_{\nu=0}^\infty \left(\frac{M_{\rm lo}}{M_{\rm hi}}\right)^\nu\,
F_\nu\!\left(\frac{Q}{M_{\rm lo}};\{\gamma^{(\nu)}\}\right)
\equiv \sum_{\nu=0}^{\infty} T^{(\nu)}(Q),
\label{Smatrixexp}
\end{equation}
where each $F_\nu$ is a calculable function parametrized by a finite set
of ``low-energy constants'' (LECs) or ``Wilson coefficients'' 
$\{\gamma^{(\nu)}\}$.
Clearly this is a paradigm to tackle any physical system,
as the variety of topics in this school attests to.
As far as we know, nature is but a pile of EFTs.

In these two lectures we will see how the paradigm fares when facing 
a very traditional field, nuclear physics, where it is hard to
come up with something that has not been tried before, 
usually without success.
The strong interactions encapsulated in QCD 
produce non-trivial structures
at the most fundamental level we can study today. Understanding
how hadrons and their bound states --- nuclei --- arise remains an 
open problem in the Standard Model of particle physics, which
hampers our ability to make predictions 
about processes involving new physics and astrophysical reactions.

It has been around 25 years since 
Weinberg \cite{Weinberg:1990rz,Weinberg:1991um}
and Rho \cite{Rho:1990cf}
proposed that EFT could
reproduce much of what was known in low-energy nuclear physics, 
while at the same time explaining some of its mysteries. 
Weinberg had earlier
articulated the EFT paradigm \cite{Weinberg:1978kz}
and was interested in general ways to set up the electroweak 
symmetry-breaking sector of the 
Standard Model \cite{Manohar:2018aog,Pich:2018ltt}.
Nuclear forces from pion exchange naturally come to mind
when pondering about how Goldstone bosons couple to matter.
The hope was that one would be able to formulate a renormalizable
theory of nuclear interactions, overcoming the 
obstacles faced since the 1950s.
``Chiral potentials'' constructed according to
Weinberg's suggestion have now become the favorite
input to ``{\it ab initio}'' methods to calculate nuclear structure
and reactions. In traditional nuclear physics, the {\it initio}
is a nonrelativistic potential among nucleons, yet 
the remarkable recent progress in Lattice QCD \cite{Beane:2010em}
means that soon
the starting point will be QCD itself \cite{Barnea:2013uqa,Contessi:2017rww}.

Unfortunately these chiral potentials produce scattering amplitudes
that do not respect RG invariance.
In the process of discovering and fixing this problem, 
other nuclear EFTs have been formulated, which apply
to various energy regimes.
Weinberg's basic insights have survived but building nuclear
EFT turned out to be much more interesting than he anticipated.
Some of the issues and advantages of explicit pion exchange are mentioned
at the end of these notes, but
I cannot possibly cover all the twists and turns of this story, not even all
nuclear EFTs. 
I will focus instead on the simplest one, Pionless EFT, and refer to
others only occasionally. 
Pionless EFT is so simple, in fact, that with small
changes it can be applied to systems of cold, neutral atoms as well.
These lectures serve also as a (somewhat idiosyncratic)  
introduction to some of 
the physics that reinvigorated the atomic field over roughly the
same time frame.
I will make no attempt to provide an extensive coverage of the literature;
only papers that are best suited to make specific points are cited.

Pionless EFT is sufficient for a sample taste of nuclear EFTs.
In contrast to many of the other lectures in the school, here we shall deal with
a situation where the leading order (LO) of the $M_{\rm lo}/M_{\rm hi}$ expansion
must be non-perturbative in order to generate poles of the $S$ matrix:
the bound states and resonances that we identify as nuclei (or molecules).
The combination of non-perturbative LO and perturbative corrections
--- relative ${\cal O}(M_{\rm lo}/M_{\rm hi})$ at next-to-leading order (NLO),
relative ${\cal O}(M_{\rm lo}^2/M_{\rm hi}^2)$ at N$^2$LO, {\it etc.} ---
is at the core of the beauty of the nuclear and atomic applications of EFT.
Pionless EFT is the ``poster EFT'' to describe such combination.
There is much regularity in the properties of nuclear and
atomic bound states and resonances,
and Pionless EFT captures a class of these regularities
that sometimes goes by the name of ``Efimov physics''.
In a magical paper almost half-a-century ago \cite{Efimov:1970zz}, 
Efimov showed that for certain nonrelativistic systems,
if a two-body bound state 
lives on the verge of non-existence, then a geometric tower of
three-body bound states exists,
with the ground state potentially quite deep.
We now know that this phenomenon is not limited to the three-body system,
but reverberates through larger clusters and even ``infinite'' matter.
As we will see, scale invariance is the key to understand 
this sort of structure.

\section{Some nuclear and atomic scales}
\label{scales}

A cursory look at the Particle Data Book \cite{Tanabashi:2018oca}
shows a bewildering variety of hadrons, which nevertheless
fall into isospin multiplets containing various charge
states with approximately the same mass. When made out of light quarks,
they have masses in the 1, 2 GeV range,
such as the proton and the neutron that can be paired in an
isospin doublet of mass $m_N\simeq 940$ MeV. 
We can infer that QCD has a characteristic scale 
$M_{\rm QCD}\sim 1$ GeV associated with its nonperturbative dynamics.

The one clear exception is the isospin triplet of light pions,
with mass $m_\pi\simeq 140$ MeV $\ll M_{\rm QCD}$. 
This low mass has long been understood
as the result of the spontaneous breaking of an approximate 
SU(2)$_{\rm L}\times$SU(2)$_{\rm R}\sim$ SO(4) chiral symmetry 
of independent rotations 
in the space of two flavors for left- and right-handed quarks. 
Because the diagonal subgroup SU(2)$_{\rm L+R}\sim$ SO(3) of isospin rotations
remains unbroken,
the three Goldstone bosons
in the coset space SO(4)/SO(3) $\sim S^3$ 
can be identified with the three pions.
Their interactions are governed by a dimensionful parameter,
the pion decay constant $f_\pi\simeq 92 \; \mathrm{MeV}\sim M_{\rm QCD}/(4\pi)$,
which is the radius of this ``chiral sphere''.
Pions are not exactly massless because
chiral symmetry is explicitly broken by the quark masses.
From perturbation theory one expects 
$m_\pi^2\sim M_{\rm QCD} \bar{m}$, 
which is in the right ballpark if the average quark mass $\bar{m}$ is a few 
MeV.
Isospin is broken by the down-up quark mass difference 
$\varepsilon \bar{m}$, where $\varepsilon \sim 1/3$ \cite{Weinberg:1977hb},
which gives rise to small splittings among isospin multiplets.
Of course, isospin is also broken explicitly by electromagnetic interactions.
All this has been understood for quite some time
and is discussed in some detail in Pich's lectures \cite{Pich:2018ltt}.

What might be somewhat surprising is that once quarks hug tight into nucleons
and pions with intrinsic sizes $\sim M_{\rm QCD}^{-1}\simeq 0.2$ fm,
nucleons form nuclei of much larger size and feebler binding. 
One proton and one neutron
bind into an isospin singlet --- the simplest nucleus, the deuteron ($^2$H) ---
with total spin $S=1$ and a binding energy $B_d\simeq 2.2$ MeV.
When $S=0$, proton and neutron are part, with two 
neutrons and two protons, 
of an isospin-triplet virtual state at $B_{d^*}\simeq 0.08$ MeV.
(A bound state is a pole of the $S$ matrix with positive imaginary
momentum, while a virtual state is a pole with negative imaginary
momentum --- both have negative energy $-B$.
A resonance consists of a pair of poles in the lower half of the
complex-momentum plane with opposite, nonzero real parts.) 
As shown in the leftmost column of Table \ref{tbl:BEs},
the binding energy per particle $B_A/A$ increases for the trinucleon isodoublet
with $S=1/2$ --- triton ($^3$H) and a slightly less-bound helion ($^3$He) ---
and even more for the isospin-singlet alpha particle ($^4$He) with $S=0$. 
The exclusion principle makes it harder for more nucleons to be together
and $B_A/A$ first decreases, but then increases again (on average) till 
it reaches a maximum of about 8 MeV for $A=56$, decreasing slowly beyond that. 
Table \ref{tbl:BEs} also shows the value of the binding energy per particle
for nuclear matter, $\lim_{A\to \infty} (B_A/A)\equiv b_\infty$.
Nuclear matter is an idealized system without surface or 
electroweak interactions,
defined by extrapolation from heavy nuclei
via the ``liquid-drop'' relation $B_A/A -b_\infty\propto A^{-1/3}$.
Consonant with such ``saturation'',
the typical nucleus size is $\sim r_{\rm nuc} A^{1/3}$, 
with $r_{\rm nuc}\simeq 1.2$ fm
--- but it can be much larger for more loosely bound nuclei,
such as light nuclei and ``halo nuclei'', which have 
a cloud of loose nucleons orbiting a more tightly bound core.
(The typical example is $^6$He, which is thought to be essentially
two neutrons around a $^4$He core.)
Halo EFT \cite{Bertulani:2002sz,Bedaque:2003wa} provides a description 
of this type of state similar to Pionless EFT, but
with the core as an additional degree of freedom. 

\begin{table}
\begin{center}
  \caption{Ground-state (first-excited) binding energies per
particle $B_A/A$ 
($B_{A^\star}/A$) of selected light nuclei and $^4$He atomic clusters,
in units of the three-body binding energy per particle,
where $B_3=8.48$ MeV and $B_3=0.1265$ K, respectively.
(To convert between K and eV, use $k=8.6 \cdot 10^{-5}$ eV K$^{-1}$.)
In the nuclear case, an entry corresponds to the deepest isobar state
with the respective nucleon number $A$,
a parenthesis indicating a virtual state.
Entries in the left column are experimental.
Entries in the middle and right columns are from Pionless EFT at LO:
from Refs. \cite{Bedaque:1999ve,Stetcu:2006ey,Contessi:2017rww,Rupak:2018gnc} 
away from unitarity
and from Ref. \cite{Konig:2016utl} at unitarity.
For $^4$He atomic clusters, the left column shows the results
\cite{kalos1981,BluGre00,HiyKam12a}
of calculations with phenomenological $^4$He-$^4$He potentials.
Entries in the middle and right columns are from Pionless EFT at LO:
from Refs. \cite{Bedaque:1998km,Platter:2004he,Bazak:2016wxm} 
away from unitarity
and from Refs. \cite{Bedaque:1998km,Deltuva:2010xd,Carlson:2017txq} 
at unitarity.
For simplicity I do not indicate 
quantities used in the construction of the interactions nor
show error estimates, which can be found in the original references.
For some entries similar numbers exist from other references.}
\vspace{0.3cm}
{\renewcommand{\arraystretch}{1.25}
\begin{tabular}
{c||c@{\hspace{3mm}} c@{\hspace{3mm}}c|c@{\hspace{3mm}}c@{\hspace{3mm}} c|}
\multicolumn{1}{c}{$3B_A/(AB_3)$ }&
\multicolumn{3}{|c|}{nucleons}&
\multicolumn{3}{|c|}{$^4$He atoms} \\
\hline 
$A$ & experiment & LO EFT & unitarity & potential & LO EFT &unitarity \\ 
\hline
\hline
$2$       & 0.39 & 0.39 & 0      & 0.0156 & 0.0152 & 0  \\
$3$  &  1 & 1  & 1 & 1 & 1 & 1 \\
$3^\star$  & (0.17) & (0.19) & 0.0019 & 0.0180 & 0.0175 & 0.0019  \\
$4$       & 2.50   & 2.6   & 3.5    & 3.3    & 3.2    & 3.46 \\
$4^\star$  & 0.71   & 0.90  & 0.75   & 0.755  & 0.759  & 0.752 \\
$5$       & 1.94   & ?     & ?      & 6.2    & 5.7    & 6.3 \\
$6$       & 1.89   & 1.4   & ?      & 9.2   & 8.2   & 8.9 \\
$\vdots$ &&&&&&\\
$16$      & 2.82 & 2.5 & ?       &  ?    & ?   & 27.4  \\
$\vdots$ &&&&&&\\
$\to \infty$   & 5.7 & ?   & ?       & 180    & ?      &  90
\\
\hline
\end{tabular}}
\label{tbl:BEs}
\end{center}
\end{table}

This disparity between QCD and nuclear scales tells us several things. 
First, nucleons are relatively far apart inside nuclei and retain their 
identity.
Second, they move slowly, that is, are approximately nonrelativistic. 
Third, their interactions must come from the exchange of the lightest 
color singlets --- the pions, giving rise to
a force of range $\sim m_\pi^{-1}\simeq 1.4$ fm ---
plus complicated mechanisms of shorter range $\sim M_{\rm QCD}^{-1}$.
Nuclei should thus be described by an
EFT with nonrelativistic nucleons and pions
(and possibly the lightest nucleon excitations)
subject to approximate chiral symmetry --- this is Chiral EFT,
whose restriction to $A=0, 1$ is 
Chiral Perturbation Theory (ChPT) \cite{Weinberg:1978kz,Pich:2018ltt}.
Moreover, in the long distances relevant for light nuclei,
even pion exchange can be considered a short-range interaction.
This is the regime of Pionless EFT,
where the only explicit degrees of freedom are 
nonrelativistic nucleons.

The situation is not totally dissimilar for some (neutral) 
{\it atoms}, like 
the boson $^4$He. Instead of QCD binding quarks into hadrons, 
QED binds a nucleus and $Z$ electrons into atoms with
energies $E_{\rm at}\sim -(Z\alpha)^2 m_e$, where $\alpha\simeq 1/137$
is the fine-structure constant and $m_e$ the electron mass. 
The atoms themselves form much more loosely bound molecules,
through the exchange of (at least) two photons and shorter-range interactions.
The former gives rise to the Van der Waals potential 
$\sim -l_{\rm vdW}^4/(2\mu r^6)$,
where $\mu$ is the reduced mass and $l_{\rm vdW}$ is the 
``Van der Waals length'', which depends on what kind of atom we are considering.
For a nice compilation of values, as well as a discussion of long-range
interactions, see Ref. \cite{Cordon:2009wh}.
For certain types of atoms, 
clusters have sizes that are
significantly larger than $l_{\rm vdW}$; in these cases,
the short-range interactions dominate and we can, up to
a point, treat the system as one with short-range forces only.
This Contact EFT is analogous to Pionless EFT, just
with a field for the atom substituted for the nucleon's.

$^4$He is a particularly interesting atom
because a macroscopic sample remains liquid at zero temperature and
exhibits the remarkable property of superfluidity.
The $^4$He dimer has been measured to have
an average separation $\langle r \rangle = 52(4)$ \AA~\cite{GriSchToe00},
which is an order of magnitude larger than the corresponding
$l_{\rm vdW}\simeq 5.4$ \AA~\cite{YanBabDal96,ZhaYanVri06}. 
Experimental numbers exist for the dimer \cite{CenPrzKom12,Zel16}
and excited-trimer \cite{Kunitski:2015qth} binding energies,
which confirm that they are loosely bound.
Sophisticated $^4$He-$^4$He potentials have been developed over the years,
which are consistent with a variety of experimental data
and allow for the prediction of the energies of larger clusters.
The results for one of these potentials 
--- dubbed the ``LM2M2 potential'' \cite{AziSla91} ---
are shown in the left $^4$He column of Table \ref{tbl:BEs}. 
Other potentials give similar results.
Apart from a huge difference in overall scale, the numbers for $A\le 4$ atoms 
have some qualitative similarity to those for nucleons.
The most obvious difference is the 
lack of exclusion principle, which translates into a monotonic 
increase in binding energy, which starts approximately as 
$(N-2)^2 B_3$ \cite{Bazak:2016wxm}. 
Just as for nucleons, though, the interaction saturates 
to a constant binding energy per particle,
the value of which was calculated with another potential
--- the ``HFDHE2 potential'' \cite{aziz1979} --- 
in Refs. \cite{kalos1981,vijay1983}.
Similarly to nuclei, cluster size first decreases, then starts to
increase till it settles into an $\sim r_{\rm at} A^{1/3}$ behavior,
with $r_{\rm at}\simeq 2.2$  \AA~\cite{vijay1983}.

These two families of systems --- nuclei and $^4$He atomic clusters ---
are peculiar for their large sizes compared
to the interaction range $R$. 
A rough estimate of the characteristic momentum 
$Q_A$ of the particles in the bound state is obtained by assuming that every
one of the $A$ particles of mass $m$ contributes the same energy $Q_A^2/(2m)$ 
to $B_A$:
\begin{equation}
Q_A\sim \sqrt{2m B_A/A}. 
\label{bindingmom}
\end{equation}
(This estimate reflects the correct location of the bound state in relative
momentum for $A=2$ and gives a finite $Q_A$ in the limit $A\to \infty$.)
For nucleons, we find $Q_3\sim 70$ MeV, while for $^4$He atoms
$Q_3\sim (12$ \AA$)^{-1}$, each one about half of the corresponding $R^{-1}$.
That is, 
particles in the three-body bound state are separated by a distance
about twice as large as the range of the interaction.
How can this be? Classically, the size of an orbit is given by the range
of the force. Thus, these are intrinsically quantum-mechanical systems,
which hold a few surprises in store for us.

\section{EFT of short-range forces}
\label{EFTSRF}

In the class of systems sometimes referred to as ``quantum halos'', 
we are interested in the $S$ matrix for processes with a typical external 
momentum $Q\ll M_{\rm hi}$, where the EFT breakdown scale $M_{\rm hi}$ is
related to the inverse of the force range, $R^{-1}$
--- $m_\pi$ or $l_{\rm vdW}^{-1}$, as the case may be.
The few-body structures we want to describe are characterized
by momenta $Q\sim Q_3$, so I will use $Q_3$ as a proxy for $M_{\rm lo}$.
The idea is to construct an expansion of the form \eqref{Smatrixexp} from 
the most general Lagrangian (density)
allowed by the symmetries supported by the relevant degrees of freedom,
exploiting the ``folk theorem'' that the resulting quantum field theory
will then generate the most general $S$ matrix allowed by the same 
symmetries \cite{Weinberg:1978kz}.

\subsection{Degrees of freedom}
\label{degrees}

In systems whose sizes are larger than the range $R$ of the force,
the constituent particles are not able to resolve
details of the potential. They feel the interaction as a contact:
the potential can be represented by a Dirac delta function and 
its derivatives. 
The only degrees of freedom we need to consider
are the constituent particles themselves --- there is no need to account 
explicitly for other particles whose exchanges are responsible for the specific
form of the potential. 

Since in the cases of interest here the particle mass $m \simge M_{hi}$, 
a nonrelativistic expansion must hold, 
which yields a much simpler field theory than in the relativistic case.
Because it takes $\simge 2m$ in energy to create a particle-antiparticle
pair, there is no need to include antiparticles
--- unless we are interested
in processes that include antiparticles to start with,
or in processes that involve particle- (baryon- and/or lepton-) number 
violation.
We can exploit this simplification by
a convenient choice of field. 
(I remind you that fields
are not directly observable and in an EFT all choices that represent the same 
degrees of freedom are equivalent \cite{Chisholm:1961tha,Kamefuchi:1961sb}.)
It is sufficient to employ a field $\psi_a(x)$ that only annihilates
particles,
\begin{equation}
\psi_a(x) \equiv\int\!\!\frac{d^3p}{(2\pi)^3}\, 
\frac{e^{-i p\cdot x}}{2p^0} \,
u_s(\vec{p}) \, a_{\vec{p}}
\label{partcreat}
\end{equation}
where $a_{\vec{p}}$ is the annihilation operator
for a particle of momentum $\vec{p}\equiv \vec{k}$,
$u_s(\vec{p})$ carries information about the spin, and
$p^0=\sqrt{\vec{p}^{\;2}+m^2}=m +\vec{p}^{\;2}/(2m)-\vec{p}^{\;4}/(8m^3)+\ldots
\equiv m +k^0$
(I am sorry, I am using the ``wrong'', West Coast metric here:
$p\cdot x=p^0 t-\vec{p}\cdot\vec{r}$.)
Propagation is only forward in time, and it is represented in Feynman
diagrams by a line going up (when time is represented as increasing upwards,
as I do here).
The absence of pair creation means that, if particle number is 
conserved, these lines go through the diagram.
The theory breaks into separate sectors, each with
a given number of particles $A$.
Life is made tremendously easier
by the fact that we can deal in turn with sectors of increasing $A$ without 
simultaneously having to worry about the feedback from larger number 
of particles.

To produce an $M_{\rm lo}/M_{\rm hi}$ expansion, we expect some form of derivative
expansion in the action. Yet derivatives of $\psi_a(x)$ contain the large
$m$ coming from the trivial factor $\exp(-imt)$ in the evolution of the
field.
In the particle's rest frame,
where the particle's four-velocity $v^\mu$, $v^2=1$, is $(1,\vec{0})$
it is convenient to remove this factor 
by defining a new ``heavy field'' \cite{Georgi:1990um}
\begin{equation}
\psi_h(x)\equiv 
e^{im v \cdot x} \psi_a (x).
\label{fieldredef1}
\end{equation}
The evolution of the new field is governed by the kinetic
energy $k^0\ll m$, since particles
will exchange small three-momenta.
Instead of the $-i p_\mu$ that contains the large $m$,
$\partial_\mu \psi_h(x)$  gives $-i k_\mu$
in momentum space.

For cases where antiparticles are present, a conjugate field
can be introduced similarly. 
Of course the possibility of pair annihilation when both particles and 
antiparticles are present allows for large momenta, and the applicability
of a $Q/m$ expansion is limited to times prior to annihilation.

\subsection{Symmetries}
\label{symmetries}

So far, no sign has been seen of violation of Lorentz invariance or
the product CPT of charge conjugation (C), parity (P), and time reversal (T).
For simplicity I will also neglect the small effects that arise from 
the violation of P, T, and baryon and lepton numbers, although they are part 
of the EFT.
All these symmetries restrict the form of the terms allowed in the action.
For example, the transformation associated with particle number is
an arbitrary phase change
\begin{equation}
\psi_h (x)\to  e^{i\alpha} \psi_h(x),
\label{partnumb}
\end{equation}
and particle-number conservation implies that $\psi_h$ enters the Lagrangian
in combination with $\psi_h^\dagger$.

For a heavy field,
the symmetry whose implementation is least obvious is Lorentz invariance.
To start with, note that
the definition of $\psi_h$ can be made in other frames where 
$v^\mu$ is not necessarily $(1,\vec{0})$. 
We introduce the residual momentum $k^\mu$
through $p^{\mu}\equiv mv^\mu +k^\mu$.
The residual momentum is only constrained by $k^2=-2m v\cdot k\ll m^2$.
There is freedom to consider a different 
residual momentum $k^\mu-q^\mu$ with $q^2=-2m v\cdot q\ll m^2$, 
and a simultaneous relabeling of the velocity,
$v^\mu\to v^\mu+q^\mu/m$. 
(You can check that $(v+q/m)^2=1$). 
Thus the theory must be invariant under this
``reparametrization invariance'' (RPI) \cite{Luke:1992cs}, 
when the field transforms as 
\begin{equation}
\psi_h (x)\to  e^{iq \cdot x} D(v+q/m, v) \, \psi_h(x),
\label{reparam}
\end{equation}
where $D$ is determined by the 
representation of the Lorentz group encoded in $u_s$.
Because of the phase, $\partial_\mu \psi_h(x)$ is not covariant
under reparametrization. 
As standard in such cases, we can introduce a
``reparametrization covariant derivative''
\begin{equation}
{\cal D}_\mu \psi_h(x)\equiv \left[\partial_\mu -imv_\mu\right]\psi_h(x)
\to  e^{iq \cdot x} D(v+q/m, v) \, {\cal D}_\mu \psi_h(x).
\end{equation}
We account for Lorentz invariance by properly contracting Lorentz indices
in the Lagrangian as usual, 
but using the reparametrization covariant derivative.

As an example, take the kinetic terms of a scalar field, 
whose RPI form is
\begin{equation}
{\cal L}_{\rm kin}=
({\cal D}^\mu \psi_h)^\dagger {\cal D}_\mu \psi_h
- m^2 \psi_h^\dagger\psi_h
= 2m \, \psi_h^\dagger \left[iv\cdot \partial - \partial^2/(2m) \right]\psi_h.
\label{Lkin1}
\end{equation}
The large mass term has disappeared, as desired, except for an overall
normalization.
The remaining terms resemble the inverse
of the usual nonrelativistic propagator, but contain also
$\partial_0^2\psi$.
To bring the propagator to the usual form, we can
define yet another field, which is canonically normalized:
\begin{equation}
\psi (x)\equiv \frac{1}{\sqrt{2m}}
\left[ 1 - \frac{i }{4m} v\cdot \partial
+ \ldots\right] \psi_h(x).
\label{fieldredef2}
\end{equation}
Substitution into Eq. \eqref{Lkin1} leads to
\begin{equation}
{\cal L}_{\rm kin}=\psi^\dagger \left\{iv\cdot \partial 
+\frac{1}{2m}\left[\left(v\cdot \partial\right)^2-\partial^2\right] 
+\ldots \right\}\psi.
\label{Lkin2}
\end{equation}
Now in the rest frame the term in square brackets reduces to the usual 
$\vec{\nabla}^2$.

\vspace{0.2cm}
\noindent
{\bf Exercise:} Consider the higher orders in $1/m$ in 
the ``$\ldots$'' of Eq. \eqref{fieldredef2}. Show that an appropriate 
field choice makes the next term
in the ``$\ldots$'' of Eq. \eqref{Lkin2} 
equal to $\vec{\nabla}^4/(8m^3)$ 
in the particle's rest frame, in line with
$k^0=\vec{k}^{\,2}/(2m)-\vec{k}^{\,4}/(8m^3)+\ldots$
\vspace{0.2cm}

The same procedure can be followed for interaction terms.
As the number of derivatives increase, we have to contend
with two or more time derivatives. 
In this case, there is a nontrivial
relation between the time derivative of the field and 
the field's conjugate momentum. 
As a result the interaction Hamiltonian is
not simply minus the interaction Lagrangian, 
but contains additional, Lorentz noncovariant
terms. It is no problem to include these interactions, 
as other noncovariant pieces arise in
covariant perturbation theory from contractions involving derivatives, 
and in a time-ordered formalism from its inherent noncovariance.
One can show explicitly \cite{Lee:1962vm} that the sum of
diagrams contributing to any given process is indeed covariant.
Alternatively, one can integrate over the field's momentum in the 
path integral arriving at an effective Lagrangian, which 
is covariant but has additional terms compared
to the classical Lagrangian we started from 
\cite{Salam:1971sp}.
For heavy fields these complications can be avoided altogether
by including in the ``$\ldots$'' of Eq. \eqref{fieldredef2}
terms with additional fields, designed to
remove $\partial_0\psi$ from interactions as well.

Accounting for a nonzero spin manifest in
the $D(v+q/m, v)$ in Eq. \eqref{reparam} is a bit more complicated.
To represent only particles, the fields obey constraints.
For example, of the four components of a Dirac spinor
only two are associated with the two spin states of
a particle with $S=1/2$. In the rest frame 
one can project onto the two upper 
(in the Dirac representation) components with the projector $(1+\gamma^0)/2$
and thereby use a Pauli spinor,
the gamma matrices reducing to the spin operator $(0, \vec{\sigma}/2)$.
In a generic frame the two ``large'' components can be selected
by the constraint $(1- \slash \!\!\!v)\psi=0$.
Spin properties are encoded in 
$S^\mu = i \gamma_5\sigma^{\mu\nu} v_\nu/2$, which satisfies 
$S\cdot v=0$ and $S^2=-3/4$, as a little gamma-matrix algebra shows.
The ``reparametrization covariant spin'' that appears in interactions is then
$\Sigma^\mu= - \gamma_5\sigma^{\mu\nu}{\cal D}_\nu/(2m)$,
which satisfies 
$\psi^\dagger \Sigma\cdot {\cal D}\psi =0$ and 
$\psi^\dagger\Sigma^2\psi=3\psi^\dagger {\cal D}^2\psi/(4m^2)$. 

At the end of the day, we simply end up with a Lagrangian that 
complies with Lorentz invariance in a $Q/m$ expansion.
Of course other, perhaps simpler, ways exist to implement 
the constraints of Lorentz invariance, for example an explicit reduction
of a fully relativistic theory.
Heavy fields with RPI are, however, more EFT-like, 
because they do not rely on detailed knowledge of the theory at $Q\sim m$.
This is particularly relevant in nuclear physics,
where, if fully relativistic nucleons
are to be considered, we need also to include mesons heavier than the pion,
for which no systematic $Q/M_{\rm hi}$ expansion is known.
RPI is thus well adapted to the expansion in $Q/M_{\rm hi}$ we seek.
More details about the heavy field formalism can be found in
the lectures by Mannel at this school \cite{Mannellectures}.
In the following, for simplicity, I will drop the interactions
that are explicitly spin-dependent and  work in the 
rest frame.

\subsection{The action}
\label{action}

The most general action for nonrelativistic particles of mass $m$ interacting
under these symmetries through short-range forces can therefore be written as
\begin{eqnarray}
{\cal S} = \int \!\! d^4x \; {\cal L} &=&\int \!\! \frac{dt}{2m} \int\!\! d^3r 
\left\{
\psi^\dagger \left( 2im\frac{\partial}{\partial t} 
+ {\vec{\nabla}}^2 
+\ldots\right)\psi
\right.
\nonumber \\
&& \left.
\qquad\qquad\qquad - 4\pi 
\left[C_0 \left(\psi^\dagger \psi \right)^2
+C_2 \left(\psi^\dagger \psi \right)
\left(\psi^\dagger {\vec{\nabla}}^2\psi +{\rm H.c.}\right)
+\ldots\right]
\right.
\nonumber \\
&& \left.
\qquad\qquad\qquad - \frac{(4\pi)^2}{3}D_0 \left(\psi^\dagger \psi \right)^3
+\ldots
\right\},
\label{L}
\end{eqnarray}
where 
$C_{0,2}$, $D_0$, {\it etc.} are the LECs
and ``\ldots'' include terms with more derivatives and/or fields.
Beware that I chose a normalization of the LECs that is not normally
used in the literature, but is very convenient for the subsequent
discussion of orders of magnitude and scale invariance.
With this choice, the mass appears in the combination $t/m$, and in observables
together with the energy $E$ as $mE \equiv k^2+\ldots$. 
As a consequence, binding energies are all $\propto 1/m$.
This is at the root of a choice of units frequently
made in atomic physics: $m=1$. I prefer to keep $m$ explicit instead.

In writing Eq. \eqref{L} I neglected spin projections in the interaction terms.
For fermions, we can use Fierz reordering,
which encodes the exclusion principle, to reduce the number
of independent terms.
For a two-state fermion ({\it e.g.} a neutron with spin up or down,
in the absence of protons), the $C_0$ interaction
operates only when the two particles are in different states.
Similarly,
the $D_0$ interaction, which requires three particles at the same point,
vanishes.
For more-state fermions, more than one $C_0$- 
and/or $D_0$-type interaction is possible, depending on the symmetries.
For example, because a nucleon consists of four states
--- isospin up (proton $p$) and down (neutron $n$), 
each with two spin states ---
four $C_0$-type interactions exist,
corresponding to the four $S$-wave channels: 
the isospin-singlet $pn$ channel with spin $S=1$ --- the deuteron channel ---
and the $pp$, $pn$, and $nn$ interactions in
the isospin-triplet channels with $S=0$.
In situations where isospin is a good approximate symmetry,
these interactions are reduced in a first approximation to two,
the one operating in the $^3S_1$ channel, and a single one in $^1S_0$.
(We commonly use the spectroscopic notation $^{2S+1}L_J$ where $S$ is the spin,
$L$ is the orbital angular momentum, and $J$ the total angular momentum.)
Similarly, there is a single
$D_0$-type interaction, operating in the nucleon-deuteron 
($Nd$) $^2S_{1/2}$ wave --- the triton/helion channel. 
(In the $^4S_{3/2}$ channel there are at least two protons
or two neutrons in the same spin state.)

\vspace{0.2cm}
\noindent
{\bf Exercise:} Show that Fermi statistics and isospin symmetry 
imply that two 
nucleons interact in two $S$ waves: 
{\it i)} $^3S_1$, which is symmetric in spin and antisymmetric in isospin, and
thus involves an $np$ pair; and
{\it ii)} $^1S_0$, which is antisymmetric in spin and symmetric in isospin, and
is split by isospin-violating interactions into $pp$, $pn$, and $nn$.
Analyze the two $S$ waves for the $Nd$ system, $^4S_{3/2}$ and $^2S_{1/2}$,
similarly.
\vspace{0.2cm}

I have not considered electromagnetic 
interactions explicitly in Eq. \eqref{L}.
The associated U(1)$_{\rm e}$ gauge symmetry
can be introduced in the usual way, that is, by requiring that
all interactions be built out of the electromagnetic covariant derivative
and the field strength. 
In the neutral-atom case, the covariant derivative is just the
usual derivative, but electromagnetic interactions still proceed through
the coupling of the atom to the field strength.
The longest-range effects arise
from two-photon exchange, the most important being the Van der Waals force.
By remaining at momenta below $l_{\rm vdW}^{-1}$, 
dimensional analysis shows that the non-analytic contributions 
from two-photon exchange enter only at ${\cal O}(Q^3)$, 
so at lower orders we can pretend to have only short-range interactions.
In the nuclear case, where the proton
is charged, one-photon exchange leads to long-range interactions, 
the most important being the well-known Coulomb force.
There is further isospin violation from shorter distances
stemming from ``hard'' photons and from the quark mass difference. 
The associated isospin violation splits various charge states. 
For nuclear ground states one can argue \cite{Konig:2015aka,Konig:2016iny} 
that isospin-breaking effects are subleading and bring no fundamental changes 
to the discussion below.

To keep the notation lean, I will ignore spin-isospin complications 
in the formulas that follow,
and pretend that only one short-range interaction of each type is important.
I will simply remark on the changes that take place when $S> 0$.

\subsection{Renormalization-group invariance and power counting}
\label{RGinvandPC}

As in any EFT, the interactions in the action \eqref{L} are local and
require regularization, that is, the introduction of a method
to suppress explicit high-momentum, or equivalently short-distance, modes.
There is an infinite number of ways of doing this.
What is common to all methods is the presence of a parameter with
dimensions of mass, which I will denote $\Lambda$.
In perturbation theory, frequently the cleanest method is dimensional
regularization because it keeps no terms that go as negative powers of
$\Lambda$.
(In most subtractions it keeps also no positive powers.)
However, dimensional regularization is difficult (impossible?) to
apply in generic nonperturbative contexts,
where we are restricted (at least for now) to momentum 
or distance regulators. 
The simplest examples of these regulators are, respectively,
a sharp momentum cutoff $\Lambda$  
and a minimum length, or lattice spacing, 
$\Lambda^{-1}$. More generally, one can use a function 
of the momentum $\vec{p}$
which vanishes smoothly as $|\vec{p}\,|$ increases beyond $\Lambda$,
say a Gaussian function of $|\vec{p}\,|/\Lambda$ with unit coefficient,
or a smooth representation of a delta function in coordinate space $\vec{r}$,
say a Gaussian function of $|\vec{r}\,|\Lambda/2$ with a coefficient that scales
as $\Lambda^3$.
{}From here on I will talk mostly about a momentum cutoff,
with only occasional remarks about dimensional regularization.

The effects of the high-momentum modes must, of course, reappear
in the LECs, which are thus dependent on $\Lambda$.
Renormalization is the process of ensuring that this dependence
is such that observables are
independent of the arbitrary regulator, or
\begin{equation}
\frac{dT(Q)}{d\Lambda}=0.
\label{RGinv}
\end{equation}
That this must be possible is a consequence of the uncertainty principle
coupled to our accounting of {\it all} interactions allowed by
symmetries: 
modes of momentum $\simge \Lambda$ can be absorbed in interactions of
range $\simle \Lambda^{-1}$ with the same symmetries. 
If $\Lambda\simge M_{\rm hi}$, this adds no extra limitation
to the EFT. If $\Lambda\simle M_{\rm hi}$, we cannot apply the EFT
all the way to its physical breakdown scale.

A central role is played in EFT by ``power counting'', that is,
a rule that organizes the infinite sequence of interactions according 
to their relevance to observables
by relating the counting index $\nu$ in Eq. \eqref{Smatrixexp} to
the properties of the various terms in the action
(number of derivatives, fields, {\it etc.}).
To preserve the insensitivity to the regulating procedure, we want each
truncation of Eq. \eqref{Smatrixexp}, 
\begin{equation}
T^{[\bar \nu]}(Q)\equiv \sum_{\nu=0}^{\bar \nu} T^{(\nu)}(Q)
\label{Smatrixtrunc}
\end{equation}
to satisfy
\begin{equation}
\frac{\Lambda}{T^{[\bar \nu]}(Q)}\frac{dT^{[\bar \nu]}(Q)}{d\Lambda}=
{\cal O}\!\left(\frac{Q}{\Lambda}\right).
\label{RGinvtrunc}
\end{equation}
In this way, the regulator does not increase the truncation error,
as long as $\Lambda\simge M_{\rm hi}$.
This places a constraint on the power counting: that it contains
enough interactions at each order to remove non-negative powers
of $\Lambda$ from observables.
If we do not include all interactions needed to ensure Eq. \eqref{RGinvtrunc},
observables are sensitive to
the arbitrary regulator --- not only its dimensionful parameter but 
also its form --- and there is no guarantee that results
reflect the low-energy limit of the underlying theory.
The canonical dimension of a LEC suggests a lower bound on its magnitude,
when we use $M_{\rm hi}$ to make it dimensionless and ${\cal O}(1)$.
However, as we are going to see, our problem involves significant 
departures from this simple dimensional analysis,
in which case RG invariance
in the form \eqref{RGinvtrunc} offers particularly useful guidance regarding
the LEC sizes. 
Regularization schemes --- such as dimensional regularization
with minimal subtraction ---
that kill positive powers of $\Lambda$ are not the most useful
in this context, because they might lead you to overlook the need 
for a LEC at the order under consideration.
(A physically relevant example is given in Ref. \cite{Bertulani:2002sz}.)

This approach is a generalization of the old concept of 
``renormalizable theory'',
where renormalization is achieved by a finite number of parameters
and at the end $\Lambda\to \infty$ is taken.
As Weinberg is fond of saying (see, {\it e.g.}, Ref. \cite{Weinberg2018}), 
``nonrenormalizable theories'' are just as renormalizable as
``renormalizable theories''.
Now we only need a finite number of parameters {\it at each order}
and $\Lambda\simge M_{\rm hi}$.
Each of the LECs obeys an RG equation
that tells us how it depends on $\Lambda$.
A generic observable will look like
\begin{equation}
O^{[\bar \nu]}(\Lambda)= O^{[\bar \nu]}(\infty) 
\left[1 + \alpha_O^{[\bar \nu]} \frac{M_{\rm lo}}{\Lambda} 
+ \beta_O^{[\bar \nu]}  \frac{M_{\rm lo}^2}{\Lambda^2} +\ldots \right],
\label{obscutdep}
\end{equation}
with $\alpha_O^{[\bar \nu]}$, $\beta_O^{[\bar \nu]}$, {\it etc.} 
numbers of ${\cal O}(1)$.
Values for a finite number of observables are used to determine the LECs
--- for these observables $\alpha_O^{[\bar \nu]} =\beta_O^{[\bar \nu]} =\ldots=0$. 
For other observables, a nonzero $\alpha_O^{(\bar \nu)}$ indicates the existence 
of at least one interaction at next order, since it generates 
a term of size ${\cal O}(M_{\rm lo}/M_{\rm hi})$ when $\Lambda \sim M_{\rm hi}$.
There is no need to take $\Lambda\to \infty$ for the theory
contains in any case errors of ${\cal O}(M_{\rm lo}/M_{\rm hi})$ from 
the truncation in the action.
But regulator cutoff variation from $\sim M_{\rm hi}$ to much larger values
does usually provide an estimate of the truncation error.
Regularization schemes
that kill {\it negative} powers of $\Lambda$,
such as dimensional regularization, deprive you of this tool.

This newer view of renormalization arose from the combination
of the traditional view with the more intuitive ``Wilsonian RG''.
The latter is usually applied to a ({\it e.g.} condensed-matter) system 
where the underlying theory is known,
and the effective theory is constructed by explicitly reducing 
$\Lambda$ starting from $M_{\rm hi}$. 
In this process all interactions allowed by symmetries are generated,
and of course depend on $\Lambda$ by construction.
For example, even if there is an underlying potential that
is mostly two-body, its expansion at large distances will contain higher-body 
components, stemming from successive two-body encounters at unresolved 
distances and times. (In most situations these higher-body forces are 
relatively small, but not always, as we will see below.)
In the Wilsonian RG, 
$\Lambda$ marks the highest on-shell momentum to which we can apply the EFT.
As $\Lambda$ is decreased it eventually crosses a physical scale 
$M_{\rm hi}'$ where the EFT needs to be reorganized,
for example due to the emergence of new degrees of
freedom (say, a Goldstone boson or another type of low-energy 
collective effect). 
If one extends this new EFT to $\Lambda$ {\it above} $M_{\rm hi}'$,
we are in the situation I described earlier: 
real momenta up to $M_{\rm hi}'$ can be considered as $\Lambda$ cuts off
virtual momenta only.
Thus, while intuitive, there is no need,
in fact no advantage, in keeping the regulator parameter $\Lambda$ below the
physical breakdown scale we are interested in. 
This is particularly true when the underlying
theory is unknown (as for the Standard Model) or known but hard to solve
explicitly (as for QCD at low energies).

Beware that applications of this more general EFT implementation of the RG
have been muddled by the multiple uses of the word ``cutoff''. 
When the Wilsonian RG is favored, typically ``cutoff'' is used for the 
regulator, which also limits the range applicability of the EFT. 
In contrast, in modern particle physics where dimensional
regularization is almost exclusively employed, 
``cutoff'' is often used for the physical breakdown scale. 
I will try to consistently distinguish the regulator cutoff $\Lambda$,
which is not physical, and the physical breakdown scale $M_{\rm hi}$.

\section{Two-body system}
\label{2body}

Now let us see how this EFT works in the simplest case, the two-body system.
Were we considering photons explicitly, we could look at the electromagnetic
properties of one particle, such as form factors (accessible
through electron scattering) or polarizabilities (through Compton scattering).
But the corresponding amplitudes are purely perturbative (except possibly
around special kinematic points), 
and my goal here is to 
illustrate the more challenging issue of building a power counting
when a subset of interactions has to be treated nonperturbatively.
That is the case when the $T$ matrix has low-energy poles.
In particular, I want to tackle the situation relevant to
the systems discussed in Sec. \ref{scales}: a shallow 
two-body bound or virtual state.

\subsection{Amplitude}
\label{2bodyamplitude}

Because the EFT splits into sectors of different $A$, we can focus on
$A=2$ without worrying, as we do in relativistic theories, about interactions
with more than four fields in the action \eqref{L}.
Of the one-body terms, all we need is the 
propagator for a particle of four-momentum $l$, 
\begin{equation}
iD_1(l^0,\vec{l})=iD_1^{(0)}(l^0,\vec{l}) 
+ iD_1^{(0)}(l^0,\vec{l}) \,
\frac{i\vec{l}^{\,4}}{8m^3} \, iD_1^{(0)}(l^0,\vec{l}) +\ldots,
\label{onebodyprop}
\end{equation}
where 
\begin{equation}
iD_1^{(0)}(l^0,\vec{l})=\frac{i}{l^0-\vec{l}^{\,2}/(2m)+i\epsilon},
\label{onebodypropLO}
\end{equation}
with $\epsilon>0$, is the LO propagator.
Because they are suppressed by ${\cal O}(Q^2/m^2)$,
relativistic corrections come at N$^2$LO or higher
depending on how we decide to count $m$ relative to $M_{\rm hi}$.
There is not much consensus about the most efficient scheme to do
this, but the issue does not appear up to NLO, which is sufficient
for these lectures.
Likewise, for an on-shell particle of momentum $\vec{p}$,
to this order we need only the first term in the expansion of the energy,
$E=k^2/(2m) - k^4/(8m^3)+\ldots $, where $k=|\vec{p}|$.
Note that I did keep the recoil term $\propto 1/m$ in 
the denominator of Eq. \eqref{onebodypropLO}. 
Were we considering light probes (such as photons) with momenta ${\cal O}(Q)$, 
they would 
deposit energies ${\cal O}(Q)$ onto the virtual particles.
In this case, we could also expand the propagator \eqref{onebodypropLO}
in powers of $Q/m$, the leading term being the static propagator 
$i/(l^0+i\epsilon)$.
Instead, here particles 
start off with $E={\cal O}(Q^2/m)$ and 
remain nearly on-shell: $l^0={\cal O}(Q^2/m)$ and thus comparable to
$\vec{l}^{\,2}/(2m)$. 
The propagator at LO is 
{\it not} static. This is the first indication
that the power counting for processes involving two or more heavy 
particles is different from that for the simpler one-body processes.

The simplest of the four-field (or in nonrelativistic parlance, two-body)
interactions in the action \eqref{L} is the $C_0$ term,
which is represented by a vertex with four legs.
It gives a momentum-independent 
tree-level contribution to the two-body $T$ matrix,
\begin{equation}
T_2^{(0,0)}(\Lambda)=-\frac{4\pi}{m} \, C_0^{(0)}(\Lambda)
\equiv -V_2^{(0)}(\Lambda),
\label{Tnoloop}
\end{equation}
where $C_0^{(0)}(\Lambda)$ denotes the LO part of $C_0$.
Things are more interesting at one-loop level,
where two $C_0$ vertices are connected by two nonrelativistic
propagators \eqref{onebodypropLO}. In the center-of-mass system,
where one incoming particle has four-momentum $(p^0, \vec{p})$
and the other $(p^0, -\vec{p})$, 
\begin{eqnarray}
T_2^{(0,1)}(k;\Lambda) &=& i \left(\frac{4\pi i}{m}C_0^{(0)}(\Lambda)\right)^2
\int \! \!\frac{d^3l}{(2\pi)^3} \! \int \!\frac{dl^0}{2\pi} 
\, D_1^{(0)}(l^0+p^0,\vec{l}+\vec{p})
\, D_1^{(0)}(-l^0+p^0,-\vec{l}-\vec{p})
\label{Toneloop1}
\nonumber
\\
&=& \left(\frac{4\pi}{m}C_0^{(0)}(\Lambda)\right)^2 
\int  \!\!\frac{d^3l}{(2\pi)^3} \, \frac{m}{\vec{l}^{\,2}-k^2-i\epsilon}.
\label{Toneloop2}
\end{eqnarray}
Here, I first integrated over the 0th component of the loop momentum,
as it is standard in nonrelativistic theories.
We can do this by contour integration: closing the contour in the upper plane,
we pick a contribution from the residue of the pole
at $p^0-(\vec{l}+\vec{p})^2/(2m)+i\epsilon$;
closing on the lower plane, a contribution from the other pole,
which gives, of course, the same result.
As first noticed in this context by Weinberg 
\cite{Weinberg:1990rz,Weinberg:1991um}, had we neglected recoil in
the one-body propagator \eqref{onebodypropLO}, 
we would have faced a pinched singularity at the
origin. The reflection of this is the appearance of the large
$m$ in the numerator of  Eq. \eqref{Toneloop2}, 
where I additionally relabeled $\vec{l}\to\vec{l}-\vec{p}$.
This form of the one-loop amplitude should come as no surprise: 
we simply have an integration over the virtual three-momentum of 
the standard Schr\"odinger propagator. 

In the form \eqref{Toneloop2}, it is clear that the integral would
diverge without regularization. The most intuitive regularization
consists of a ``non-local'' regulator that depends
separately on the incoming and outgoing nucleon momenta:
a function $F(l/\Lambda)$ with the properties
that $F(0)= 1$ (to preserve the physics at low momentum) and
$F(x\to \infty)= 0$ (to kill high momenta):
\begin{equation}
\int \! \! \frac{d^3l}{(2\pi)^3}\, \frac{m}{{\vec l}^{\,2}-k^2-i\epsilon}
\equiv  \frac{m}{2\pi^2}\int_0^\infty \!\! dl \, \frac{l^2}{l^2-k^2-i\epsilon} 
F(l/\Lambda)
\equiv I_1(k;\Lambda)
\end{equation}
The simplest choice is a step function, $F(x)=\theta(1-x)$:
\begin{eqnarray}
I_1(k;\Lambda) &=& \frac{m}{2\pi^2}
\int_0^\Lambda \!\! dl \, \frac{l^2}{l^2-k^2-i\epsilon}
\nonumber\\
&=& 
\frac{m}{2\pi^2}\left(\Lambda 
+ \frac{k^2}{2} 
\int_{-\Lambda}^\Lambda \!\! dl \, \frac{1}{l+k+i\epsilon}\, \frac{1}{l-k-i\epsilon}
\right)
\nonumber\\
&=& 
\frac{m}{4\pi} \left(\frac{2}{\pi} \Lambda 
+ ik
-\frac{2 k^2}{\pi\Lambda}
+{\cal O}(k^4/\Lambda^3)\right),
\label{I13}
\end{eqnarray}
where I redefined $\epsilon \to 2k\epsilon >0$ 
and again used contour integration, now in the complex-$l$ plane.
More generally,
\begin{equation}
I_1(k;\Lambda) =
\frac{m}{4\pi} \left(\theta_1 \Lambda + ik
+\theta_{-1} \frac{k^2}{\Lambda} +\ldots \right),
\label{I14}
\end{equation}
where $\theta_{1-2n}$, $n=0, 1, \ldots$, are numbers that depend on the specific 
form of the regulator \cite{vanKolck:1998bw}. 

\vspace{0.2cm}
\noindent
{\bf Exercise:} Show that $\theta_{1-2n}=0$ in dimensional regularization
with minimal subtraction.
While a subtraction of the pole in two spatial dimensions 
(``power divergence subtraction'', or PDS) \cite{Kaplan:1998tg,Kaplan:1998we}
retains the linear divergence,
subtracting instead all poles leads to a result
identical to that of a momentum-cutoff regulator \cite{Phillips:1998uy}.
\vspace{0.2cm}

We then arrive at 
\begin{equation}
T_2^{(0,1)}(k;\Lambda) = \frac{4\pi}{m} C_0^{(0)2}(\Lambda)
\left(\theta_1 \Lambda + ik
+\theta_{-1} \frac{k^2}{\Lambda} +\ldots \right).
\label{Toneloop4}
\end{equation}
As we will show shortly, the regulator-dependent terms can be eliminated.
As in any EFT, the meaningful loop term is the non-analytic (in energy) $ik$,
the ``unitarity term''.
Relative to the tree-level \eqref{Tnoloop}, it is ${\cal O}(C_0 Q)$.
This is to be contrasted with the analogous situation in ChPT
\cite{Pich:2018ltt},
where the one-loop diagrams are down with respect to the tree 
by $Q^2/(4\pi f_\pi)^2$ \cite{Weinberg:1978kz,Manohar:1983md}. 
The difference is that, whereas the relativistic
loop gives $Q^2/(4\pi)^2$, the nonrelativistic loop gives $mQ/(4\pi)$. 
There are {\it two} enhancements for heavy particle processes:
an infrared enhancement by $m$ \cite{Weinberg:1990rz,Weinberg:1991um}
and an ``angular'' enhancement of $4\pi$ 
\cite{vanKolck:1997ut,Kaplan:1998tg,Kaplan:1998we,vanKolck:1998bw}.
We can think of $mQ/(4\pi)$ as arising from the simple rules:
\begin{eqnarray}
&&{\rm heavy \; particle \; propagator} \to m/Q^2,
\nonumber \\
&&{\rm loop \; integral} \to Q^5/(4\pi m),
\nonumber \\
&&{\rm derivative} \to Q.
\label{heavyrules}
\end{eqnarray}

If we make the dimensional guess
$C_0^{(0)}={\cal O}(M_{\rm hi}^{-1})$, we see that the loop is suppressed 
by one order of the expansion parameter, ${\cal O}(Q/M_{\rm hi})$.
The theory is perturbative and any pole of $T_2$ can only have
binding momentum of ${\cal O}(M_{\rm hi})$ or higher,
which is outside the EFT.
In order to accommodate a shallow pole with $Q_2={\cal O}(M_{\rm lo})$, 
we must assume instead that $C_0^{(0)}={\cal O}(M_{\rm lo}^{-1})$. 
I will discuss this assumption in Sec. \ref{unitarity};
for now, let us see what it implies.

\begin{figure}[t]
\begin{center}
\includegraphics[scale=.8]{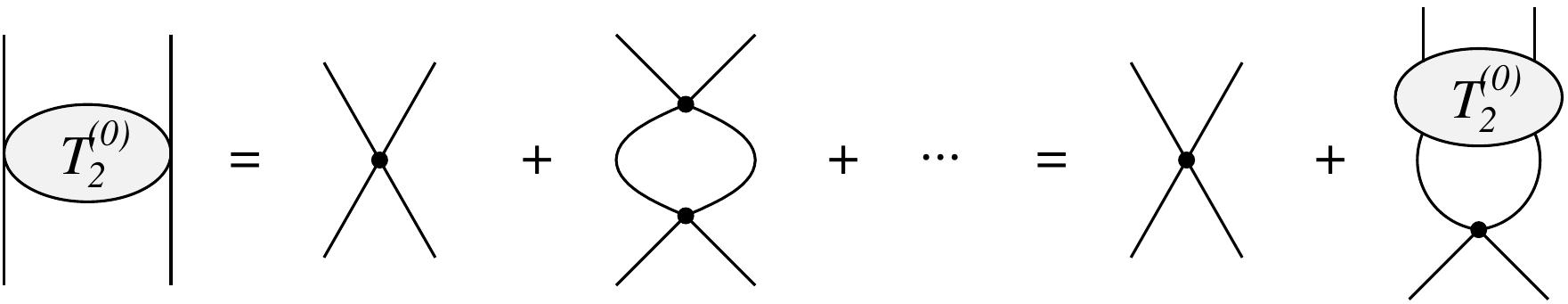}
\end{center}
\caption{Two-body $T$ matrix in Pionless EFT at LO, $T_2^{(0)}$. 
Solid lines denote particle propagation up, Eq. \eqref{onebodypropLO}. 
The dotted vertex stands for
the $C_0^{(0)}$ contact interaction, Eq. \eqref{Tnoloop}.
The first equality represents Eq. \eqref{Tsum};
the second, Eq. \eqref{LSeq}.}
\label{figLO2bodyamp}       
\end{figure}

Under this assumption the one-loop diagram is ${\cal O}(Q/M_{\rm lo})$
compared to the tree diagram.
It is not difficult to see that the $n$-loop diagram $T_2^{(0,n)}$ 
is proportional to the $n$th power of one-loop diagram
and its magnitude is ${\cal O}(Q^n/M_{\rm lo}^n)$ compared to the tree.
For $Q\simge M_{\rm lo}$, the whole geometric series must be resummed,
\begin{equation}
T_2^{(0)}(k;\Lambda)\equiv \sum_{n=0}^{\infty} T_2^{(0,n)}(k;\Lambda) 
=-\frac{4\pi}{m} 
\left(\frac{1}{C_0^{(0)}(\Lambda)}+ \theta_1\Lambda
+ik+ \theta_{-1}\frac{k^2}{\Lambda}+ \ldots\right)^{-1}.
\label{Tsum}
\end{equation}
This LO amplitude is shown in Fig. \ref{figLO2bodyamp}.
Even though we derived it by explicitly summing up the individual
diagrams, it can be obtained directly from an integral equation,
the  Lippmann-Schwinger (LS) equation, also shown in Fig. \ref{figLO2bodyamp}.
Quite generally,
in a way that also applies to other EFTs for heavy particles,
we can define the potential $V$ as the sum of diagrams that cannot be
split by cutting only heavy particle lines
\cite{Weinberg:1990rz,Weinberg:1991um}.
In Pionless EFT $V$ reduces to the sum of all
tree diagrams, but in Chiral EFT it includes also loop diagrams with pions,
as we will discuss in Sec. \ref{longrange}.
Apart from a sign, the potential gives the $T$ matrix in first 
Born approximation, 
one-loop amplitude diagrams represent the second Born approximation, and so on.
In our case here, the LO potential is given by Eq. \eqref{Tnoloop}
and the LS equation 
is, after integrating over the 0th component of the loop momentum,
\begin{equation}
T_2^{(0)}(k,\Lambda)=-V_2^{(0)}(\Lambda)
-\int \!\!\frac{d^3l}{(2\pi)^3} \, T_2^{(0)}(k,\Lambda) 
\, \frac{m}{\vec{l}^{\,2}-k^2-i\epsilon} \, V_2^{(0)}(\Lambda).
\label{LSeq}
\end{equation}
More generally the $T$ matrix depends on the incoming $\vec{p}$
and outgoing $\vec{p}\,'$ relative momenta, as well as the energy. 
The LS equation then involves the ``half-off-shell'' $T$ matrix, 
since one of the momenta is integrated over.
In contrast, our LS equation \eqref{LSeq} with a non-local or ``separable''
regulator can be solved easily
with the {\it Ansatz} that $T_2^{(0)}$ depends only on the energy:
by taking $T_2^{(0)}(k,\Lambda)$ and $V_2^{(0)}(\Lambda)$ out of the integral
in the right-hand side and combining this term with the left-hand side,
we obtain Eq. \eqref{Tsum} directly from Eqs. \eqref{LSeq} and \eqref{Tnoloop}.
When the regulator is chosen to be ``local'' or ``non-separable''
--- a function solely of the momentum transfer $\vec{p}\,'-\vec{p}$,
which translates into a function of the spatial coordinate $\vec{r}$ ---
the loops do not form a simple geometric series and
we are usually forced to solve Eq. \eqref{LSeq} numerically.

Equation \eqref{Tsum} still contains explicit $\Lambda$ dependence
which, if not controlled, will lead to regulator dependence in observables.
For example, there is a pole in Eq. \eqref{Tsum} at imaginary momentum,
or equivalently negative energy;
if $C_0^{(0)}$ were $\Lambda$ independent, then the corresponding binding
energy would be $\propto \Lambda^2/m$.
To avoid such a disaster, the $\Lambda$ dependence of $C_0^{(0)}(\Lambda)$
must cancel the term linear in $\Lambda$,
\begin{equation}
\frac{1}{C_0^{(0)}(\Lambda)}= -\theta_1\Lambda
+\frac{1}{C_{0R}^{(0)}} \left(1+ \frac{\zeta_{-1}}{C_{0R}^{(0)}\Lambda}
+\ldots\right),
\label{C00}
\end{equation}
with $C_{0R}^{(0)}$ --- the renormalized LEC --- a constant to be determined
from experimental data or from matching to the underlying theory,
and $\zeta_{-1-2n}$, $n=0,1,\ldots$, 
a set of arbitrary numbers related to the choice of input.
In terms of the renormalized LEC the LO amplitude becomes
\begin{equation}
T_2^{(0)}(k;\Lambda)= -\frac{4\pi}{m} 
\left[\frac{1}{C_{0R}^{(0)}} +ik
+\frac{1}{\Lambda} 
\left(\frac{\zeta_{-1}}{C_{0R}^{(0)2}}+\theta_{-1} k^2\right) 
+ \ldots
\right]^{-1}.
\label{TLOren}
\end{equation}
Now the LO amplitude has a physical pole at $k=i\kappa^{(0)}$, where the binding
momentum $\kappa^{(0)}=1/C_{0R}^{(0)} + \ldots$.
This is a bound state if $C_{0R}^{(0)} >0$ and a virtual state 
if $C_{0R}^{(0)} <0$.
In either case the binding energy is 
\begin{equation}
B_2^{(0)}(\Lambda)
= \frac{1}{mC_{0R}^{(0)2}}
\left[1+ \frac{2}{C_{0R}^{(0)}\Lambda}\left(\zeta_{-1}-\theta_{-1}\right) +\ldots
\right].
\label{B2LO}
\end{equation}

The renormalized LEC is obtained from one datum.
For example, we could use the value of the amplitude at 
a chosen momentum $k=\mu$: in order to have $T_2^{(0)}(\mu;\Lambda)=T_2(\mu)$
$\Lambda$ independent, we take
$\zeta_{-1}=-\theta_{-1} C_{0R}^{(0)2}\mu^2$ {\it etc.},
and the 
renormalized LEC is fixed by $C_{0R}^{(0)}=-[4\pi/(mT_2(\mu))+i\mu]^{-1}$.
If we choose $\mu=i\sqrt{mB_2}$, then 
$B_2^{(0)}(\Lambda)=B_2$ is $\Lambda$ independent,
but other observables, such as the amplitude at zero
momentum, will have residual regulator dependence.
For other choices of $\mu$,
the binding energy contains nonvanishing $\Lambda^{-1}$ terms.
We might as well choose to fit $C_{0R}^{(0)}$ 
to several low-energy data simultaneously, each with a certain weight,
in which case $C_{0R}^{(0)}$ does not necessarily depend 
on a single fixed momentum $\mu$.

Regardless of the choice of input observable(s),
we have eliminated the dangerous regulator dependence: 
the amplitude \eqref{TLOren} satisfies \eqref{RGinvtrunc},
only negative powers of $\Lambda$ appear in observables,
and they can be made arbitrarily small by increasing $\Lambda$.
The bare parameter $C_{0}^{(0)}(\Lambda)$ has disappeared as well.
Its size does not matter;
it is of course the surviving, renormalized LEC that has size
$C_{0R}^{(0)}={\cal O}(M_{\rm lo}^{-1})$,
giving $T_2^{(0)}(k;\Lambda)={\cal O}(4\pi/(m M_{\rm lo}))$
and $\kappa^{(0)}={\cal O}(M_{\rm lo})\ll M_{\rm hi}$.
Equation \eqref{B2LO}, for example, is exactly of the generic form 
\eqref{obscutdep}.
Renormalization at LO has been completed: the simple contact interaction
is renormalizable at the two-body level.

Now we can describe the physics of the shallow state in a controlled
expansion. Let us see how this works at NLO.
We expect nonvanishing NLO corrections on the basis that the
residual-regulator effects in observables such as Eq. \eqref{B2LO}
can be as large as ${\cal O}(M_{\rm lo}/M_{\rm hi})$.
Indeed, we can write the LO amplitude as 
\begin{equation}
T_2^{(0)}(k;\Lambda)=-\frac{4\pi}{m} 
\left(\frac{1}{C_{0R}^{(0)}}+ik\right)^{-1} 
+\delta T_2^{(1)}(k;\Lambda)
+\ldots,
\label{TLO}
\end{equation}
where
\begin{equation}
\delta T_2^{(1)}(k;\Lambda)\equiv \frac{m} {4\pi\Lambda} 
\left[T_2^{(0)}(k;\infty)\right]^2 
\left(\frac{\zeta_{-1}}{C_{0R}^{(0)2}}+\theta_{-1}k^2\right)
\label{TNLOfromsum}
\end{equation}
has the size of an NLO term:
$\delta T_2^{(1)}(k;M_{\rm hi})= {\cal O}(4\pi/(m M_{\rm hi}))$.
The form of this correction 
suggests that the NLO interaction that removes such a 
residual regulator dependence
is the $C_2$ term in Eq. \eqref{L}, plus a perturbative correction to $C_0$. 
At tree level,
\begin{equation}
T_2^{(1,0)}(\vec{p}\,', \vec{p};\Lambda)=-\frac{4\pi}{m}
\left(C_0^{(1)}(\Lambda)- C_2^{(1)}(\Lambda)\, \frac{\vec{p}\,'\,^2+\vec{p}\,^2}{2}
\right) =-V_2^{(1)}(\vec{p}\,', \vec{p};\Lambda),
\label{TNLOnoloop}
\end{equation}
where I denoted the NLO parts of $C_{0,2}$ by $C_{0,2}^{(1)}(\Lambda)$. 
We expect 
$T_2^{(1,0)}(\vec{p}\,', \vec{p}; M_{\rm hi})\sim \delta T_2^{(1)}(k;M_{\rm hi})$, 
which is a perturbative correction to the LO amplitude.

\begin{figure}[t]
\begin{center}
\includegraphics[scale=.8]{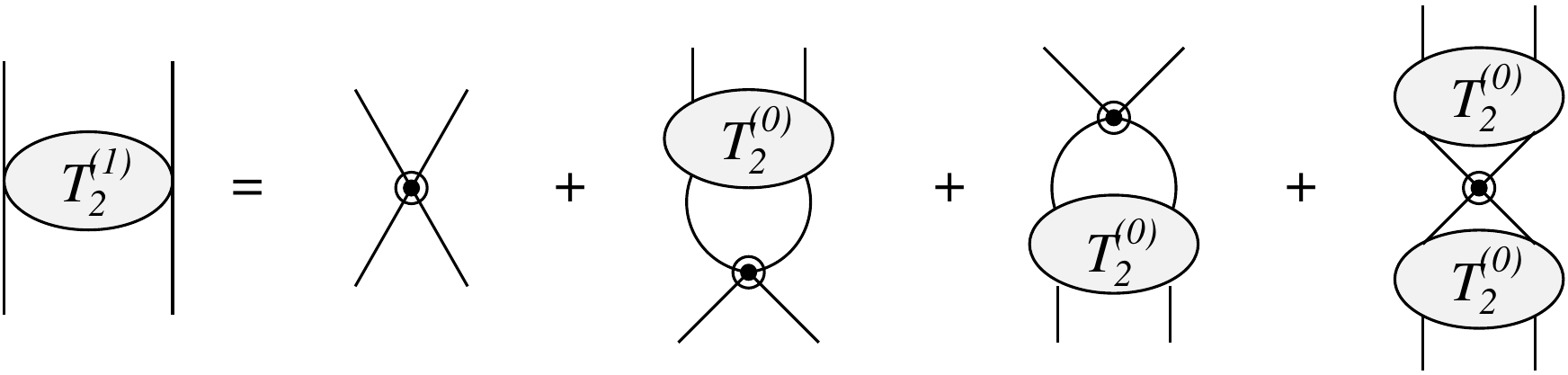}
\end{center}
\caption{Two-body $T$ matrix in Pionless EFT at NLO, $T_2^{(1)}$. 
The circled vertex stands for
the $C_0^{(1)}$ and $C_2^{(1)}$ contact interactions, Eq. \eqref{TNLOnoloop}.}
\label{figNLO2bodyamp}       
\end{figure}

However, adding a loop connecting $V_2^{(1)}$ to $T_2^{(0)}$ 
through a Schr\"odinger propagator gives a contribution of the same size.
Thus, at NLO we have only one vertex representing NLO interactions,
but it is dressed in all possible ways by LO interactions
--- see Fig. \ref{figNLO2bodyamp}.
In the language of the LS equation,
Eq. \eqref{TNLOnoloop} is an NLO correction to the potential.
The subleading potentials are solved in what is called
{\it distorted-wave} Born approximation, which differs from
the ordinary Born approximation in that the LO is not just
a plane wave.

\vspace{0.2cm}
\noindent
{\bf Exercise:} Show that when the external particles
are on-shell, $|\vec{p}\,'|=|\vec{p}|=k$, the amplitude 
in Fig. \ref{figNLO2bodyamp} is \cite{vanKolck:1998bw}
\begin{equation}
T_2^{(1)}(k;\Lambda)=-\frac{m}{4\pi}
\left[\frac{T_2^{(0)}(k;\Lambda)}{C_0^{(0)}(\Lambda)} \right]^2
\left[C_0^{(1)}(\Lambda)
- \left(-\frac{4}{\pi}\, \theta_3 \Lambda^3 \, C_0^{(0)}(\Lambda) 
+ k^2\right) C_2^{(1)}(\Lambda)  +\ldots \right],
\label{TNLO}
\end{equation}
where $\theta_3$ is a regulator-dependent number.
\vspace{0.2cm}

Combining Eqs. \eqref{TNLO} and \eqref{TNLOfromsum}, we can write the 
full NLO amplitude as
\begin{equation}
T_2^{(1)}(k;\Lambda)+\delta T_2^{(1)}(k;\Lambda)
=-\frac{m}{4\pi}\left[\frac{T_2^{(0)}(k;\infty)}{C_{0R}^{(0)}}\right]^2
\left(C_{0R}^{(1)}- C_{2R}^{(1)}\, k^2 \right)+\ldots,
\label{TNLOall}
\end{equation}
if we impose that the NLO LECs be given by
\begin{eqnarray}
\frac{C_2^{(1)}(\Lambda)}{C_0^{(0)2}(\Lambda)} &=&
\frac{C_{2R}^{(1)}}{C_{0R}^{(0)2}}-\frac{\theta_{-1}}{\Lambda}
+\ldots,
\label{C21}\\
\frac{C_0^{(1)}(\Lambda)}{C_0^{(0)2}(\Lambda)} &=&
-\frac{4}{\pi}\theta_3 \,\Lambda^3 \, 
\, \frac{C_2^{(1)}(\Lambda)}{C_0^{(0)}(\Lambda)} 
+\frac{C_{0R}^{(1)}}{C_{0R}^{(0)2}}
+ \frac{\zeta_{-1}}{C_{0R}^{(0)2}\Lambda} +\ldots,
\label{C01}
\end{eqnarray}
in terms of the renormalized LECs
$C_{0,2R}^{(1)}={\cal O}((M_{\rm lo}^2M_{\rm hi})^{-1})$. 
We have succeeded in renormalizing the NLO amplitude.

There is only one new physical parameter at NLO, $C_{2R}^{(1)}$,
which is related to the energy dependence of the amplitude. 
The LEC $C_{0R}^{(1)}$ is introduced for convenience only,
so as to allow us to keep
the observable used to fix $C_{0}^{(0)}(\Lambda)$ unchanged.
For example, if we want to keep the energy-independent part 
of the amplitude unchanged, we take $C_{0R}^{(1)}=0$,
but more generally $C_{0R}^{(1)}$ depends on $C_{0R}^{(0)}$ and $C_{2R}^{(1)}$.
In any case, the leading regulator dependence from Eq. \eqref{TLO}
gets replaced by the physical effect from $C_{2R}^{(1)}$.
In particular, $\zeta_{-1}$ disappeared, showing 
that the different choices of input observable at LO are an NLO effect. 
Similarly, the freedom we have now to fix $C_{2R}^{(1)}$ is 
a higher-order effect. 

To see how $C_{2R}^{(1)}$ affects the bound or virtual state 
it is convenient to write the amplitude up to NLO as
\begin{eqnarray}
T_2^{[1]}(k;\Lambda) &=&
T_2^{(0)}(k;\infty)\left[1 
-\frac{m}{4\pi}\, \frac{T_2^{(0)}(k;\infty)}{C_{0R}^{(0)2}}
\left(C_{0R}^{(1)} -C_{2R}^{(1)} \, k^2\right)
+\ldots\right]
\nonumber\\
&=&-\frac{4\pi}{m} \left(\frac{1}{C_{0R}^{(0)}}+ik
-\frac{C_{0R}^{(1)}}{C_{0R}^{(0)2}}+\frac{C_{2R}^{(1)}}{C_{0R}^{(0)2}} \, k^2\right)^{-1}
\left[1+ {\cal O}\left(\frac{M_{\rm lo}^2}{M_{\rm hi}^2}, 
\frac{M_{\rm lo}^2}{\Lambda M_{\rm hi}}\right)\right].
\label{TLO+NLO}
\end{eqnarray}
Now the binding energy becomes
\begin{equation}
B_2^{[1]}(\Lambda)
= \frac{1}{mC_{0R}^{(0)2}}
\left[1-2\left(\frac{C_{0R}^{(1)}}{C_{0R}^{(0)}}
+\frac{C_{2R}^{(1)}}{C_{0R}^{(0)3}}\right)+\ldots\right].
\label{B2NLO}
\end{equation}
If we choose to fit $B_2$ at LO and want it unchanged at NLO,
then we choose $C_{0R}^{(1)}=-C_{2R}^{(1)}/C_{0R}^{(0)2}$.
If we choose to fit two other observables at NLO
--- say the energy-independent part of the amplitude 
and its linear dependence on the energy ---
the binding energy will have a physical shift 
${\cal O}(M_{\rm lo}/M_{\rm hi})$ instead of 
the regulator dependence ${\cal O}(M_{\rm lo}/\Lambda)$
in Eq. \eqref{B2LO}.
The remaining regulator dependence in the ``$\ldots$''
is ${\cal O}(M_{\rm lo}^2/(\Lambda M_{\rm hi}))$ and
no larger than N$^2$LO as long as $\Lambda \simge M_{\rm hi}$:
the binding energy is predicted to a better precision.

The procedure can be generalized to higher orders 
\cite{vanKolck:1998bw,Chen:1999tn}.
At N$^2$LO, for example, we have to consider two insertions of $C_2$ and
one of a four-derivative, four-field operator, both acting on $S$ waves only.
At the same order in the two-nucleon case there is also a tensor
operator that mixes $^3S_1$ and $^3D_1$ 
\cite{Kaplan:1998tg,Kaplan:1998we,Chen:1999tn}. The first contributions to
higher-wave phase shifts enter at N$^3$LO via two-derivative four-field 
operators. The simple rule 
\cite{vanKolck:1997ut,Kaplan:1998tg,Kaplan:1998we,vanKolck:1998bw,Chen:1999tn}
is that --- under the assumption that
$M_{\rm lo}$ does not contaminate other waves --- an operator
gets an $(M_{\rm hi}/M_{\rm lo})^n$  enhancement over dimensional analysis,
where $n=1,2$ is the number of $S$ waves it connects. For example,
since $C_2$ connects $S$ to $S$ waves, 
$C_{2R}^{(1)}={\cal O}((M_{\rm lo}^2M_{\rm hi})^{-1})$ instead of
$C_{2R}^{(1)}={\cal O}(M_{\rm hi}^{-3})$.

\subsection{Connection to the effective-range expansion}
\label{ERE}

The form of Eq. \eqref{TLO+NLO} should be familiar from scattering in 
quantum mechanics. 
Recall that $S$ waves should dominate at low energies for their lack of 
centrifugal barrier, and that the corresponding phase shift  $\delta_2(k)$ 
is defined in terms of the $S$ matrix as $S_2(k)=\exp(2i \delta_2(k))$.
Thus, 
\begin{equation}
T_2(k) = -\frac{2\pi i}{m k} \left[S_2(k) -1 \right]
=-\frac{4\pi}{m}\left[-k\cot\delta_2(k)+ik\right]^{-1}.
\label{phaseshift}
\end{equation}
At very low energies, 
the effective-range expansion (ERE) \cite{Bethe:1949yr} is known to hold,
\begin{equation}
k \cot\delta_2(k) = -1/a_2 +r_2 \, k^2/2 - P_2^3 \, k^4/4+ \ldots,
\end{equation}
where $a_2$, $r_2$, $P_2$, $\ldots$ are real parameters 
--- respectively,
the scattering length, effective range, shape parameter, {\it etc.},
all with dimensions of distance.
Comparison with Eq. \eqref{TLO+NLO} gives
\begin{equation}
a_2 = C_{0R}^{(0)}+C_{0R}^{(1)}+\ldots
= {\cal O}(M_{\rm lo}^{-1}), \qquad
r_2 = -2\frac{C_{2R}^{(1)}}{C_{0R}^{(0)2}} + \ldots
= {\cal O}(M_{\rm hi}^{-1}), 
\qquad
\ldots
\label{renLECsandEREparams}
\end{equation}

The scattering length is just the amplitude at zero energy
and fixes $C_{0R}^{(0)}$ at LO if we choose
the renormalization scale $\mu=0$ in the discussion below Eq. \eqref{B2LO}.
The effective range provides a simple, explicit
example of the usefulness of a high regulator cutoff.
Because we chose a regulator for which we could iterate
$C_0^{(0)}$ analytically, we could separate 
$\delta T_2^{(1)}(k;\Lambda)$ from $T_2^{(0)}(k;\infty)$
in Eq. \eqref{TLO}. 
Since $\Lambda$ is not physical,
we might as well take the amplitude at a given order to be its 
$\Lambda\to \infty$ limit, as in old times.
Alas, in a numerical calculation --- which is necessary
for this EFT with a local regulator that
depends on the transferred momentum, or for $A\ge 3$, or for other EFTs ---
this cannot be done. An LO calculation will have
subleading pieces buried in them. In the case considered here,
an effective range $-\theta_1 \Lambda^{-1}$ is induced.
For the error of the LO calculation not to exceed
its truncation error, which is
$\pm {\cal O}(M_{\rm hi}^{-1})$ for $r_2$, we have to take 
$\Lambda \simge M_{\rm hi}$. Even though the theory is renormalizable
at LO, a ``Wilsonian cutoff'' $\Lambda \simle M_{\rm hi}$ would lead,
in the absence of an explicit integration of modes in a known underlying
theory, to excessive errors.
Moreover, the magnitude of the change in the
LO amplitude upon variation in $\Lambda$ from $M_{\rm hi}$, 
when $|\theta_1| \Lambda^{-1}={\cal O}(M_{\rm hi}^{-1})$,
to much larger values,
when $\theta_1 \Lambda^{-1}\simeq 0$, gives an
estimate of the actual size of $r_2$ and thus of the LO error.

\begin{figure}[t]
\begin{center}
\includegraphics[scale=.6]{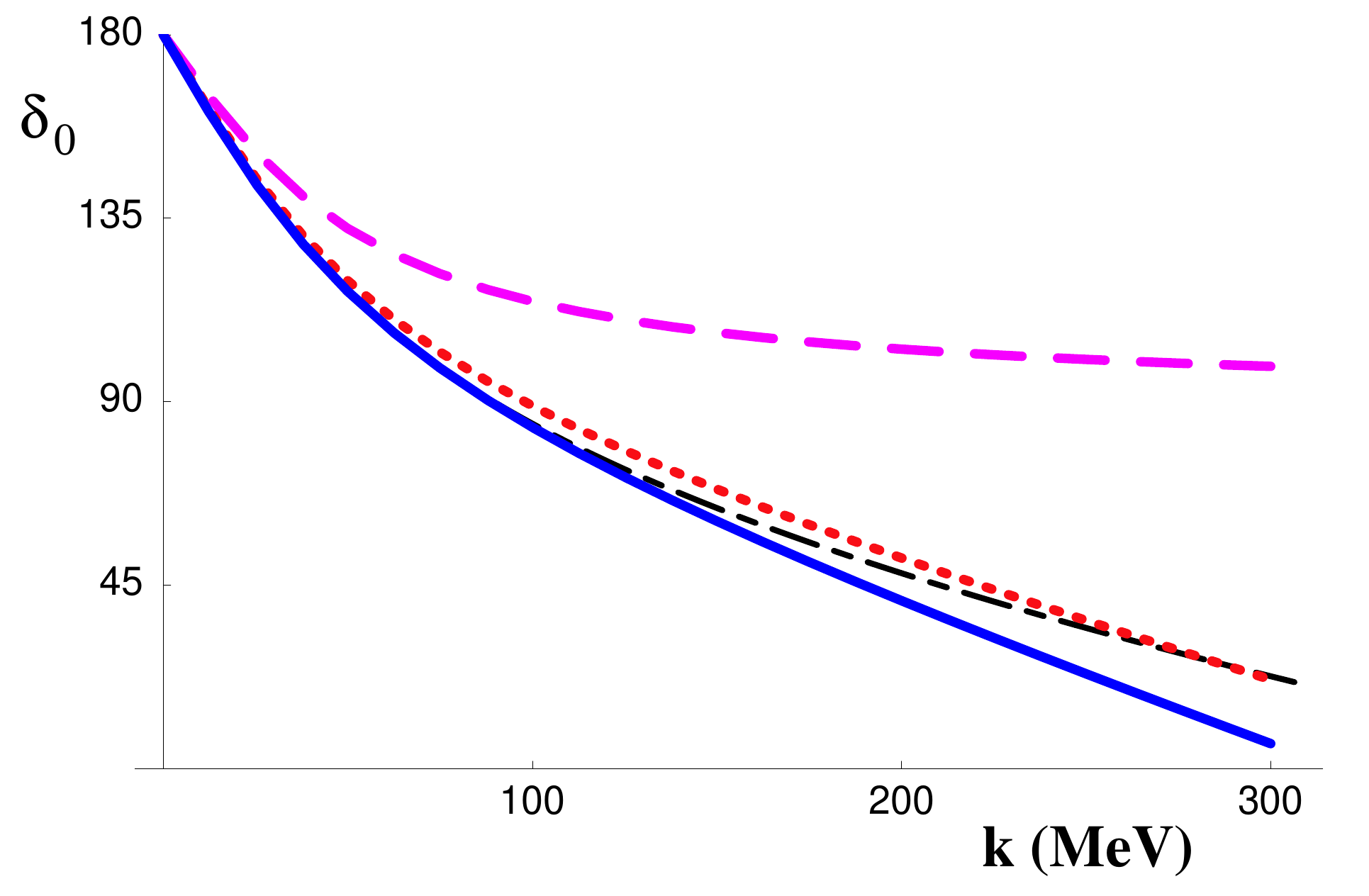}
\end{center}
\caption{$^3S_1$ two-nucleon phase shift $\delta_0$ (in degrees)
as function of the center-of-mass momentum $k$ (in MeV) in Pionless EFT:
LO, (magenta) dashed line; NLO, (red) dotted line; 
and N$^2$LO, (blue) solid line. 
The Nijmegen potential result
\cite{Stoks:1994wp} is given by the (black) dash-dotted line.
Reprinted from Ref. \cite{Chen:1999tn} with permission from Elsevier.
}
\label{Sphasenopi}       
\end{figure}

Given the direct relation between renormalized LECs and 
ERE parameters in Eq. \eqref{renLECsandEREparams},
the latter are often used as input data in Pionless EFT.
Once the LECs present at each order are determined from an equal number 
of data points, the full phase shifts
can be predicted within the truncation error, as long as 
$\Lambda \simge M_{\rm hi}$. As an example, the phase shifts
in the $^3S_1$ two-nucleon system at lowest orders obtained in 
Ref. \cite{Chen:1999tn} are shown in Fig. \ref{Sphasenopi}.

You might be disappointed that, after so much work,
we found that Pionless EFT in the two-body system is essentially
just the ERE, known for 70 years.
In fact, one can show \cite{vanKolck:1998bw} that this EFT is also equivalent 
to even older approaches:
short-distance boundary conditions \cite{BethePeierls:19351,BethePeierls:19352}
and Fermi's pseudopotential \cite{Fermi:1936}.
The EFT can be seen as a derivation of these older results,
once EFT's general framework is deployed to this particular class 
of nonrelativistic
systems with short-range interactions. As such,
it is the proverbial cannon to kill a fly.
However, a cannon can kill more.
The calculation above can be generalized (numerically)
to Chiral EFT, where it provides a guide
for much more complicated renormalization.
Moreover, Pionless EFT applies 
to systems which are outside the scope of the ERE.
For example, in the nuclear case we can look at the electromagnetic properties
of the deuteron \cite{Chen:1999tn}, 
thanks to the inclusion of
consistent one- and two-nucleon electromagnetic operators.
I will instead
describe how Pionless EFT allows us to generalize the ERE to more 
than two particles, where amazing new structures arise.

\section{Fine tuning, unitarity, and scale invariance}
\label{unitarity}

Before we consider more bodies, a few remarks about the EFT in 
the two-body system.

One would naively expect all ERE parameters to be 
comparable to the force range $R$ and indeed,
for most parameter values of most finite-range potentials,
that is what one finds, with bound states or resonances that are not
particularly shallow. 
In this case, for $k\ll R^{-1}$ each term in the ERE is larger than the next
and the corresponding EFT is purely perturbative.
But by dialing one or more potential parameters one can make
$|a_2|\gg R$ with other ERE parameters still of ${\cal O}(R)$.
For $k\ll |a_2|^{-1}$, the EFT is still perturbative but as we have just seen,
if we want to continue to describe physics of momentum up to $R^{-1}$,
we have to resum (only!) $C_0^{(0)}$.
By {\it fine tuning} we can reach the extreme point,
the ``unitarity limit'' where $|a_2|\to \infty$
and $T_2(k)$ takes (up to range corrections) its maximum value 
$\propto k^{-1}$.

\vspace{0.2cm}
\noindent
{\bf Exercise:} Consider a three-dimensional spherical well
with dimensionless depth $\alpha$,
\begin{equation}
V_2(\vec{r}) = - \frac{\alpha^2}{m R^2} \, \theta(R-r).
\label{squarewell}
\end{equation}
Solve the Schr\"odinger equation for the $S$ wave in the usual way, {\it i.e.},
by matching inside and outside solutions. 
Obtain $\delta_2(k)$ from the outside solution 
(asymptotically $\psi_2(r)\propto [\sin(kr + \delta_2(k))]/r$) 
and show that 
\begin{equation}
T_2(k)= -\frac{2\pi i}{m k}\left[ e^{-2ikR} \;
\frac{\sqrt{\alpha^2 +(kR)^2}\cot\sqrt{\alpha^2 +(kR)^2}+ikR}
     {\sqrt{\alpha^2 +(kR)^2}\cot\sqrt{\alpha^2 +(kR)^2}-ikR} -1 \right],
\end{equation}
and thus
\begin{equation}
a_2 = R \left( 1- \frac{\tan\alpha}{\alpha}\right),
\quad
r_2= R \left( 1- \frac{R}{\alpha^2a_2}-\frac{R^2}{3a_2^2}\right),
\quad
\ldots
\label{EREparamsquarewell}
\end{equation}
For generic values of $\alpha$, we see that
$a_2\sim r_2\sim \ldots \sim R$.
However, when $\alpha \simeq (2n+1)\pi/2 \equiv \alpha_c$, while 
still $r_2\sim \ldots \sim R$, 
$a_2\simeq R/[\alpha_c(\alpha - \alpha_c)]\gg R$.
For $\alpha$ just below $\alpha_c$, there is a shallow virtual state;
as the attraction increases past $\alpha_c$, a shallow bound state appears. 
In this example, $M_{\rm hi}\sim 1/R$ and 
$M_{\rm lo}\sim \alpha_c|\alpha - \alpha_c|/R$ \cite{vanKolck:1998bw}.
\vspace{0.2cm}

As we saw in the previous section, we incorporate such fine tuning
in the EFT by allowing the LECs to scale with the small $M_{\rm lo}$.
Although I used the methods of quantum field theory (the proverbial cannon),
we can also use standard quantum mechanics, supplemented by regularization
and renormalization.
The LO interaction at tree level, Eq. \eqref{Tnoloop}, is just a constant
in momentum space, which in coordinate space is
a delta function. 
The appearance of the Schr\"odinger propagator at one loop,
Eq. \eqref{Toneloop2}, 
reveals that the iteration of this interaction
is equivalent to solving the Schr\"odinger equation with 
a delta-function potential. 
In fact, one can show on general grounds that the LS equation is equivalent
to the Schr\"odinger equation. 
Even if we prefer to solve the Schr\"odinger equation, 
regularization and renormalization are still necessary. 
For example, the LO potential in coordinate
space can be written as
\begin{equation}
V_2^{(0)}(\vec{r};\Lambda) =\frac{4\pi}{m} \, C_0^{(0)}(\Lambda) 
\, \delta_\Lambda^{(3)}(\vec{r}),
\label{regdeltafunc}
\end{equation}
where $\delta_\Lambda^{(3)}(\vec{r})$ is a regularization
of the three-dimensional delta function, 
that is, a smearing over distances $r\simle \Lambda^{-1}$ with
$\lim_{\Lambda\to \infty}\delta_\Lambda^{(3)}(\vec{r}) =\delta^{(3)}(\vec{r})$.
Whatever the underlying potential is,
we can use, say, the spherical well \eqref{squarewell}
as such regularization.
In this guise, the range of the well functions as a regulator parameter,
$\Lambda \equiv 1/R$, and $\alpha = \alpha (R)$ can be adjusted
so that
$a_2$ in Eq. \eqref{EREparamsquarewell} is $R$ independent
and reproduces a given experimental value.
Another popular regularization for its analytical 
simplicity is a delta-shell potential.
For many-body calculations a Gaussian regularization is particularly
convenient for its smoothness.

\vspace{0.2cm}
\noindent
{\bf Exercise:} Solve the Schr\"odinger equation for 
the potential \eqref{regdeltafunc}.
{\it Hint}: Fourier transform to momentum space and choose a sharp 
momentum regulator.
Show that the LO results of the previous section are reproduced and,
in addition, 
the negative-energy wavefunction is
\begin{equation}
\psi_2^{(0)}(r)\propto \frac{1}{r}\exp(-r/a_2)
\end{equation}
as $\Lambda \to \infty$.
Thus $a_2$, which is basically the scattering
amplitude at $k=0$, is a measure of the size of the system.
A real bound (virtual) state has a (non-)normalizable wavefunction 
and corresponds to $a_2> 0$ ($a_2< 0$).
\vspace{0.2cm}

The bare LECs change with the regulator parameter
$\Lambda$. At LO, renormalization requires Eq. \eqref{C00}.
This is the solution of the RG equation
\begin{equation}
\mu \frac{d}{d\mu} \left(\mu C_0^{(0)}(\mu)\right) = 
\mu C_0^{(0)}(\mu) 
\left(1+ \mu C_0^{(0)}(\mu)\right),
\label{C0RGE}
\end{equation}
with $\mu \equiv \theta_1 \Lambda$.
This equation admits two fixed points: 
a ``trivial'' $\mu C_0^{(0)}(\mu)=0$
and the ``non-trivial'' $\mu C_0^{(0)}(\mu)=-1$ \cite{Weinberg:1991um}.
When $\mu\ll |a_2|^{-1}$, we are near the trivial point where
perturbation theory holds.  
On the other hand, for  $\mu\gg |a_2|^{-1}$, 
\begin{equation}
C_0^{(0)}(\Lambda; C_{0R}^{(0)})= -\frac{1}{\theta_1\Lambda}
\left[1+\frac{1}{\theta_1\Lambda C_{0R}^{(0)}}
+{\cal O}\!\left(\frac{1}{\Lambda^{2} C_{0R}^{(0)2}}\right)
\right],
\label{C00prime}
\end{equation}
the flow is close to the non-trivial fixed point, and
all diagrams containing only this vertex should be resummed. 
The unitarity limit corresponds to the non-trivial fixed point,
\begin{equation}
C_0^{(0)}(\Lambda; \infty)= -\frac{1}{\theta_1\Lambda}.
\label{C00unitarity}
\end{equation}

The fine tuning needed to produce the unitarity limit
can be carried out experimentally for cold atoms through the
mechanism of Feshbach resonances \cite{Kohler:2006zz}
--- this was one of the 
reasons for the explosion of interest in these systems.
The mechanism works when the system has two coupled channels with
different spins and thresholds --- ``open'' and ``closed'' channels ---
and the relative position of a bound state in the closed channel 
can be changed by an external magnetic field.
In the open channel, the scattering length 
varies and diverges as the energy of the bound state crosses the open threshold.
A short-range EFT for this situation is discussed in 
Ref. \cite{Cohen:2004kf}.
In contrast, the $^4$He dimer just happens to be close to the unitarity limit
even in the absence of a magnetic field. The scattering length and
effective range calculated with the LM2M2 potential are \cite{JanAzi95} 
$a_2\simeq 100$ \AA~$\simeq 18 \, l_{\rm vdW}$ and 
$r_2\simeq 7.3$ \AA~$\simeq 1.3 \, l_{\rm vdW}$,
with similar values for other sophisticated potentials
--- even though, of course, $a_2$ is very sensitive to potential
details because of fine tuning.

Nucleons are not as close to unitarity:
for $np$ in the $^3S_1$ (deuteron) channel,
$a_{2,S=1}\simeq 5.4$ fm~$\simeq 3.8 \, m_{\pi}^{-1}$ and 
$r_{2,S=1}\simeq 1.8$ fm~$\simeq 1.3 \, m_{\pi}^{-1}$.
However, in the $^1S_0$ channel, where there is a shallow virtual state, 
the relative magnitudes of $np$ parameters are not very different from 
atomic $^4$He,
$a_{2,S=0}\simeq -23.7$ fm~$\simeq -17 \, m_{\pi}^{-1}$ and 
$r_{2,S=0}\simeq 2.7$ fm~$\simeq 1.9 \, m_{\pi}^{-1}$.
In QCD, the only free parameters are the quark masses,
and we can imagine alternative worlds where 
the interactions are fundamentally unchanged but explicit 
chiral-symmetry breaking is larger and the range of the pion-exchange
force is, consequently, smaller.
As the quark masses change, so do nuclear binding energies.
Because heavier quarks 
are easier to evolve in imaginary time in a four-dimensional
space-time lattice, Lattice QCD has provided so far only
``alternative facts''
about light nuclei \cite{Beane:2010em}.
The situation is still in flux, with different methods of
signal extraction leading to contradictory results, but
in the majority of calculations it seems that nuclei at 
larger $\bar{m}$ 
are more bound versions of their counterparts in our 
world \cite{Wagman:2017tmp}.
At large quark masses, where there is no Chiral EFT, Pionless EFT
offers the only viable description of these nuclei. 
One can take few-body observables calculated in Lattice QCD as input
to Pionless EFT, thus bypassing the need for experimental data,
and use Pionless EFT to calculate the structure of heavier nuclei 
\cite{Barnea:2013uqa,Contessi:2017rww}. 
A possible scenario \cite{Beane:2001bc} for quark-mass
variation is one in which the deuteron
and the $^1S_0$ virtual state become, respectively, unbound below and bound 
above, but near, the physical point. 
If this is the case, then the mechanism of fine tuning in QCD is
parallel to that of Feshbach resonances
for atoms, with the magnetic field replaced by the quark masses.

Even when $|a_2|\gg R$ is finite, as for nucleons 
at physical quark masses and $^4$He atoms,
the unitarity limit is useful:
in the ``unitarity window'' $|a_2|^{-1} \ll k \ll R^{-1}$,
$T_2(k)$ is close to the maximum value $\propto k^{-1}$:
\begin{equation}
T_2(|a_2|^{-1} \ll k \ll R^{-1})= \frac{4\pi}{m} \left(ik\right)^{-1}
\left[1+{\cal O}\!\left(kM_{\rm hi}^{-1},k\Lambda^{-1},(ka_2)^{-1}\right)\right].
\label{T2}
\end{equation}
When we retain only the first term, 
there is no dimensionful parameter other than $k$ itself.
The vanishing of the binding energy then is a reflection
of scale invariance. Under a change of scales \cite{Hagen:1972pd}
with parameter $\alpha>0$,
\begin{equation}
r\to \alpha r, 
\qquad
t/m\to \alpha^2 t/m, 
\qquad
\Lambda\to \alpha^{-1} \Lambda,
\qquad
\psi \to \alpha^{-3/2} \psi,
\label{scale}
\end{equation}
the first two terms in Eq. \eqref{L}
--- the nucleon bilinear and the $C_0^{0)}$ contact interaction --- 
are invariant on account of Eq. \eqref{C00unitarity}.
Under a scale change, $mE\to \alpha^{-2}mE$, but in the unitarity limit
there is no scale, so $B_2=0$. 
In this limit the $A=2$ system is also conformally invariant 
\cite{Mehen:1999nd}.

Away from the unitarity limit, 
scale symmetry is explicitly broken by the dimensionful parameter
$C_{0R}^{(0)}=a_2$
in Eq. \eqref{C00prime}.
At subleading orders scale symmetry is also broken by $M_{\rm hi}$ in
the form of the higher ERE parameters.
The dependence of $B_2$ on dimensionful parameters may be determined
with the ``spurion field'' method \cite{dEspagnat:1956},
which is designed to exploit the consequences of an approximate
symmetry.
The idea is that {\it if} under scale invariance these parameters changed
according to their canonical dimension, then the system would remain 
invariant. For example, if $a_2$ changed to
$\alpha a_2$, then the first two terms in Eq. \eqref{L} would still
be invariant.
In that case, the energy after the transformation
should equal the transformed energy: $B_2(\alpha a_2)=\alpha^{-2}B_2(a_2)$.
This implies $B_2(a_2)\propto (ma_2^2)^{-1}$, see Eq. \eqref{B2LO}.
Now, since $a_2$ is actually fixed, $B_2(a_2)$ reflects the
specific way in which $a_2$ breaks scale invariance.
In this particular case the spurion method is just dimensional
analysis, since by allowing $a_2$ to vary we are 
changing all dimensionful quantities appearing to this order
according to their (inverse mass) dimension. 
And, of course, this relation
was obtained earlier directly from Eq. \eqref{TLOren},
but the spurion method illustrates how considerations of symmetry
underlie dynamical results.

The message is that we are dealing with fine-tuned systems, where 
for $A=2$ we are close to the non-trivial fixed point associated
with unitarity and scale invariance. 
There are significant departures from naive dimensional analysis, but
renormalization provides a useful guide to infer 
the corresponding enhancements.

\section{Three-body system}
\label{3body}

Whatever the reason for fine tuning, one can ask what structures
it produces. The first surprise comes when $A=3$.

Many-body forces are not forbidden by any symmetry, and yet we are
used to think of them as small. 
That is a consequence of their high dimensionality.
From dimensional analysis we
might expect them to be highly suppressed, 
{\it e.g.} $D_0={\cal O} (M_{\rm hi}^{-4})$.
If this is the case, the properties
of many-body systems are determined, to a very good
approximation, by two-body interactions.
The simplest connected diagram for three particles 
consists of an LO interaction between
two particles (say 1 and 2), followed by propagation of one particle 
(say 2), and its interaction with the third particle. 
In the next simplest diagram, either particle 2 or 3
further propagates till it interacts with particle 1,
giving rise to a loop.
Using the power-counting rules \eqref{heavyrules},
the expected size of the latter diagram relative to the former is
\begin{equation}
\frac{{\cal O} \left((4\pi C_0^{(0)}/m)^3\, (Q_3^5/(4\pi m))\, (m/Q_3^2)^3\right)}
{{\cal O} \left((4\pi C_0^{(0)}/m)^2 \, (m/Q_3^2)\right)}
={\cal O} (C_0^{(0)} Q_3).
\label{3Nloop}
\end{equation}
This counting extends straightforwardly to diagrams with more loops.
For $Q_3\simge C_{0R}^{(0)-1}= {\cal O}(M_{\rm lo})$, 
the arguments of Sec. \ref{2body} apply to any of the three two-body subsystems,
meaning the LO interactions $C_0^{(0)}$
must resummed into the LO two-body $T$ matrix $T_2^{(0)}$.
Subleading corrections are treated perturbatively.
This argument applies also to the scattering of one particle
on a two-body bound state, such as $nd$ scattering
or particle-dimer scattering in the atomic lingo.
From the corresponding 
three-body $T$ matrix, $T_3$, one can find the three-body bound states.
The issue now is, is $T_3$
properly renormalized up to N$^4$LO, when we naively expect the appearance 
of the first three-body force?

\subsection{Auxiliary field}
\label{auxfields}

In the systems we want to describe, where there are shallow $T$-matrix poles,
it is often times convenient to introduce auxiliary fields with the quantum
numbers of these poles. They can be thought of as ``composite'' fields for
the corresponding states, which are not essential but do 
simplify the description of the larger systems,
particularly reactions involving the bound state.

Most useful is a field for the dimer, the ``dimeron'', 
which I denote by $d$, with the quantum number of the $A=2$ pole
--- first introduced in this context in Ref. \cite{Kaplan:1996nv}.
From the evolution of this field, whose mass is defined as
$2m-\Delta$, $2m$ is removed by a field redefinition.
The corresponding action is obtained by replacing the Lagrangian ${\cal L}$
in Eq. \eqref{L}
by 
\begin{eqnarray}
{\cal L}_d&=&\psi^\dagger \left(i \frac{\partial}{\partial t} 
+ \frac{{\vec{\nabla}}^2}{2m}  +\ldots\right)\psi
+d^\dagger\left[\Delta + \sigma\left(i \frac{\partial}{\partial t} 
+ \frac{{\vec{\nabla}}^2}{4m}  +\ldots\right)\right]d 
\nonumber \\
&&
- \frac{g_0}{\sqrt{2}} 
\left(d^\dagger\,\psi\psi + \psi^\dagger \psi^\dagger\, d\right)
- h_0 \, d^\dagger d \,\psi^\dagger\psi +\ldots,
\end{eqnarray}
where $\sigma=\pm 1$ and $g_0$, $h_0$, $\ldots$ are LECs. 
In particle-dimer scattering with a relative momentum $Q\simle Q_2$, 
when the dimer cannot be broken up,
a lower-energy, Halo EFT can be constructed where 
$d$ is an ``elementary'' field. For that, take $g_0=0$ and $\sigma=+1$,
with $h_0$ being the leading contact interaction between particle and dimer
analogous to $C_0$ in Eq. \eqref{L}.
In the case we are interested in here, $Q\simge Q_2$, 
the coupling $g_0\ne 0$ to two particles ensures the composite
nature of the dimeron field.
The particle-particle interaction proceeds through the dimeron propagator,
and $h_0$ represents a three-body force.
Integrating out the $d$ field brings back Eq. \eqref{L}. 

The power counting of Sec. \ref{2body}
is reproduced if $\Delta={\cal O}(M_{\rm lo})$ and 
$g_0={\cal O}(\sqrt{4\pi/m})$. In this case the kinetic terms
of the dimeron are subleading. 
The full dimeron propagator at LO is the sum of bubbles,
\begin{eqnarray}
iD_2^{(0)}(p^0,\vec{p};\Lambda) &=& \frac{i}{\Delta^{(0)}+i\epsilon} \left[
1 
- \frac{g_0^2}{\Delta^{(0)}+i\epsilon}I_1(k;\Lambda)
+ \left(\frac{g_0^2}{\Delta^{(0)}+i\epsilon}I_1(k;\Lambda)\right)^2 
+\ldots
\right]
\nonumber\\
&=&i \frac{4\pi}{mg_0^2}\left[
\frac{4\pi}{m}\left(\frac{\Delta^{(0)}}{g_0^2} + I_1(k;\Lambda)\right)
+i\epsilon
\right]^{-1},
\label{dimeronpropLO}
\end{eqnarray}
and the NLO correction is
\begin{equation}
iD_2^{(1)}(p^0,\vec{p};\Lambda)= 
i\left(\Delta^{(1)}+\frac{\sigma^{(1)} k^2}{m}\right)
\, \left(iD_2^{(0)}(p^0,\vec{p};\Lambda)\right)^2,
\label{dimeronpropNLO}
\end{equation}
where in these expressions $k\equiv \sqrt{mp^0-{\vec p}^{\,2}/4 +i\epsilon}$.
The dimeron can be thought to represent the bound- or virtual-state
propagator once we multiply numerator and denominator 
by $\Delta^{(0)}/g_0^2 - I_1(k;\Lambda)$ in order to remove
the square root of the energy from the denominator. Expanding around the
pole one can obtain the residue, that is, the wavefunction renormalization,
\begin{equation}
Z_2^{-1}= \left.\frac{\partial}{\partial p^0} 
\left(D_2(p^0,\vec{p})\right)^{-1}\right|_{p_0=-B_2}.
\label{wf}
\end{equation}

Attaching two external particle legs to the dimeron propagator
multiplies it by $-g_{0}^2$ and gives $iT_2$.
Equation \eqref{dimeronpropLO} shows that there is 
only one parameter, $\Delta^{(0)}/g_{0}^2$, at LO. 
This redundancy is frequently eliminated with a redefinition of the auxiliary
field to make \cite{Griesshammer:2004pe}
\begin{equation}
g_{0}\equiv \sqrt{\frac{4\pi}{m}},
\label{haraldschoice}
\end{equation} 
which elevates
$\sigma$ to a full-blown LEC $\sigma(\Lambda)$ rather
than a just sign.
With this choice and 
the renormalization scale $\mu=0$ for simplicity, 
the $T$ matrix has the forms 
\eqref{TLO} and \eqref{TNLOall}
with
\begin{equation}
\Delta_R^{(0)}\equiv \Delta^{(0)}(\Lambda) +\theta_1 \Lambda 
=\frac{1}{C_{0R}^{(0)}}=\frac{1}{a_2},
\qquad
\Delta_R^{(1)}\equiv \Delta^{(1)}(\Lambda) =0,
\label{haraldschoiceconseq1}
\end{equation} 
and
\begin{equation}
\sigma_R^{(1)}\equiv \sigma^{(1)}(\Lambda) + \theta_{-1}\frac{m}{\Lambda}
= \frac{mC_{2R}^{(1)}}{C_{0R}^{(0)2}}= - \frac{mr_2}{2}.
\label{haraldschoiceconseq2}
\end{equation}
Note that $r_2>0$, as in most situations, requires $\sigma <0$, that is,
$d$ is a ghost field --- and yet the two-body amplitude is perfectly fine.
Renormalization is different than before, however,
because the dimeron induces energy-dependent corrections
instead of the momentum-dependent $C_{2}$ corrections.
Thus, no $\Lambda^3$ divergence appears at NLO.
In this case one can resum the NLO corrections without
running into problems with the RG \cite{vanKolck:1998bw},
contrary to the case of momentum-dependent corrections \cite{Beane:1997pk}.

\subsection{Amplitude}
\label{3bodyamplitude}

\begin{figure}[t]
\begin{center}
\includegraphics[scale=.8]{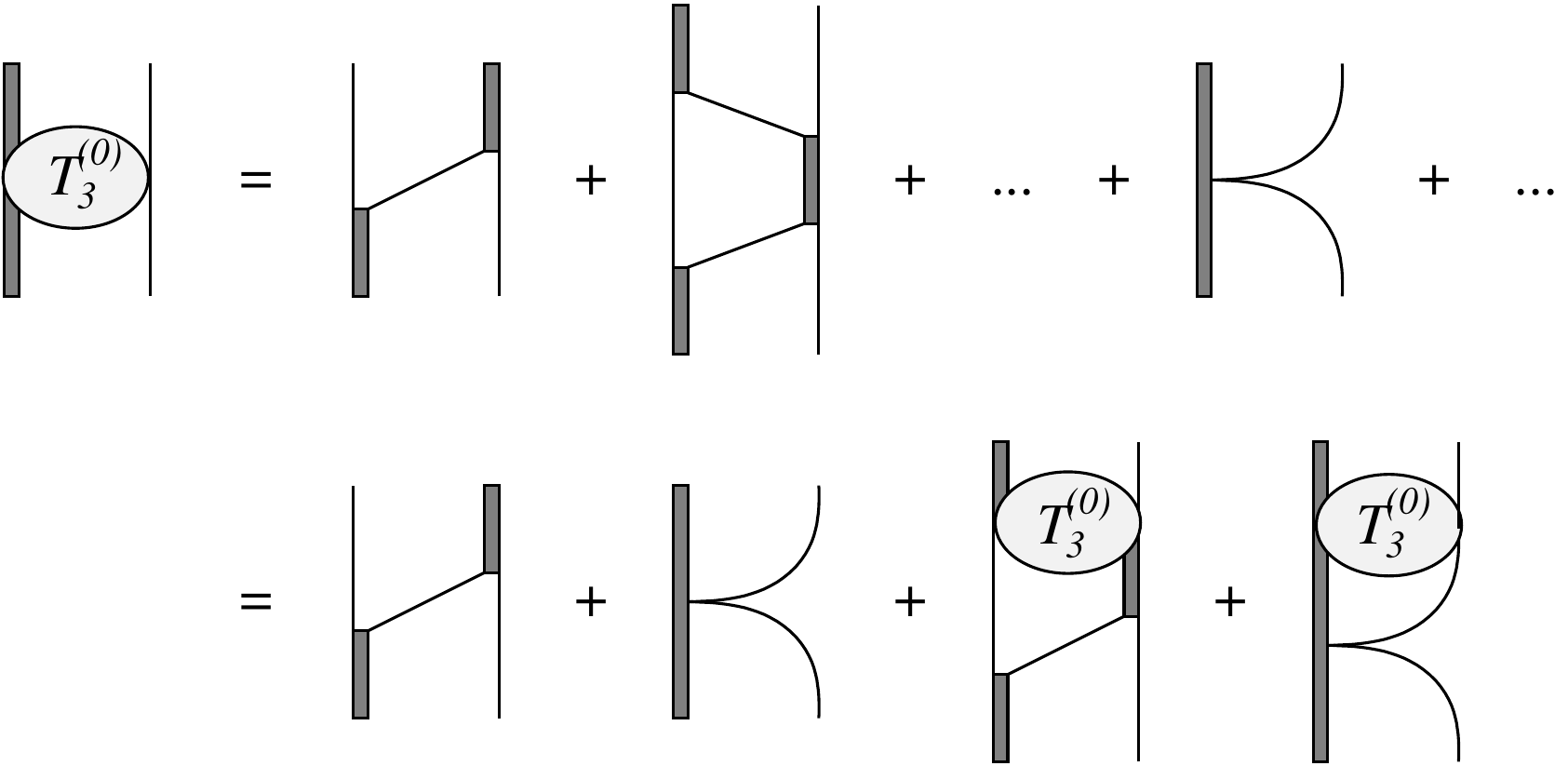}
\end{center}
\caption{Three-body $T$ matrix in Pionless EFT at LO, $T_3^{(0)}$. 
The filled double line represents the full dimeron propagator 
\eqref{dimeronpropLO}, when internal,
or the corresponding wavefunction renormalization \eqref{wf}, 
when external.
The two-particle--dimeron vertex
stands for the $g_0$ interaction, Eq. \eqref{haraldschoice},
while the particle-dimeron contact is the three-body force $h_0^{(0)}$, 
Eq. \eqref{3bf2}.}
\label{figLO3bodyamp}       
\end{figure}

In terms of the auxiliary field, we can represent the scattering
of a particle on a dimer at LO through the ``one-particle''
exchange diagrams shown in Fig. \ref{figLO3bodyamp}.
The whole series of ``pinball'' diagrams with multiple such exchanges
needs to be included on account of Eq. \eqref{3Nloop},
giving rise to an integral equation
known as the Skorniakov--Ter-Martirosian equation.
In these diagrams the dimeron propagator is the LO 
propagator \eqref{dimeronpropLO}.
At NLO, one includes one insertion of the NLO propagator
\eqref{dimeronpropNLO} in all possible ways,
and analogously for higher orders.

Let us work again in the center-of-mass frame, where at LO
the incoming (outgoing) 
dimer has energy $k^2/(4m) -B_2^{(0)}$ ($k^2/(4m) -B_2^{(0)} +E'$)
and momentum $\vec{p}$ ($\vec{p}\,'$),
and the incoming (outgoing) 
particle has energy $k^2/(2m)$ ($k^2/(2m) -E'$)
and momentum $-\vec{p}$ ($-\vec{p}\,'$).
The total energy is $E=3k^2/(4m) -B_2^{(0)}$.
We can take the initial state to be on-shell, $|\vec{p}|=k$. 
For simplicity, we take 
the $\Lambda\to \infty$ limit in the dimer propagator.
The integration over the 0th component of
the loop momentum is similar to the one done in Sec. \ref{2bodyamplitude}:
we pick a pole from, say, the particle propagator, and 
are left with a three-dimensional integral involving the dimeron
propagator.
At this point we can set $E'=(k^2-\vec{p}^{\; '2})/(2m)$, which holds
when the final state is on-shell.
With the choice \eqref{haraldschoice},
we find for the half-off-shell amputated amplitude 
\begin{equation}
t_3(\vec{p}\,', \vec{p}\,)
= -v_3(\vec{p}\,', \vec{p}\,)
-\lambda \int \!\! \frac{d^3l}{(2\pi)^3}
\, \frac{t_3(\vec{p}\,',\vec{l}\,)\, v_3(\vec{l},\vec{p}\,)}
{-1/a_2+\sqrt{3\vec{l}^{\,2}/4-mE}},
\label{t3}
\end{equation}
where 
\begin{equation}
v_3(\vec{p}\,', \vec{p}\,)= 
\frac{8\pi}{mE-\vec{p}\,'^{\,2}-\vec{p}^{\;2}-\vec{p}\,'\cdot\vec{p}}.
\label{v3}
\end{equation}
The above equation with $\lambda=1$ is derived from the EFT for bosons  
\cite{Bedaque:1998kg,Bedaque:1998km}.
For three nucleons with total spin $S=3/2$, the
equation takes the same form but with $\lambda=-1/2$ 
\cite{Bedaque:1997qi,Bedaque:1998mb}.
Instead, when $S=1/2$ one finds a pair of coupled integral equations.
In the ultraviolet (UV) limit where scattering length and binding energy can be 
discarded, these equations decouple \cite{Bedaque:1999ve}
into a pair of equations like Eq. \eqref{t3},
one with $\lambda=1$, the other with $\lambda=-1/2$.
Thus, even though I consider here the single equation \eqref{t3}, the lessons
learned from different values of $\lambda$ can be applied to nucleons
in the triton channel as well.

Now, for simplicity we focus on the most important, $S$ wave.
We can project on it by integrating over the angle between $\vec{p}\,'$
and $\vec{p}$. 
Performing the integration also over the angles in the loop integral,
the equation simplifies to
\begin{equation}
t_{3,0}(p', k)= -v_{3,0}(p', k)
-\frac{\lambda}{2\pi^2}\int_0^\Lambda \!\!dl \, l^2 \,
\frac{t_{3,0}(p', l)\, v_{3,0}(l, k)}
{-1/a_2+\sqrt{3l^{\,2}/4-mE}},
\label{t3S}
\end{equation}
where I took a sharp (three-body) cutoff for definiteness and
\begin{equation}
v_{3,0}(p', k)= 
\frac{4\pi}{{p'} k} 
\ln\left(\frac{{p'}^{2}-p' k+k^2-mE}{{p'}^{2}+p' k+k^2-mE}\right).
\label{v3S}
\end{equation}
The on-shell scattering amplitude is obtained by making $p'=k$
and accounting for wavefunction renormalization,
\begin{equation}
T_{3,0}(k)= \sqrt{Z_2^{(0)}} \, t_{3,0}(k,k) \, \sqrt{Z_2^{(0)}}.
\label{T3fromt3}
\end{equation}

As in the two-body case, the first step to solve the integral equation is
to look at the UV region, $p'\gg k\simge 1/a_2$, 
where the equation reduces to
\begin{equation}
t_{3,0}(p'\gg k)= 
\frac{4\lambda}{\sqrt{3}\pi}
\int_0^\Lambda \!\!\frac{dl}{p'} \,  
\ln\left(\frac{{p'}^{2}+p'l+l^2}{{p'}^{2}-p'l+l^2}\right)
\,  t_{3,0}(l\gg k).
\label{t3UV}
\end{equation}
This equation is homogeneous so it cannot fix the overall
normalization of $t_{3,0}(p'\gg k)$, but it does determine
the dependence on $p'$ in the region $\Lambda > p'\gg k$.
Scale invariance \eqref{scale} suggests the {\it Ansatz}
$t_{3,0}(p'\gg k)\propto {p'}^{-(s+1)}$, which works if $s$ obeys
\begin{equation}
\frac{8\lambda}{\sqrt{3}s}
\frac{\sin(\pi s/6)}{\cos (\pi s/2)} =1.
\label{seq}
\end{equation}
This relation is analyzed in detail in Ref. \cite{Griesshammer:2005ga}.
Because of the additional inversion symmetry $p'\to 1/p'$, the roots
come in pairs.
For $-1/2\le \lambda\le \lambda_c \equiv 3\sqrt{3}/(4\pi) 
\simeq 0.4135$, the roots are real. 
The root with $\Re(s)> -1$ ensures that $t_{3,0}(p'\gg k)$ goes to zero,
in which case the amplitude has no essential sensitivity
to the regulator and predictions about the three-body system can
be made at LO.
In particular, for three nucleons with $S=3/2$, 
when $\lambda=-1/2$, one finds that $t_{3,0}(p'\gg k)\propto {p'}^{-3.17}$,
which is softer than the ${p'}^{-2}$ behavior expected in perturbation
theory from $v_{3,0}(p'\gg k)\propto {p'}^{-2}$. 
The numerical solution of Eq. \eqref{t3S} gives a low-energy
amplitude in good agreement with phenomenology, which improves
at subleading orders \cite{Bedaque:1997qi,Bedaque:1998mb}. 
Because of the good UV behavior of the LO amplitude,
one can resum higher-order terms to make calculations
easier without jeopardizing RG invariance.
As an example, the $S=3/2$ $nd$ scattering length $a_{nd, S=3/2}
= 5.09 + 0.89 + 0.35 + \ldots$ fm $= 6.33 \pm 0.05$ fm 
\cite{Bedaque:1997qi}, to be compared with the experimental
value $6.35 \pm 0.02$ fm \cite{Dilg:1971gqb}.

In contrast, for other $\lambda$ values
the solutions are complex, and for 
$\lambda> \lambda_c \equiv 3\sqrt{3}/(4\pi)$ the roots are imaginary.
In particular, for the bosonic case $\lambda=1$ there is a pair of
imaginary solutions $s = \pm i s_0$, with $s_0 \simeq 1.00624$. The two
solutions are equally acceptable (or actually unacceptable...) and
lead to an asymptotic behavior of
the half-off-shell amplitude of the form
\begin{equation}
t_{3,0} (p' \gg k) \propto
\cos \left(s_0 \ln (p'/\Lambda) +\delta \right),
\label{t3Sasym}
\end{equation}
where 
$\delta$ is a dimensionless, $p'$-independent number. 
A numerical solution of Eq. \eqref{t3S} confirms this oscillatory
behavior with $\delta=0.76\pm 0.01$ \cite{Bedaque:1998kg,Bedaque:1998km}.
Small changes in $\Lambda$ propagate to lower momenta
and lead to dramatic changes in the observable $t_{3,0}(k, k)$,
but the changes are periodic.
One can show that this solution, first found in Ref. \cite{Danilov1961}, 
supports a sequence of bound states that appear with
the same periodicity as $\Lambda$ increases,
with the binding energy of each state growing as 
$\Lambda^2/m$ \cite{Thomas:1935zz}.
This solution is obviously unacceptable: the first two terms
in Eq. \eqref{L} are not renormalizable beyond $A=2$.

How can we maintain RG invariance? The only possibility is a three-body
force, and the one provided by $h_0$ is the simplest. 
For this fix to work, 
this force has to be assumed to be LO, but even then it is not obvious that 
it can remove the regulator dependence when iterated.
Upon including
\begin{equation}
h_0^{(0)}(\Lambda) \equiv 8\pi \frac{H(\Lambda)}{\Lambda^2},
\label{3bf2}
\end{equation}
where $H(\Lambda)$ is dimensionless,
we have additional diagrams, also shown in
Fig. \ref{figLO3bodyamp}. The LO amplitudes
$t_3^{(0)}(\vec{p}\,', \vec{p})$ 
and $t_{3,0}^{(0)}(p', k)$ still satisfy Eqs. \eqref{t3} and \eqref{t3S}
but with, respectively,
\begin{eqnarray}
v_3(\vec{p}\,', \vec{p}\,)&\to& v_3(\vec{p}\,', \vec{p}\,) +h_0^{(0)}(\Lambda)
\equiv v_3^{(0)}(\vec{p}\,', \vec{p};\Lambda),
\label{v3LO}
\\
v_{3,0}(p', k)&\to& v_{3,0}(p', k) +h_0^{(0)}(\Lambda) 
\equiv v_{3,0}^{(0)}(p', k;\Lambda).
\label{v3SLO}
\end{eqnarray}
The asymptotic equation \eqref{t3UV} now becomes, for 
the physically relevant $\lambda=1$,
\begin{equation}
t_{3,0}^{(0)}(p'\gg k)= 
\frac{4}{\sqrt{3}\pi}
\int_0^\Lambda \!\!\frac{dl}{p'} \,  
\left[\ln\left(\frac{{p'}^{2}+p'l+l^2}{{p'}^{2}-p'l+l^2}\right)
- 2 \,\frac{p'l}{\Lambda^2} \, H(\Lambda)\right]
\,  t_{3,0}^{(0)}(l\gg k).
\label{t3UVcorrect}
\end{equation}
Only for $p'\sim \Lambda$ is the three-body force important.
In the region $p'\ll \Lambda$, the behavior \eqref{t3Sasym} still holds,
but now $\delta$ is determined by $H(\Lambda)$. 
We can define the physical, dimensionful parameter $\Lambda_\star$
through
\begin{equation}
\delta(H(\Lambda))=s_0 \ln (\Lambda/\Lambda_\star).
\label{Lambdastar}
\end{equation}
Since
\begin{equation}
t_{3,0}^{(0)} (p' \gg k) \propto
\cos \left(s_0 \ln (p'/\Lambda_\star)\right)
\label{t3Sasymcorrect}
\end{equation}
is now essentially cutoff independent, so will the low-energy
on-shell amplitude.
$\Lambda_\star$ can then be
determined from low-energy data or matching to the underlying theory.
Again, numerical experimentation shows \cite{Bedaque:1998kg,Bedaque:1998km}
that this can indeed be achieved.
One can also show that bound states now accrete periodically from below
(that is, from very large binding energies)
as the regulator cutoff becomes large enough to accommodate them.
As $\Lambda$ increases their binding energies approach constants.
With the addition of the three-body force, the EFT is renormalizable
at LO for $A=3$.

\begin{figure}[t]
\begin{center}
\includegraphics[scale=.6]{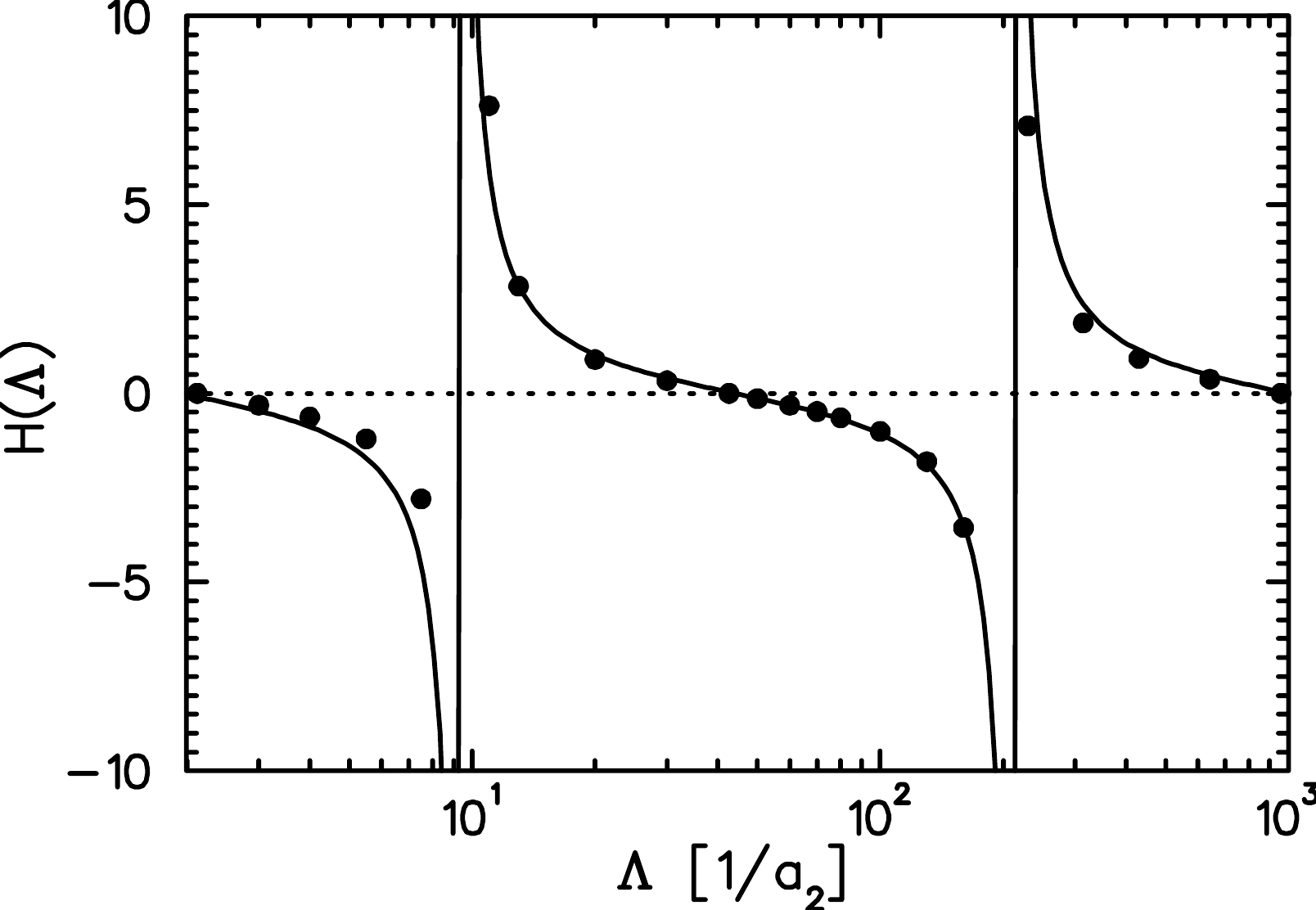}
\end{center}
\caption{Dimensionless three-body force $H$ in Pionless EFT at LO,
Eq. \eqref{3bf2},
as a function of the regulator cutoff $\Lambda$ (in units of $a_2^{-1}$):
numerical solution for $a_3=1.56 \, a_2$, dots; and Eq. \eqref{H}
with $\Lambda_\star=19.5 \, a_2^{-1}$, solid line.
Reprinted from Ref. \cite{Bedaque:1998km} with permission from Elsevier.
}
\label{figLO3bodyforce}       
\end{figure}

An approximate form can be obtained for $H(\Lambda)$ 
by going back to Eq. \eqref{t3S} for two values of
the regulator cutoff, $\Lambda$ and $\Lambda'>\Lambda$.
Imposing that the two equations agree in the region $\Lambda> p\gg k$, 
and making the approximation \eqref{t3Sasymcorrect} for the $\Lambda'$
solution also when $p'\sim \Lambda'$, one finds 
\cite{Bedaque:1998kg,Bedaque:1998km}
\begin{equation}
H(\Lambda) \simeq c \, \frac
{\sin\left(s_0 \ln (\Lambda/\Lambda_\star)-\tan^{-1}(s_0^{-1})\right)}
{\sin\left(s_0 \ln (\Lambda/\Lambda_\star)+\tan^{-1}(s_0^{-1})\right)},
\label{H}
\end{equation}
where $c\simeq 1$.
$H(\Lambda)$ can also be extracted purely numerically by demanding
that one low-energy datum (for example, the particle-dimer scattering length 
$a_3$)
be reproduced at any value of $\Lambda$.
The agreement between approximate and numerical results is
shown in Fig. \ref{figLO3bodyforce}. 
The best fit gives $c=0.879$ \cite{Braaten:2011sz}.

At tree level, the $h_0$ particle-dimeron interaction generates
the three-particle force 
\begin{equation}
D_0^{(0)}(\Lambda)=
\frac{3h_0^{(0)}(\Lambda)}{4\pi\Delta^{(0)2}(\Lambda)}
\propto \frac{1}{\Lambda^4} \,
\frac{\sin \!\left(s_0 \ln(\Lambda/\Lambda_\star) -\tan^{-1}(s_0^{-1})\right)}
{\sin \!\left(s_0 \ln(\Lambda/\Lambda_\star) +\tan^{-1}(s_0^{-1})\right)}
\left[1+ {\cal O}\left((C_{0R}^{(0)}\Lambda)^{-1}\right)\right].
\label{D0}
\end{equation}
In coordinate space, the corresponding potential is
\begin{equation}
V_3^{(0)}(\vec{r}_{12},\vec{r}_{23};\Lambda) = \frac{(4\pi)^2}{m} 
D_0^{(0)}(\Lambda)\;
\delta^{(3)}_\Lambda(\vec{r}_{12})\delta^{(3)}_\Lambda(\vec{r}_{23}).
\label{Vijk}
\end{equation}
where $\vec{r}_{ij}$ is the position of particle $i$ with respect to
particle $j$.

The argument above applies directly to bosons and indirectly to
nucleons with $S=1/2$.
Therefore a
three-nucleon force is needed for RG invariance \cite{Bedaque:1999ve}, 
consistently
with the fact that the $D_0$ force has a non-vanishing projection onto
the $^2S_{1/2}$ channel.
In channels with angular momentum $l>0$ similar equations are obtained 
with the logarithm replaced by Legendre polynomials of the second kind
\cite{Griesshammer:2005ga}. 
An equation for $s$ analogous to \eqref{seq} is obtained, involving
a hypergeometric function. For both $\lambda=1$ and $\lambda=-1/2$,
$s\simeq l+1$ in fair agreement with the expectation
from perturbation theory, which can be shown from the Legendre polynomials 
to be $t_{3,l}(p'\gg k)\propto p^{-(l+2)}$.
There is therefore no need for additional three-body forces at LO.

The perturbative NLO correction that accounts for two-body range effects 
induces a finite change in the three-body system and 
does not require an additional three-body force for RG invariance
\cite{Bedaque:1998km,Hammer:2001gh};
a correction $D_0^{(1)}(\Lambda)$ is sufficient.
A two-derivative three-body force does enter,
however, at N$^2$LO \cite{Bedaque:2002yg}.
Thus, while the LO three-body force is
enhanced by $(M_{\rm hi}/M_{\rm lo})^4$ over simple dimensional
analysis, three-body force corrections, which are amenable to
perturbation theory, seem to be suppressed by the expected relative factors
of $M_{\rm lo}/M_{\rm hi}$.

\subsection{Bound states and correlations}
\label{phillips}

The EFT produces a 
series of discrete bound states whose spacing depends
on the two-body scattering length.
The three-body binding momenta quickly exceed $M_{\rm hi}$,
so that only a finite number of states 
($\sim \ln(|a_2|/R)/\pi$ for an underlying potential of range 
$R$ \cite{Amado:1992vn,Bedaque:1999ve})
are within the range of applicability of the EFT.
For atomic $^4$He, for example,
both the ground \cite{Schoellkopf:1996} and 
first-excited \cite{Kunitski:2015qth}
states have been detected, with a ratio of binding energies 
of about 60, see Table \ref{tbl:BEs}. 
For nucleons only the triton (and helion, separated
from triton only by small isospin-breaking effects) is observed, 
but there is a {\it virtual} $nd$ state 
with a binding energy about 6 times smaller.

\begin{figure}[t]
\begin{center}
\includegraphics[scale=.9]{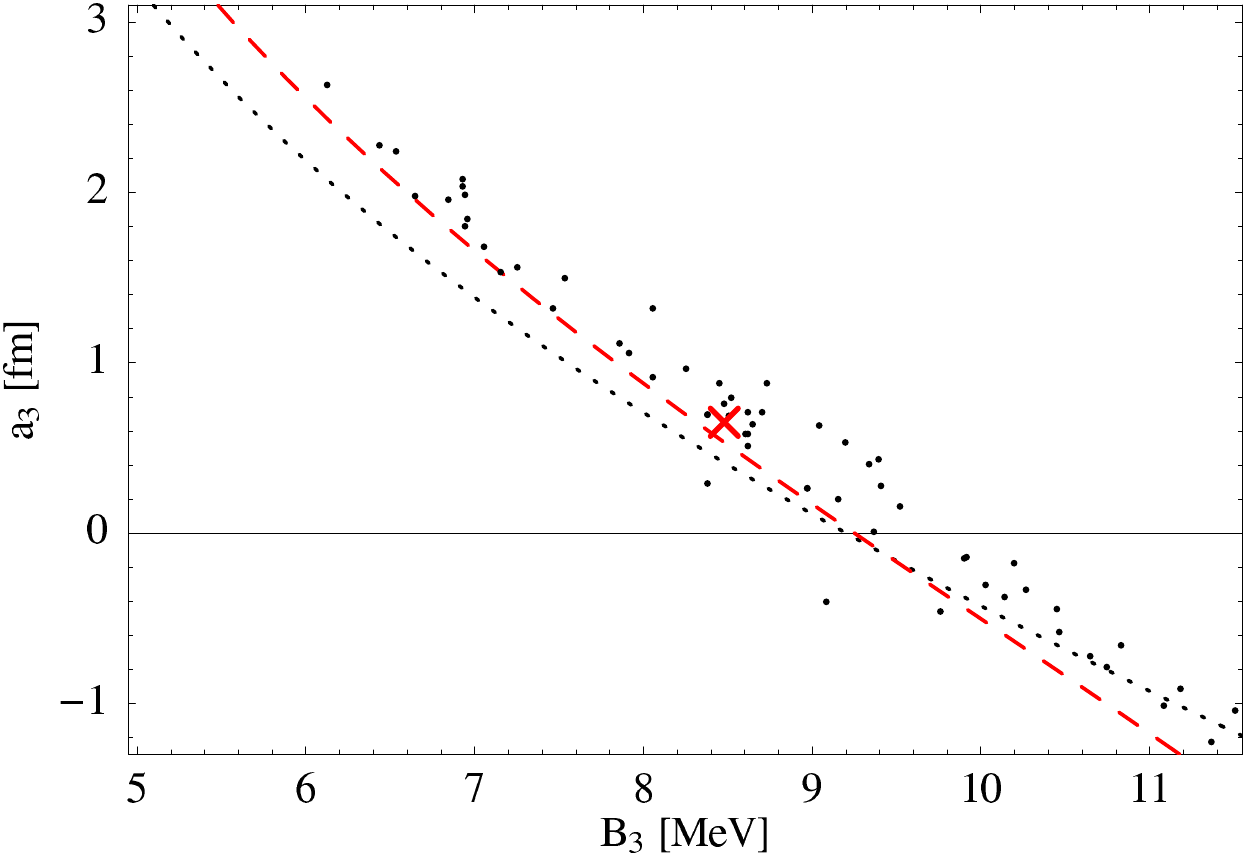}
\end{center}
\caption{Phillips line in the plane of doublet $nd$ scattering
length $a_3\equiv a_{nd,S=1/2}$ (in fm) 
and triton binding energy $B_3\equiv B_t$ (in MeV),
in Pionless EFT: LO, (black) dotted line; and NLO, (red) dashed line.
The dots represent a variety of nuclear potentials with the same
two-nucleon scattering lengths and effective ranges \cite{Efimov1988}.
The cross is the experimental result.
Reprinted from Ref. \cite{Bedaque:2002yg} with permission from Elsevier.
}
\label{figphillipsline}       
\end{figure}

Because a single parameter emerges in the three-body force up to NLO,
one expects correlations among these binding energies and phase
shifts in channels not affected by the exclusion principle.
The classic example is the Phillips line \cite{Phillips:1968zze}: 
a line in the plane spanned by the triton binding energy $B_t$
and the $S=1/2$ $nd$ scattering length $a_{nd,S=1/2}$.
This correlation was first discovered empirically, 
as a line formed by points representing various phenomenological 
potentials, which describe two-nucleon data up to relatively high
momenta. 
From the potential-model perspective, this line is a mystery:
one would expect results to form an amorphous cluster 
around the experimental point.
From the EFT point of view, instead, this line is indication
that these potentials differ by one relevant parameter 
not determined by two-body physics.
As $\Lambda_\star$ is varied, the LO EFT also produces a 
line \cite{Bedaque:1998km,Bedaque:1999ve},
which lies close not only to the experimental point
but also to the phenomenological line.
At NLO the line position changes \cite{Bedaque:2002yg}, 
approaching models and experiment,
see Fig. \ref{figphillipsline}.
Taking the EFT error into account, the line is actually a band.
This generalizes an earlier (regulator-dependent) explanation 
\cite{Efimov1988}.

As one would expect \cite{Bedaque:1998km}, 
this property is generic and $^4$He potentials also fall on 
a Phillips line \cite{Roudnev2012}. 
Other correlations can be understood similarly.
This means that the various phenomenological potentials, 
with their many parameters and varied forms,
are basically equivalent to the same EFT with different values 
of $\Lambda_\star$. 
For $A\ge 3$, Pionless EFT is definitely not just the ERE.

The proximity of the EFT Phillips line
to the experimental point means that once one datum is used to determine
$\Lambda_\star$ at LO and NLO, other three-body data can be predicted
or postdicted in agreement with experiment.
For example, if we fit $a_{nd,S=1/2}$, the triton binding energy 
$B_t= 8.0 + 0.8 + \ldots$ MeV \cite{Bedaque:1999ve,Hammer:2001gh},
compared with the measured $8.48$ MeV. Alternatively,
one can use the experimental value of $B_t$ as input.
Agreement with phase shifts
is good already at LO \cite{Bedaque:1999ve},
improves at NLO \cite{Hammer:2001gh},
and improves further still at N$^2$LO \cite{Bedaque:2002yg},
where a second three-body parameter is needed as input.
The three-body amplitude and various
observables have been calculated fully perturbatively
up to N$^2$LO for bosons in Refs. \cite{Ji:2011qg,Ji:2012nj}
and for nucleons in Ref. \cite{Vanasse:2013sda}.
Reviews of the state-of-the-art three-body calculations in Pionless EFT
can be found in Refs. \cite{Ji:2015jgm,Vanasse:2016jtc}.

\section{Limit cycle and discrete scale invariance}
\label{DIS}

The three-body force \eqref{D0} has a surprising cutoff dependence.
$H(\Lambda)$ is the solution of an unusual RG equation,
\begin{equation}
\mu \frac{d}{d\mu} H(\mu) \simeq
2\left(1+H^{2}(\mu)\right)
\label{D0RGE}
\end{equation}
and is log-periodic, taking the same
value for $\Lambda$ and  $\alpha_l\Lambda$, with
\begin{equation}
\alpha_l = e^{l\pi/s_0}\simeq (22.7)^{l},
\label{disscale}
\end{equation}
$l$ an integer.
This is an RG limit cycle. 
The possibility of such a cycle in QCD had been 
conjectured \cite{Wilson:1974sk},
and the three-body system provided the first example
in a field theory \cite{Wilson:2004de}.
Not many such examples exist ---
for a short review, see Ref. \cite{Bulycheva:2014twa}.

This force  appears at LO 
not only for small $|a_2^{-1}|$ but also in the unitarity limit.
In this limit there is no $A=2$ dimensionful parameter at LO, 
yet renormalization for $A=3$ forces on us a dimensionful parameter 
$\Lambda_\star$. This is an example of ``dimensional transmutation'':
the scale invariance present in the unitarity limit is ``anomalously'' broken.
The limit cycle signals a remaining discrete
scale invariance (DSI) \cite{Bedaque:1998kg,Bedaque:1998km,Bedaque:1999ve}. 
Because of the characteristic dependence on $\Lambda$
in Eq. \eqref{D0}, the three-body term in Eq. \eqref{L} 
is invariant under the transformation
\eqref{scale}, but only for the discrete values \eqref{disscale}.
Other examples of the anomalous breaking of scale invariance
and of DSI with its characteristic log-periodicity 
are discussed in Refs. \cite{Camblong:2003mz}
and \cite{Sornette:1997pb}, respectively.

The first consequence of the breaking of scale invariance is that 
$\Lambda_\star$ offers a dimensionful scale for binding energies.
By dimensional analysis, the three-body binding energies 
can be written as
\begin{equation}
B_{3;n} = \frac{\kappa_\star^2}{m} \; 
\beta_{3;n}\!\left((a_2\kappa_\star)^{-1}\right)
=\frac{\kappa_\star^2}{m} \;\left[\beta_{3;n}(0)
+\frac{\beta_{3;n}'(0)}{a_2\kappa_\star}+\dots\right],
\quad
\beta_{3;0}\!\left(0\right)=1,
\label{B3n}
\end{equation}
where $n\ge 0$ ($n=0$ denoting the deepest state within the EFT),
$\beta_{3;n}((a_2\kappa_\star)^{-1})$ are universal, dimensionless functions, 
and $\beta_{3;n}(0)$, $\beta_{3;n}'(0)$, {\it etc.} are pure numbers
arising from an expansion in $(a_2\kappa_\star)^{-1}$.
Because $\Lambda_\star$ is only defined up to a factor
$\exp(n_\star\pi/s_0)$, with $n_\star$ an integer, it was traded above
by a fixed scale $\kappa_\star$ defined from the ground-state
energy at unitarity:
\begin{equation}
\ln(\kappa_\star)=\ln(\beta \, \Lambda_\star) \mod\pi/s_0 , 
\label{kappastar}
\end{equation}
with $\beta \simeq 0.383$ \cite{Braaten:2004rn}.

DSI manifests itself in the spectrum.
The energy of a bound state after a discrete scale transformation
should equal the transformed energy but not necessarily of the same level,
so that 
\begin{equation}
\beta_{3;n+l}(0)= \alpha_l^{-2}\, \beta_{3;n}(0) 
\quad
\Leftrightarrow
\quad
\beta_{3;n}(0)= e^{-2n\pi/s_0} \beta_{3;0}(0).
\label{beta3nunit}
\end{equation}
Thus discrete scale invariance leads to a geometric tower of states
extending up to threshold, 
with successive states having a ratio of binding energies
\begin{equation}
B_{3;n+1}/B_{3;n}=\exp(-2n\pi/s_0)\simeq 1/515.
\end{equation}
This amazing structure was first predicted by Efimov \cite{Efimov:1970zz}
and its signals have now been seen in cold-atom systems around
Feshbach resonances, see for example Refs. \cite{grimm,grimm2}.

Away from unitarity DSI is only an approximate symmetry, even at LO.
Although the deep spectrum might be little affected,
a finite $a_2^{-1}$ distorts the spectrum in the infrared (IR).
Using the spurion field method, the deviation from unitarity
due to the two-body scattering length can be followed,
\begin{equation}
\beta_{3;n+l}\!\left((\alpha_l a_2\kappa_\star)^{-1}\right)
= \alpha_l^{-2}\,\beta_{3;n}\!\left((a_2\kappa_\star)^{-1}\right).
\label{beta3n}
\end{equation}
This relation gives information about how Efimov's
tower evolves as $|a_2^{-1}|$ grows. For example, 
taking a derivative and expanding
in $(a_2\kappa_\star)^{-1}$, we see the leading effect
of tower deformation:
\begin{equation}
\beta_{3;n+l}'(0)= \alpha_l^{-1}\, \beta_{3;n}'(0) 
\quad
\Leftrightarrow
\quad
\beta_{3;n}'(0)=  e^{-n\pi/s_0} \beta_{3;0}'(0),
\label{B3nexpanded}
\end{equation}
where $\beta_{3;0}'(0)\simeq 2.11$ \cite{Braaten:2004rn}.
Note that here the spurion method is not simply
dimensional analysis because $\Lambda_\star$ is kept fixed.
It instead tracks how the two-body scattering length
explicitly breaks DSI.
Equation \eqref{beta3n} is only valid to the extent that
the three-body force retains DSI except
for $(a_2\Lambda)^{-1}$ terms --- that is, as long as Eq. \eqref{D0} contains
no $a_2\Lambda_\star$ dependence, which would require in the spurion
method that we scaled $\Lambda_\star$ as well.

As $a_2^{-1}>0$ grows the two-body bound state moves away
from threshold quadratically in $a_2^{-1}$. Progressively more excited Efimov 
states fail to have energy below the particle-dimer threshold,
disappearing as virtual states \cite{Adhikari:1982zz,Rupak:2018gnc}.
We have the counter-intuitive situation
where {\it fewer} three-body states survive 
as the two-body attraction increases.
For $a_2^{-1}<0$, three-body bound states exist even though there are
no two-body bound states --- the system is said to be Borromean 
in reference to the coat of arms of the Borromeo family,
which displays three rings with the property that, when one is removed,
the other two are free.
The properties of the Efimov spectrum are discussed in detail in
Ref. \cite{Braaten:2004rn}.

NLO corrections from the two-body effective range can be handled similarly.
Since Eq. \eqref{D0} contains no $r_2\Lambda_\star$ dependence,
the coefficients of linear
corrections should scale with $\alpha_l^{-3}$ \cite{Platter:2008cx},
as can be easily verified with the spurion method.
(However, an explicit calculation \cite{Platter:2008cx}
indicates that these coefficients vanish.)
The generalization to higher orders is obvious.

In the nuclear case,
the existence of a single three-body force at LO leads to an additional
approximate symmetry, which is exact in the unitarity window:
independent rotations in spin and isospin,
which form the SU(4)$_{\rm W}$ group proposed by Wigner \cite{Wigner:1936dx}
to explain some of the properties of heavier nuclei.
Away from unitarity, the symmetry is broken by the difference 
in inverse scattering lengths between $^3S_1$ and $^1S_0$ channels,
in ranges at NLO, {\it etc.}
The approximate SU(4)$_{\rm W}$ symmetry of this EFT was elaborated upon
in Refs. \cite{Mehen:1999qs,Vanasse:2016umz}.

\section{More-body systems}
\label{morebody}

Let us summarize the EFT so far.
At LO, the action is given by one-body kinetic, two-body $C_0$ and 
three-body $D_0$ terms in Eq. \eqref{L},
and at NLO by the two-body $C_2$ term. 
Other interactions need to be accounted for at N$^2$LO
--- including another three-body force and, for nucleons, 
a two-body tensor force --- and higher orders.
At two-body unitarity, there is a single scale
at LO, $\Lambda_\star$ 
(or equivalently the $\kappa_\star$ of Eq. \eqref{kappastar}).
The crucial issue now is whether higher-body forces appear at LO.
If they do, new scales will be introduced in an essential,
nonperturbative way. 
If they do not, all low-energy properties for $A\ge 4$ can be predicted at LO,
and the issue becomes whether any of these higher-body forces 
show up at NLO, or at another relatively low order that causes sizable
distortions to the LO predictions.

A difficulty we face in answering these questions is the complexity of
$A\ge 4$ calculations. 
In perturbative EFTs, where there is no fine tuning to dramatically 
enhance LO, the size of interactions can be guessed by naive dimensional
analysis \cite{Manohar:1983md}. This rule is inferred by looking at the
regulator dependence of loops in perturbation theory.
In our case, we need instead to look at the
regulator dependence of an integral equation at LO
and of the distorted Born approximation in subleading orders,
as we have just done for $A=3$.

All $A\ge 4$ Pionless EFT calculations that I am aware of
are based on the numerical
solution of (some version of) the many-body Schr\"odinger equation.
At LO this is done with the potential
\begin{equation}
V^{(0)}= \sum_{\{ij\}}\; V_2^{(0)}(\vec{r}_{ij};\Lambda)
+\sum_{\{ijk\}}\; V_3^{(0)}(\vec{r}_{ij},\vec{r}_{jk};\Lambda),
\label{sumVijVijk}
\end{equation}
where 
${\{ij\}}$ and ${\{ijk\}}$ denote doublets and triplets, respectively,
while $V_2^{(0)}$ and $V_3^{(0)}$
are given by Eqs. \eqref{regdeltafunc} and  \eqref{Vijk}.
If a many-body force is missing,
many-body observables will not be renormalized properly ---
they will fail to converge as the regulator cutoff $\Lambda$
increases.
If there is one regulator for which lack of convergence is seen,
renormalizability is disproved. 

A hand-waving argument suggests that higher-body forces are {\it not}
needed at LO for RG invariance. 
The two-body potential \eqref{regdeltafunc} is singular but
$C_0^{(0)}(\Lambda)$ in Eq. \eqref{C00prime}
has an overall $\Lambda^{-1}$ 
so that the potential scales at high momentum just as the kinetic repulsion. 
Smaller terms $\propto (C_{0R}^{(0)}\Lambda)^{-1}$ are adjusted to give
enough attraction for the two-body
state to be slightly bound, or slightly virtual.
When we embed the two-body potential in an $A$-body system,
the number $A(A-1)/2$ of doublets grows faster than the number 
$A-1$ of kinetic terms (one term goes into the kinetic energy of the center
of mass), so the system collapses \cite{Thomas:1935zz}.
An effectively repulsive three-body force --- Eq. \eqref{Vijk},
which at high momentum scales just as the two-body potential thanks
to Eq. \eqref{D0} ---
is needed to keep
$A=3$ stable. Because the number $A(A-1)(A-2)/6$ of triplets 
grows even faster than doublets, $A\ge 4$ systems should not collapse
but have instead well-defined limits for $\Lambda \simge M_{\rm hi}$.

This argument, simplistic as it is, seems to be borne out by
explicit calculations. 
The pioneering $A=4$ calculations of
Refs. \cite{Platter:2004he,Platter:2004zs,Hammer:2006ct} 
have found convergence --- at least in the range of cutoff values examined ---
in the binding energies of the ground state 
for bosons and nucleons, 
as well as of the first excited state for bosons.
This result has been confirmed several times afterwards with
various regulators, for example Refs. 
\cite{Kirscher:2009aj,Kirscher:2015yda,Bazak:2016wxm,Konig:2016utl,
Contessi:2017rww}.
Similarly, the ground-state energies of $A=6,16$ nucleons
converge without many-nucleon forces \cite{Stetcu:2006ey,Contessi:2017rww}.
The four spin-isospin nucleon states require
five- and more-body forces to include derivatives, which should 
suppress them.
This exclusion-principle suppression 
is absent for bosons, but calculations of $A=5,6$ 
ground-state energies \cite{Bazak:2016wxm}
showed no evidence of the need for those forces, either.
In fact, binding energies show  
just the behavior \eqref{obscutdep} expected from a
properly renormalized order.
Discussions found in the literature regarding this issue are summarized
in Ref. \cite{Kolck:2017zzf}.

The absence of LO higher-body forces 
has fundamental implications for the physics of $A\ge 4$ systems.
One is that there are
correlations among low-energy many-body observables
through $\Lambda_\star$, similar to the Phillips line.
The simplest example is the Tjon line \cite{Tjon:1975sme} 
in the plane of the four- and
three-body ground-state energies, $B_{4;0}$ and $B_{3;0}$.
As with the Phillips line, this correlation was discovered empirically
by plotting results of phenomenological nuclear potentials.
It also exists for $^4$He atoms \cite{NakAkaTanLim78}.
It materializes in EFT as a variation of $\Lambda_\star$
at fixed two-body input
\cite{Platter:2004he,Platter:2004zs,Hammer:2006ct}.
The EFT line at LO again is close to both the phenomenological line
and the experimental point, suggesting the EFT might converge
for $A=4$ as well.
There is at least a very large class of potentials that 
do not seem to have an extra, essential parameter introduced
by a four-body force.
The absence of higher-body forces at LO 
further implies the existence of ``generalized Tjon lines'' in the 
planes spanned by other ground-state 
energies, for example \cite{Bazak:2016wxm}
$B_{5;0}$ or $B_{6;0}$ for $A=5,6$, and $B_{3;0}$.
Again, such correlations had been discovered earlier in the context of
potential models \cite{NakLimAkaTan79,LimNakAkaTan80}.

Table \ref{tbl:BEs} summarizes existing results for binding energies of nuclei
and $^4$He atoms at LO in EFT. 
For nuclei \cite{Bedaque:1999ve,Stetcu:2006ey,Contessi:2017rww,Rupak:2018gnc}, 
one can see agreement at the level
expected from an expansion where one of the parameters is 
$r_2/a_2 \sim 30\%$ (in the $^3S_1$ channel). 
Results for atomic $^4$He obtained with
potential models \cite{kalos1981,BluGre00,HiyKam12a}
and with LO EFT \cite{Bedaque:1998km,Platter:2004he,Bazak:2016wxm}
are also given in Table \ref{tbl:BEs}.
Here the discrepancy is no larger than $\simeq 10 \%$
for $A\le 6$, which is surprising because
an estimate of the binding momentum, Eq. \eqref{bindingmom},
suggests that $Q_6\, l_{\rm vdW} \sim 1.3$.
Perhaps Eq. \eqref{bindingmom} for $Q_A$ is an overestimate.

Until recently, calculations for $A\ge 4$ that went beyond LO 
included a resummation of the NLO two-body interaction.
Although they show improved results over LO for 
$A=4, 16, 40$ nuclei \cite{Kirscher:2009aj,Lensky:2016djr,Bansal:2017pwn},
they are limited to cutoff values $\Lambda\simle M_{\rm hi}$.
A test of RG invariance requires a perturbative treatment of subleading
corrections, which was carried out for bosons at NLO in 
Ref. \cite{Bazak:2018qnu}.
Surprisingly, a four-body force is necessary and sufficient 
for renormalization of the $A\ge 4$ energies. 
Once it is fixed to $B_{4;0}$, 
$B_{5;0}^{[1]}$ and $B_{6;0}^{[1]}$ come out well,
strengthening the case that Pionless EFT converges better than expected.
The limit of validity of Pionless EFT remains an open question.

Implications of the absence of other LO forces
are even stronger at unitarity, where DSI is expected
to hold: except for small corrections, 
all states within the validity of the EFT, {\it i.e.},
those states that are insensitive to the details of physics
at distances $r\simle R$, are fixed by 
a {\it single} parameter $\Lambda_\star$.
Reference \cite{Hammer:2006ct} discovered that for bosons
an $A=3$ state spawns two $A=4$ states,
one very close to the $A=3$ threshold and one about four times deeper. 
According to the accurate calculation of Ref. \cite{Deltuva:2010xd},
for the two lowest $A=4$ states at unitarity,
$B_{4;0;0}/B_{3;0}\simeq 4.611$ and $B_{4;0;1}/B_{3;0}\simeq 1.002$.
These states have been spotted in atomic systems \cite{Ferlaino:2009zz}.

Remarkably, potential-model calculations show that
this doubling process continues with increasing number of {\it bosons} 
\cite{Ste10,Gattobigio:2011ey,vonStecher:2011zz,Gattobigio:2012tk},
so that for a given $A$ there are $2^{(A-3)}$ ``interlocking'' towers of states.
Generalizing Eq. \eqref{B3n},
\begin{equation}
B_{A;n;\{i\}} = \frac{\kappa_\star^2}{m} \; 
\beta_{A;n;\{i\}}\!\left((a_2\kappa_\star)^{-1}\right),
\label{BNn}
\end{equation}
by labeling each state with the $A=3$ ancestor state ($n$) and
a set $\{i\}=\{i_1,i_2,...,i_{A-3}\}$ 
tracking the doubling, with $i_j=0$ ($1$) denoting the lower (higher) state
in the $j$-th doubling. 
Just as before, the dimensionless functions
$\beta_{A;n;\{i\}}$ of $(a_2\kappa_\star)^{-1}$
reduce at unitarity to pure numbers $\beta_{A;n;\{i\}}(0)$,
which obey
\begin{equation}
\beta_{A;n+l;\{i\}}(0)= \alpha_l^{-2}\, \beta_{A;n;\{i\}}(0)
\quad
\Leftrightarrow
\quad
\beta_{A;n;\{i\}}(0)= e^{-2n\pi/s_0} \,\beta_{A;0;\{i\}}(0).
\label{betaNnunit}
\end{equation}
Again the spurion method gives
\begin{equation}
\beta_{A;n+l;\{i\}}\!\left((\alpha_l a_2\kappa_\star)^{-1}\right)
= \alpha_l^{-2}\, \beta_{A;n;\{i\}}\!\left((a_2\kappa_\star)^{-1}\right),
\label{betaNn}
\end{equation}
with similar implications as for $A=3$.

I am not aware of an explanation for the doubling,
which has a topological interpretation \cite{Horinouchi:2016gqg},
but the replicating towers are a reflection of the surviving DSI.
For $A\ge 4$, all but the two lower states appear as resonances
in the scattering of a particle on the $(A-1)$-particle ground state.
Because of the tower structure, we can focus on 
these two lower states.
The higher one is near the ground state
of the system with one less particle,
\begin{equation}
\beta_{A;0;{0,...,0,1}}(0)\simeq \beta_{A-1;0;{0,...,0}}(0),
\label{betaN01}
\end{equation}
and thus can be thought of as a two-body system of a particle and an
$(A-1)$ cluster. 
This state and its cousins up the tower
are examples of ``halo states'', such as observed in halo nuclei
--- states that have a clusterized structure 
in which a certain number of ``valence'' particles 
orbits a tight cluster of the remaining particles.

The ground states close to unitarity get
deeper and deeper as $A$ increases. We can write
\begin{equation}
\frac{B_A}{A}\equiv 
\frac{B_{A;0;\{0\}}}{A}\simeq \frac{3}{A} \, \beta_{A;0;\{0\}}(0)\, \frac{B_{3;0}}{3}
\equiv \kappa_A \frac{B_3}{3}.
\label{BNntoB3n}
\end{equation}
where the set of numbers $\kappa_A$ 
encapsulates the dynamical information about the 
ground states at unitarity.
Relation \eqref{BNntoB3n} expresses
the generalized Tjon lines \cite{Bazak:2016wxm} at unitarity.

\begin{figure}[t]
\begin{center}
\includegraphics[scale=0.5]{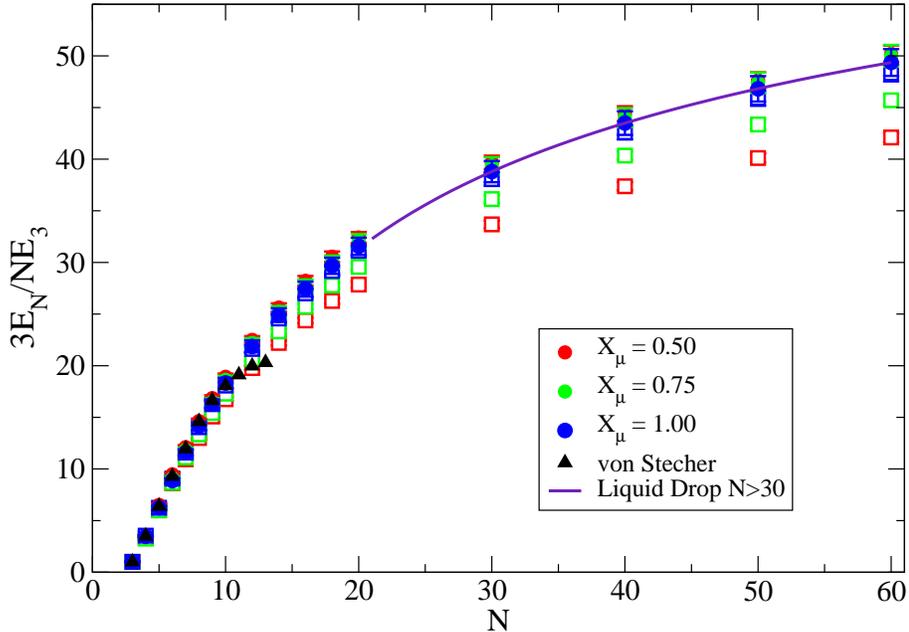}
\end{center}
\caption{Binding energy per particle at unitarity scaled by
the three-body binding energy per particle, $\kappa_N=3E_N/(NE_3)$, as 
function of the number $N$ of bosons in a cluster. 
Pionless EFT results at LO for various regulator cutoffs are shown
as open and filled (colored) symbols. Black triangles show potential-model 
results up to $N=15$ from Ref. \cite{Ste10}. 
The (blue) solid line shows a fit to the $N\ge 30$ (blue) points, 
corresponding to the largest cutoff values, 
using the liquid-drop formula \eqref{BinftytoB3n}.
Reprinted figure with permission from Ref. \cite{Carlson:2017txq}.
Copyright 2017 by the American Physical Society.
}
\label{figboseclusterenergies}       
\end{figure}

The $\kappa_A$ for $A\le 60$ bosons have been calculated recently
using Monte Carlo techniques to solve the Schr\"odinger equation
\cite{Carlson:2017txq},
with selected results shown in Table \ref{tbl:BEs}.
At small $A$, $\kappa_A$ is approximately linear in $A$, as obtained
earlier 
\cite{Ste10,vonStecher:2011zz,Nicholson:2012zp,Kievsky:2014dua,YanBlu15,Bazak:2016wxm},
but eventually saturation sets in, where 
the growth tapers off --- see Fig. \ref{figboseclusterenergies}. 
This change in behavior is fitted well by
a ``liquid-drop'' formula,
\begin{equation}
\kappa_A = \kappa \left( 1+ \eta A^{-1/3} + \ldots \right),
\label{BinftytoB3n}
\end{equation}
with $\kappa = 90\pm 10$ and $\eta=-1.7\pm 0.3$, respectively,
the dimensionless ``volume'' and ``surface'' terms.
The factor of $\simeq 90$ is large but still well below the $\simeq 515$
that provides an upper bound for the EFT breakdown scale.

Atomic $^4$He is close enough to unitarity to sustain two trimers,
an excited tetramer not too far from the ground-state trimer,
and a ground-state tetramer about 4.4 deeper than the ground-state trimer.
As shown in Table \ref{tbl:BEs},
$A=5,6$ systems have energies close to unitarity values as well.
And an equation of the type \eqref{BinftytoB3n} also
describes calculated ground-state energies \cite{vijay1983}, yielding
$\kappa \simeq 180$ and $\eta\simeq -2.7$.
The energy of the bulk is thus $\sim 2$ away from unitarity.
It is possibly beyond an EFT approach \cite{Kievsky:2017mjq}.

For $A>4$ {\it multi-state fermions} at unitarity
the spectral pattern is not clear, as no calculations are available,
but towers must also exist.
For the ground states, Eq. \eqref{BNntoB3n} still holds,
but with a different set of numbers $\kappa_A$.
The four-state fermion system, for example, reduces to
a bosonic system \cite{Platter:2005},
similarly to the three-nucleon system \cite{Bedaque:1999ve}.
$\kappa_4$ is then the same in both cases, but
$\kappa_{A\ge 5}$ must differ on account of
the exclusion principle and shell structure.

Although it can be seen from Table \ref{tbl:BEs}
that the two-nucleon system is not as close to unitarity as 
two $^4$He atoms, the first excited state of the triton is 
almost bound,
the alpha particle has an excited level close to the triton,
and the alpha-particle ground state is about 3 times deeper.
We argued recently \cite{Konig:2016utl} that nuclei are close enough
to unitarity for a perturbative expansion in $(a_2\kappa_\star)^{-1}$ to converge.
Reference \cite{Konig:2016utl} shows, for example, that the
Tjon line can be obtained in perturbation theory
from the Tjon line at unitarity. 
It would be extremely gratifying if it turns out that
one can indeed devise a theory
of nuclear physics based on a single parameter, plus perturbative 
corrections.
For more speculation along these lines, see Ref. \cite{Kolck:2017zzf}.

\section{Long-range forces} 
\label{longrange}

Pionless EFT acquired a life of its own. But it was born to address
the nonperturbative renormalization issues befalling Chiral EFT. 
Despite Pionless EFT's successes, many feel Chiral EFT should 
be better suited to nuclear physics, where traditionally 
the pion has been thought to be an important ingredient.
And indeed, the binding momenta estimated 
with Eq. \eqref{bindingmom} are not far
below the pion mass. (How seriously one should take factors
of 2 or 3 in Eq. \eqref{bindingmom}, 
which are important in this comparison, remains unclear.)
So let me end these lectures 
with a taste of the basic issues that confront Chiral EFT. 
These challenges are more basic than pion exchange, as
they arise from the renormalization of singular interactions and
affect also Van der Waals forces. 
But, since little has been done for
the latter, my main focus here will be on pion exchange.
Much more complete accounts of Chiral EFT can be found
in various reviews, {\it e.g.}, Refs. 
\cite{Bedaque:2002mn,Epelbaum:2008ga,Machleidt:2011zz}.

Chiral EFT is an EFT for momenta $Q\sim m_\pi\simeq 140$ MeV, when pions 
appear as explicit degrees of freedom. 
(One might want to include also the lowest-lying nucleon excitations
such as the Delta isobar,
which is separated in mass from the nucleon by only $m_\Delta-m_N \simeq 300$ 
MeV.)
As pseudo-Goldstone bosons of chiral symmetry, pions can be included 
through a non-linear realization of SO(4),
when the isospin doublet of nucleon fields $N$ 
transforms as under the SO(3) subgroup
of isospin, but with ``parameters'' that depend on the pion fields.
The isospin triplet of pion fields $\boldsymbol{\pi}$ itself
parametrizes the coset space $S^3$ of radius $f_\pi$, 
and always appear in the combination $\boldsymbol{\pi}/f_\pi$.
One defines chiral covariant derivatives of pion and nucleon (and maybe
nucleon-excitation) fields,
and constructs all interactions that are isospin symmetric,
which then will be automatically chiral invariant.
Quark masses break chiral symmetry (including isospin)
as components of certain SO(4) vectors, 
so we add to the action all interactions that break the symmetry as tensor
products of the corresponding vectors.
This procedure ensures that chiral symmetry is broken in Chiral EFT
just as in QCD. 
Details can be found in Ref. \cite{Weinberg:1996kr}.

For our purposes here, the most important terms are a subset
of the leading Lagrangian: pion kinetic and mass terms, and the dominant
pion-nucleon interaction, that is,
\begin{equation}
\Delta {\cal L}_{\boldsymbol{\pi}} = \frac{1}{2}
\left(\partial_\mu\boldsymbol{\pi}\, \partial^\mu\boldsymbol{\pi}
- m_\pi^2\boldsymbol{\pi}^2\right)
+ \frac{g_A}{2f_\pi} \; N^\dagger \left(\vec{S} \cdot
\boldsymbol{\tau \cdot} \vec{\nabla}\boldsymbol{\pi}\right)N +\ldots,
\label{DeltaL}
\end{equation}
where 
$\boldsymbol{\tau}$ are the Pauli matrices is isospin space
and $g_A\simeq 1.27$ is a LEC, the axial-vector coupling constant. 
I relegate to ``\ldots'' interactions with more fields, derivatives
and powers of quark masses, including the chiral partners of the terms
shown explicitly. 
The Chiral EFT Lagrangian consists of the Lagrangian
in Eq. \eqref{L} (with $\psi\to N$) supplemented by
its chiral partners and the additional terms in Eq. \eqref{DeltaL}. 
However, since this is a different theory, the
LECs of short-range interactions in Eq. \eqref{L} have new values:
they run differently and their renormalized values are, in general, 
also different --- we have to repeat the renormalization procedure 
to relate them to data in the presence of pions.

The EFT still splits into sectors of different nucleon number $A$,
but now there is more interesting physics at $A=0,1$ (the domain of ChPT), 
some of which is covered in Pich's lectures \cite{Pich:2018ltt}.
Still, the sectors with $A\ge 2$ are richer because they 
involve {\it two} types of loops 
\cite{Weinberg:1990rz,Weinberg:1991um}: the ``reducible'' loops we have seen
already in the context of Pionless EFT,
where we can separate a diagram in at least two parts by cutting horizontally
through heavy particle lines only; and ``irreducible'' loops, where
we must cut also through one or more pion lines.
In the first type of loop, we pick a pole from
one heavy particle propagators, the typical energy is
$\sim Q^2/m$, and the magnitude of the contribution can be estimated
by the rule \eqref{heavyrules}. In the second type of loop,
the pion propagators give energies $\sim Q$, so that nucleon
propagators are in a first approximation static --- recoil is
suppressed by a relative ${\cal O}(Q/m)$ and can be included
perturbatively.
For irreducible loops the power counting rules are, instead, those used
in ChPT \cite{Weinberg:1978kz},
\begin{eqnarray}
&&{\rm pion \; propagator} \to Q^{-2},
\nonumber \\
&&{\rm heavy \; particle \; propagator} \to Q^{-1},
\nonumber \\
&&{\rm loop \; integral} \to Q^4/(4\pi)^2,
\nonumber \\
&&{\rm derivative, pion \; mass} \to Q.
\label{lihgtrules}
\end{eqnarray}
Only the second type of loop is present for $A=0,1$.
An extra irreducible loop amounts to ${\cal O}(Q^2/(4\pi f_\pi)^2)$ 
\cite{Weinberg:1978kz,Manohar:1983md}, so an expansion is possible for
$Q\sim m_\pi \ll 4\pi f_\pi$.  By demanding RG invariance,
one can infer the scaling of LECs known as NDA \cite{Manohar:1983md}. 
For consistency, it requires 
$4\pi f_\pi={\cal O}(M_{\rm QCD})=m_N$, so that both
the loop and nonrelativistic expansions are part
of the $m_\pi/M_{\rm QCD}$ expansion. 

To isolate irreducible loops when $A\ge 2$, we define the potential as the sum 
of these irreducible loops. 
In addition to the contact interactions of Pionless EFT,
we have also pion exchange.
The power counting \eqref{lihgtrules} indicates
that one-pion exchange (OPE) 
between two nucleons is the dominant long-range potential, with
corrections starting {\it two} orders down the 
$M_{\rm lo}/M_{\rm hi}$ expansion \cite{Ordonez:1992xp}.
A crucial difference with respect to $A=0,1$
is that the sum of irreducible loops 
does not generate the full $T$ matrix,
is not directly observable, and is {\it not} RG invariant
in itself. The logic we follow for $T$ is the same as for $A=0,1$, but we now
have to infer the sizes of LECs taking into account the reducible loops,
as we do in the absence of pions.
Since these loops generate regulator dependence,
the potential itself cannot be regulator independent.

Challenges start already at the level of OPE:
in momentum space
\begin{equation}
\Delta V_{\boldsymbol{\pi}}(\vec{p}\,', \vec{p})
= -\frac{4\pi}{m_N M_{N\!N}} \, \boldsymbol{\tau_1\cdot \tau_2}\, 
\frac{\vec{\sigma}_1\cdot \vec{q}\; \vec{\sigma}_2\cdot \vec{q}}
{\vec{q}^{\; 2}+m_\pi^2}+\ldots,
\label{OPEmom}
\end{equation}
and in coordinate space
\begin{eqnarray}
\Delta V_{\boldsymbol{\pi}}(\vec{r}) 
&=& \frac{\boldsymbol{\tau_1\cdot \tau_2}}{m_N M_{N\!N}} 
\left\{S_{12}(\hat{r})
\left[1+ m_\pi r+\frac{(m_\pi r)^2}{3}\right] \frac{e^{-m_\pi r}}{r^3}
\right.
\nonumber\\
&&\left.\qquad\qquad
+\frac{\vec{\sigma}_1\cdot \vec{\sigma}_2}{3}
\left(-4\pi \, \delta^{(3)}(\vec{r})+ \frac{m_\pi^2}{r} \,  e^{-m_\pi r}\right)
\right\}+\ldots
\label{OPEcoord}
\end{eqnarray}
Here the subscripts $_{1,2}$ label nucleons,
$\vec{q}\equiv \vec{p}-\vec{p}\, '$ is the transferred three-momentum
and $q^{02}\ll \vec{q}^{\; 2}$ was neglected in the pion propagator
since it is a higher-order effect.
The tensor operator, defined as
$S_{12}(\hat{r})\equiv 3\, \vec{\sigma}_1\cdot\hat{r}\, 
\vec{\sigma}_2\cdot\hat{r} -\vec{\sigma}_1\cdot \vec{\sigma}_2$,
can be shown to vanish for total spin $S=0$.
For $S=1$ it mixes waves of $l=j\pm 1$, where it has one positive and 
one negative eigenvalue, except for $^3P_0$ where it is diagonal
with a negative eigenvalue.
It also acts on states with $l=j$, where it has a positive eigenvalue.
The ``tensor force'', which is highly singular,
is attractive in some uncoupled waves like $^3P_0$ and $^3D_2$, and 
in one of the eigenchannels of each coupled wave.
In contrast, $\vec{\sigma}_1\cdot \vec{\sigma}_2$ takes the value
$3$ or $-1$ when $S=0$ or $S=1$, respectively. 
(An analogous remark holds for $\boldsymbol{\tau_1\cdot \tau_2}$ 
and total isospin $I=0,1$.)
This ``central force'' has two components:
{\it i}) a contact term that can be absorbed in the two $C_0^{(0)}$
contact interactions;
and
{\it ii}) a long-range component with Yukawa form, which
is attractive in isovector (isoscalar)
channels for $S=0$ ($S=1$).
In the ``$\dots$'' one finds higher orders, which are more singular still. 

In order to bring the numerical factor in Eq. \eqref{OPEmom} 
to the same form I used for contact interactions, I introduced 
\begin{equation}
M_{N\!N}\equiv \frac{16\pi f_\pi^2}{g_A^2m_N},
\label{MNN}
\end{equation}
following Refs. \cite{Kaplan:1998tg,Kaplan:1998we}.
From the rule \eqref{heavyrules} one expects 
\begin{equation}
\frac{{\cal O}((4\pi/(m_NM_{N\!N}))^2 \,(m_NQ/(4\pi)))}
{{\cal O}(4\pi/(m_NM_{N\!N}))}
={\cal O}(Q/M_{N\!N})
\end{equation}
for the magnitude of the ratio between the non-analytic part
of once-iterated OPE and single OPE.
By NDA, $M_{N\!N}={\cal O}(f_\pi)$, but numerically $M_{N\!N}\simeq 290$ MeV,
or about three times larger. There are also other numbers of ${\cal O}(1)$
floating around. 
If we take $M_{N\!N}= {\cal O}(M_{\rm hi})$, then OPE
is an NLO effect in an $m_\pi/M_{N\!N}$ expansion
\cite{Kaplan:1998tg,Kaplan:1998we}. 
In this case LO would be formally identical to
LO in Pionless EFT, perhaps explaining why the latter seems to work well even
for the alpha particle where the binding momentum \eqref{bindingmom}
is not very small compared to $m_\pi$.
More generally, one can show \cite{Kaplan:1998tg,Kaplan:1998we} that 
in this case the same power
counting of Pionless EFT applies, with the pion mass counting
as a derivative. All pion exchanges are perturbative.
Results at NLO show the expected improvement over LO,
but unfortunately at N$^2$LO, where the first iteration of OPE appears,
all hell breaks loose: in channels where the tensor force is attractive,
no convergence is found for momenta $Q\simge 100$ MeV \cite{Fleming:1999ee}.

The inference is that the purely numerical factors do not in general
help convergence and the NDA estimate $M_{N\!N}={\cal O}(f_\pi)$ is realistic.
For quark masses such that $m_\pi \simle M_{N\!N}$, pions are perturbative for 
$Q\sim m_\pi$, but for larger masses one has to consider OPE as an LO
potential. This has the virtue of providing a scale for binding momenta
which is related to chiral symmetry: the amplitude is a series of the 
schematic form \cite{Bedaque:2002mn}
\begin{equation}
T_2^{(0)}(Q)\sim \frac{4\pi}{m_NM_{N\!N}} 
\left[1+ \frac{Q}{M_{N\!N}} + \left(\frac{Q}{M_{N\!N}}\right)^2+ \ldots \right]
\sim \frac{4\pi}{m_N} \frac{1}{M_{N\!N}-Q},
\end{equation}
allowing for a pole at $Q\sim M_{N\!N} ={\cal O}(f_\pi)\ll M_{\rm QCD}$. 
The existence of shallow nuclei, except perhaps the very lightest where 
binding momenta are even smaller,
is then tied to spontaneous chiral-symmetry breaking.
 
Taking OPE as LO means that we need to deal with its singular nature,
as for delta functions in Pionless EFT. 
For large momentum, the potential \eqref{OPEmom} is not more singular 
than \eqref{Tnoloop} and, on the surface, does not seem to offer further 
challenges. However, the more intricate 
angular dependence leads to an $r^{-3}$ behavior
at short distances instead of $\delta^{(3)}(\vec r)$,
with contributions also to waves higher than $S$.
Stronger singularities such as $r^{-5}$ appear at higher orders.
Things remain similar to atomic systems, in that we need to deal with
the $r^{-6}$ singularity of 
the Van der Waals interaction if we are interested in momenta
$Q\sim l_{\rm vdW}^{-1}$. And smaller but more singular components
exist in this context as well \cite{Cordon:2009wh}.

The quantum mechanics of singular interactions has a long history
\cite{Frank:1971xx},
when it was more or less agreed that an attractive singular interaction 
in itself
is not sufficient to define the solution of the Schr\"odinger equation.
The reason is that for a potential of the type 
$-\alpha^2/(2\mu r^n)$, with $\mu$ the reduced mass and $n\ge 2$, 
the two allowed solutions both oscillate with
decreasing amplitude as the distance decreases, in contrast with a regular
potential for which there are two clearly distinct solutions --- regular 
and irregular. As a consequence, there is an undetermined phase, just like
in Eq. \eqref{t3Sasym}: the zero-energy $l=0$ wavefunction, for example,
can be shown to be given at small $r$ by 
\begin{equation}
\psi(r) \propto r^{n/4 -1} 
\cos\left(\frac{\alpha}{n/2 -1}\, r^{1-n/2}  +\delta_n\right)
+ \ldots
\label{wfr-n}
\end{equation}
for $n>2$. For $n=2$, 
$\alpha r^{1-n/2}/(n/2 -1) \to \sqrt{\alpha^2-1/4} \, \ln (r/R)$.
Equation \eqref{wfr-n} is parallel to Eq. \eqref{t3Sasym}. 
(In fact, Efimov \cite{Efimov:1970zz}
first arrived at his geometric spectrum by consideration of
the three-particle equation in coordinate space.)
The phase $\delta_n$ can be fixed by a single
counterterm regardless of the value of $n$ \cite{Beane:2000wh}.
For example, with a spherical-well regularization of the 
corresponding delta function (see Sec. \ref{unitarity}),
progressively more unphysical bound states cross threshold 
as the regulator distance $R$ decreases --- similarly to the Thomas collapse in 
the three-body system --- unless
the depth is adjusted to produce a value of the wavefunction at $R$
that gives the phase one wants.
The phase in turn determines the low-energy properties of the
scattering amplitude.
Just like in the three-body case above, the LEC oscillates as 
$\Lambda\sim R^{-1}$ increases.
However, only for $n=2$, when the classical system is scale invariant,
is the dependence periodic in $\ln \Lambda$.
The RG analysis of singular potentials is discussed in 
Ref. \cite{PavonValderrama:2007nu}.
From the EFT point of view, the quantum mechanics of
singular potentials is just the renormalization of the LO amplitude.

For $l>0$, the centrifugal barrier effectively suppresses 
the effect of the long-range potential on the amplitude by factors of $l^{-1}$.
Only for lower waves does the long-range potential need to be iterated
in the low-energy region where the EFT applies 
\cite{Nogga:2005hy,Birse:2005um,Long:2007vp}.
At subleading orders where the potential gets more singular,
renormalization can still be carried out with further contact interactions
\cite{Long:2007vp},
at least as long as corrections are treated in perturbation theory
as done for Pionless EFT in Sec. \ref{2body}.

In the nuclear case there are complications arising from the 
tensor and spin operators:
\begin{itemize}

\item The $^3S_1$ wave gets mixed with $^3D_1$ by
the OPE tensor force. The tensor force has one attractive eigenvector,
which is finite in the chiral limit,
and the $C_0$ LEC in this channel,
expected on the basis of NDA, is sufficient for renormalization at LO
\cite{Frederico:1999ps,Beane:2001bc}.

\item In spin-triplet channels where the tensor force is repulsive,
OPE can be iterated without RG problems \cite{Nogga:2005hy}.
In contrast, when it is attractive, RG invariance is destroyed by
iteration \cite{Nogga:2005hy,PavonValderrama:2005uj}. 
One expects an angular-momentum suppression 
similar to the one seen for a central force, so that beyond a critical
angular momentum $l_{\rm cr}$ OPE can be considered subleading.
However, in lower waves like $^3S_1$-$^3D_1$  and $^3P_0$, one can argue
\cite{Nogga:2005hy,Birse:2005um,Wu:2018lai}
that OPE needs to be iterated in the low-energy region, in agreement
with the findings of Ref. \cite{Fleming:1999ee}.
In these waves, additional contact interactions with derivatives,
which would be expected by NDA only at higher orders,
are necessary and sufficient for renormalization at LO \cite{Nogga:2005hy}.

\item In the simplest spin-singlet channel, $^1S_0$, 
OPE takes the form of an attractive 
Yukawa interaction $\propto m_\pi^2$, 
which by itself generates a finite amplitude
but is far from providing enough binding to explain the virtual state.
A contact $C_0$ interaction must still be present at LO. 
However, the interference between the two interactions gives
rise to a $m_\pi^2\ln \Lambda$ term --- in perturbation theory, it comes from
a diagram where OPE takes place between two contact interactions,
but the same regulator dependence is seen nonperturbatively 
\cite{Kaplan:1996xu}.
Renormalization requires a chiral-symmetry-breaking contact interaction
with LEC $m_\pi^2D_2$ at LO \cite{Kaplan:1996xu}.
Thus this contact interaction is also enhanced with respect to NDA,
and there is no
straightforward chiral expansion of the contact interactions.

\item In all other spin-singlet channels, the absence of a contact
interaction in LO means OPE can be iterated without RG problems
\cite{Nogga:2005hy}.
However, as the angular momentum $l$ increases factors of $l^{-1}$  
suppress its contribution, and OPE is really perturbative and thus subleading
\cite{PavonValderrama:2016lqn}.

\item Residual cutoff dependence indicates the need at NLO for
a single two-nucleon correction from the $^1S_0$ $C_2$ contact interaction, 
treated in perturbation theory \cite{Long:2012ve}
--- in the same way as in Pionless EFT (Sec. \ref{2body}).

\item 
Since OPE changes the asymptotic
behavior of the two-nucleon amplitude, it is not immediately obvious whether
a three-body force is needed for renormalization
at LO or perhaps NLO. Explicit calculations 
\cite{Nogga:2005hy,Song:2016ale} show it is not.
\end{itemize}

During the period while nuclear EFTs were being formulated, rapid progress
was achieved in the development of methods to solve the Schr\"odinger
equation for increasingly higher $A$ with a given potential.
While they first used purely phenomenological potentials,
eventually most calculations became based on 
Weinberg's original suggestion \cite{Weinberg:1990rz,Weinberg:1991um}
to use ``chiral potentials'', where:
{\it i)} contact interactions are assumed to have sizes 
given by NDA;
{\it ii)} the expansion of irreducible diagrams is truncated at a certain 
order; and
{\it iii)} the truncated potential is treated exactly.
The resulting amplitudes are not renormalizable and much work
goes into finding the ``best'' regulator to fit data with.
In contrast, a properly renormalized EFT has only been explored beyond NLO
in the two-nucleon sector where it has given promising results
\cite{Valderrama:2009ei,Valderrama:2011mv,Long:2011qx,Long:2011xw,Long:2012ve}.
There is still much to learn about implementing corrections
in perturbation theory.
Despite its age, this is a field with plenty of open problems.

\section{Conclusion}
\label{conc}

EFT is not only a tool for inferring new degrees of freedom and symmetries,
but also for understanding the emergence of new structures.
I hope to have given you a flavor of this latter aspect of EFT's power 
in the context of nuclear and atomic EFTs.
Through Pionless EFT, I described how renormalization 
in a nonperturbative setting can be very different 
from perturbation theory,
yet sufficiently tractable for us to observe
the emergence of new phenomena:
the non-trivial fixed point of two-body unitarity,
the limit cycle of three-body physics,
and the description of larger structures from a single essential
parameter.
The remaining discrete scale invariance allows for many-body 
spectra reminiscent of Russian dolls, 
for ground states that saturate as the number of particles grows very large,
and for a (quantum) liquid. 
How such a picture can be matched with Chiral EFT,
where equally bizarre renormalization takes place, is a 
question for you to tackle.

\section*{Acknowledgments}

I am grateful to C\'ecile DeWitt-Morette for teaching me, among
other things, the value of passion and determination in science.
I first met Paolo Gambino when I attended the Carg\`ese school 25 years ago 
upon her recommendation. I thank him for the invitation to Les Houches.
I greatly appreciated the hospitality of the directors of the school,
in particular Sacha Davidson, who gently oversaw everything including,
with extraordinary diligence and patience, the completion of these lecture 
notes.
This work was supported in part 
by the U.S. Department of Energy, Office of Science, Office of Nuclear Physics, 
under award DE-FG02-04ER41338
and by the European Union Research and Innovation program Horizon 2020
under grant No. 654002.


\begin{thebibliography}{64}

\bibitem{Manohar:2018aog}
  A.V.~Manohar,
  ``Introduction to Effective Field Theories'',
  lectures at this school,
  arXiv:1804.05863 [hep-ph].

\bibitem{Weinberg:1990rz}
  S.~Weinberg,
  Phys.\ Lett.\ B {\bf 251} (1990) 288.

\bibitem{Weinberg:1991um}
  S. Weinberg,
  Nucl. Phys. B {\bf 363} (1991) 3.

\bibitem{Rho:1990cf}
  M.~Rho,
  Phys.\ Rev.\ Lett.\  {\bf 66} (1991) 1275.

\bibitem{Weinberg:1978kz}
  S.~Weinberg,
  Physica A {\bf 96} (1979) 327.

\bibitem{Pich:2018ltt}
  A.~Pich,
  ``Effective Field Theory with Nambu-Goldstone Modes'',
  lectures at this school,
  arXiv:1804.05664 [hep-ph].

\bibitem{Beane:2010em}
  S.R.~Beane, W.~Detmold, K.~Orginos, and M.J.~Savage,
  Prog.\ Part.\ Nucl.\ Phys.\  {\bf 66} (2011) 1
  [arXiv:1004.2935 [hep-lat]].

\bibitem{Barnea:2013uqa}
  N.~Barnea, L.~Contessi, D.~Gazit, F.~Pederiva, and U.~van Kolck,
  Phys.\ Rev.\ Lett.\  {\bf 114} (2015) 
  052501
  [arXiv:1311.4966 [nucl-th]].

\bibitem{Contessi:2017rww}
  L.~Contessi, A.~Lovato, F.~Pederiva, A.~Roggero, J.~Kirscher, 
  and U.~van Kolck,
  Phys.\ Lett.\ B {\bf 772} (2017) 839
  [arXiv:1701.06516 [nucl-th]].
  
\bibitem{Efimov:1970zz}
  V.~Efimov,
  Phys.\ Lett.\  {\bf 33B} (1970) 563.

\bibitem{Tanabashi:2018oca} 
  M.~Tanabashi {\it et al.} [Particle Data Group],
  Phys.\ Rev.\ D {\bf 98} (2018) 
  030001.

\bibitem{Weinberg:1977hb}
  S.~Weinberg,
  Trans.\ New York Acad.\ Sci.\  {\bf 38} (1977) 185.

\bibitem{Bertulani:2002sz}
  C.A. Bertulani, H.-W. Hammer, and U. van Kolck,
  Nucl. Phys. A {\bf 712} (2002) 37
  [nucl-th/0205063].

\bibitem{Bedaque:2003wa}
  P.F. Bedaque, H.-W. Hammer, and U. van Kolck,
  Phys. Lett. B {\bf 569} (2003) 159
  [nucl-th/0304007].

\bibitem{Bedaque:1999ve}
  P.F. Bedaque, H.-W. Hammer, and U. van Kolck,
  Nucl. Phys. A {\bf 676} (2000) 357
  [arXiv:nucl-th/9906032].

\bibitem{Stetcu:2006ey}
  I. Stetcu, B.R. Barrett, and U. van Kolck,
  Phys. Lett. B {\bf 653} (2007) 358
  [nucl-th/0609023].

\bibitem{Rupak:2018gnc}
  G.~Rupak, A.~Vaghani, R.~Higa, and U.~van Kolck,
  arXiv:1806.01999 [nucl-th].

\bibitem{Konig:2016utl}
  S. K\"onig, H.W. Grie{\ss}hammer, H.-W. Hammer, and U. van Kolck,
  Phys.\ Rev.\ Lett.\  {\bf 118} (2017) 
  202501,
  [arXiv:1607.04623 [nucl-th]].

\bibitem{kalos1981}
M.H. Kalos, M.A. Lee, P.A. Whitlock, and G.V. Chester,
Phys. Rev. B {\bf 24} (1981) 115. 

\bibitem{BluGre00} 
  D. Blume and C.H. Greene,
  J. Chem. Phys. {\bf 112} (2000) 8053.

\bibitem{HiyKam12a} 
  E. Hiyama and M. Kamimura,
  Phys. Rev. A {\bf 85} (2012) 022502
  [arXiv:1111.4370 [physics.atom-ph]].

\bibitem{Bedaque:1998km}
  P.F. Bedaque, H.-W. Hammer, and U. van Kolck,
  Nucl. Phys. A {\bf 646} (1999) 444
  [nucl-th/9811046].

\bibitem{Platter:2004he}
  L.~Platter, H.-W.~Hammer, and U.-G.~Mei{\ss}ner,
  Phys. Rev. A {\bf 70} (2004) 052101
  [cond-mat/0404313].

\bibitem{Bazak:2016wxm}
  B. Bazak, M. Eliyahu, and U. van Kolck,
  Phys. Rev. A {\bf 94} (2016) 052502
  [arXiv:1607.01509 [cond-mat.quant-gas]].

\bibitem{Deltuva:2010xd}
  A. Deltuva,
  Phys. Rev. A {\bf 82} (2010) 040701
  [arXiv:1009.1295 [physics.atm-clus]].

\bibitem{Carlson:2017txq}
  J.~Carlson, S.~Gandolfi, U.~van Kolck, and S.A.~Vitiello,
  Phys. Rev. Lett. {\bf 119} (2017) 
  223002
  [arXiv:1707.08546 [cond-mat.quant-gas]].

\bibitem{Cordon:2009wh}
  A. Calle Cord\'on and E.~Ruiz Arriola,
  Phys. Rev. A {\bf 81} (2010) 044701
  [arXiv:0912.1714 [cond-mat.other]].

\bibitem{GriSchToe00} 
  R.E. Grisenti {\it et al.},
  Phys. Rev. Lett. {\bf 85} (2000) 2284.

\bibitem{YanBabDal96} 
  Z.-C. Yan, J.F. Babb, A. Dalgarno, and G.W.F. Drake,
  Phys. Rev. A {\bf 54} (1996) 2824
  [atom-ph/9607002].

\bibitem{ZhaYanVri06} 
  J.-Y. Zhang, Z.-C. Yan, D. Vrinceanu, J.F. Babb, and H.R. Sadeghpour,
  Phys. Rev. A {\bf 74} (2006) 014704
  [physics/0603232].

\bibitem{CenPrzKom12} 
  W. Cencek, M. Przybytek, J. Komasa, J.B. Mehl, 
  B. Jeziorski, and K. Szalewicz,
  J. Chem. Phys. {\bf 136} (2012) 224303.

\bibitem{Zel16} 
  S. Zeller {\it et al.},
  Proc. Natl. Acad. Sci. U.S.A. {\bf 113} (2016) 14651
  [arXiv:1601.03247 [physics.atom-ph]].

\bibitem{Kunitski:2015qth} 
  M. Kunitski {\it et al.},
  Science {\bf 348} (2015) 551
  [arXiv:1512.02036 [physics.atm-clus]].

\bibitem{AziSla91} 
  R.A. Aziz and M.J. Slaman,
  J. Chem. Phys. {\bf 94} (1991) 8047.

\bibitem{aziz1979}
R.A. Aziz, V.P.S. Nain, J.S. Carley, W.L. Taylor, and G.T. McConville,
J. Chem. Phys. {\bf 70} (1979) 4330.

\bibitem{vijay1983}
V.R. Pandharipande, J.G. Zabolitzky, S.C. Pieper, R.B. Wiringa, and 
U. Helmbrecht,
Phys. Rev. Lett. {\bf 50} (1983) 1676.
  
\bibitem{Chisholm:1961tha}
  J.S.R. Chisholm,
  Nucl. Phys. {\bf 26} (1961) 
  469.

\bibitem{Kamefuchi:1961sb}
  S. Kamefuchi, L. O'Raifeartaigh, and A. Salam,
  Nucl. Phys. {\bf 28} (1961) 529.

\bibitem{Georgi:1990um}
  H.~Georgi,
  Phys. Lett. B {\bf 240} (1990) 447.

\bibitem{Luke:1992cs}
  M.E.~Luke and A.V.~Manohar,
  Phys. Lett. B {\bf 286} (1992) 348
  [hep-ph/9205228].

\bibitem{Lee:1962vm}
  T.D.~Lee and C.N.~Yang,
  Phys. Rev. {\bf 128} (1962) 885.

\bibitem{Salam:1971sp}
  A.~Salam and J.A.~Strathdee,
  Phys. Rev. D {\bf 2} (1970) 2869.

\bibitem{Mannellectures}
T. Mannel, ``Effective Field Theories for Heavy Quarks'',
lectures at this school.

\bibitem{Konig:2015aka}
  S. K\"onig, H.W. Grie\ss hammer, H.-W. Hammer, and U. van Kolck,
  J. Phys. G {\bf 43} (2016) 
  055106
  [arXiv:1508.05085 [nucl-th]].

\bibitem{Konig:2016iny}
  S.~K\"onig,
  J. Phys. G {\bf 44} (2017) 
  064007
  [arXiv:1609.03163 [nucl-th]].

\bibitem{Weinberg2018}
S. Weinberg,
Phys. Rev. Lett. {\bf 121} (2018) 220001.
  
\bibitem{vanKolck:1998bw}
  U. van Kolck,
  Nucl. Phys. A {\bf 645} (1999) 273
  [arXiv:nucl-th/9808007].

\bibitem{Kaplan:1998tg}
  D.B. Kaplan, M.J. Savage, and M.B. Wise,
  Phys. Lett. B {\bf 424} (1998) 390
  [nucl-th/9801034].

\bibitem{Kaplan:1998we}
  D.B. Kaplan, M.J. Savage, and M.B. Wise,
  Nucl. Phys. B {\bf 534} (1998) 329
  [nucl-th/9802075].

\bibitem{Phillips:1998uy}
  D.R.~Phillips, S.R.~Beane, and M.C.~Birse,
  J. Phys. A {\bf 32} (1999) 3397
  [hep-th/9810049].

\bibitem{Manohar:1983md}
  A.~Manohar and H.~Georgi,
  Nucl.\ Phys.\ B {\bf 234} (1984) 189.

\bibitem{vanKolck:1997ut} 
  U. van Kolck,
  Lect. Notes Phys. {\bf 513} (1998) 62
  [hep-ph/9711222].

\bibitem{Chen:1999tn}
  J.W.~Chen, G.~Rupak, and M.J.~Savage,
  Nucl. Phys. A {\bf 653} (1999) 386
  [nucl-th/9902056].

\bibitem{Bethe:1949yr}
  H.A. Bethe,
  Phys. Rev. {\bf 76} (1949) 38.

\bibitem{Stoks:1994wp}
  V.G.J.~Stoks, R.A.M.~Klomp, C.P.F.~Terheggen, and J.J.~de Swart,
  Phys. Rev. C {\bf 49} (1994) 2950
  [nucl-th/9406039].

\bibitem{BethePeierls:19351}
H.A. Bethe and R. Peierls,
Proc. Roy. Soc. Lond. A {\bf 148} (1935) 146.

\bibitem{BethePeierls:19352}
H.A. Bethe and R. Peierls,
Proc. Roy. Soc. Lond. A {\bf 149} (1935) 176.
 
\bibitem{Fermi:1936}
E. Fermi, 
Ric. Scientifica {\bf 7} (1936) 13.

\bibitem{Kohler:2006zz}
  T.~K\"ohler, K.~Goral, and P.S.~Julienne,
  Rev. Mod. Phys. {\bf 78} (2006) 1311
  [cond-mat/0601420].

\bibitem{Cohen:2004kf}
  T.D. Cohen, B.A. Gelman, and U. van Kolck,
  Phys. Lett. B {\bf 588} (2004) 57
  [nucl-th/0402054].

\bibitem{JanAzi95} 
  A.R. Janzen and R.A. Aziz,
  J. Chem. Phys. {\bf 103} (1995) 8626.

\bibitem{Wagman:2017tmp}
  M.L. Wagman, F. Winter, E. Chang, Z. Davoudi, W. Detmold, K. Orginos, 
  M.J. Savage, and P.E. Shanahan,
  Phys. Rev. D {\bf 96} (2017) 
  114510
  [arXiv:1706.06550 [hep-lat]].

\bibitem{Beane:2001bc}
  S.R. Beane, P.F. Bedaque, M.J. Savage, and U. van Kolck,
  Nucl. Phys. A {\bf 700} (2002) 377
  [nucl-th/0104030].

\bibitem{Hagen:1972pd}
  C.R. Hagen,
  Phys. Rev. D {\bf 5} (1972) 377.

\bibitem{Mehen:1999nd}
  T. Mehen, I.W. Stewart, and M.B. Wise,
  Phys. Lett. B {\bf 474} (2000) 145
  [hep-th/9910025].

\bibitem{dEspagnat:1956}
B. d'Espagnat and J. Prentki,
Nuovo Cim. {\bf 3} (1956) 1045.

\bibitem{Kaplan:1996nv}
  D.B. Kaplan,
  Nucl. Phys. B {\bf 494} (1997) 471
  [nucl-th/9610052].

\bibitem{Griesshammer:2004pe}
  H.W. Grie{\ss}hammer,
  Nucl. Phys. A {\bf 744} (2004) 192
  [nucl-th/0404073].

\bibitem{Beane:1997pk}
  S.R. Beane, T.D. Cohen, and D.R. Phillips,
  Nucl. Phys. A {\bf 632} (1998) 445
  [nucl-th/9709062].

\bibitem{Bedaque:1998kg}
  P.F. Bedaque, H.-W. Hammer, and U. van Kolck,
  Phys. Rev. Lett. {\bf 82} (1999) 463
  [nucl-th/9809025].

\bibitem{Bedaque:1997qi}
  P.F. Bedaque and U. van Kolck,
  Phys. Lett. B {\bf 428} (1998) 221
  [nucl-th/9710073].

\bibitem{Bedaque:1998mb}
  P.F. Bedaque, H.-W. Hammer, and U. van Kolck,
  Phys. Rev. C {\bf 58} (1998) R641
  [nucl-th/9802057].

\bibitem{Griesshammer:2005ga}
  H.W.~Grie{\ss}hammer,
  Nucl. Phys. A {\bf 760} (2005) 110
  [nucl-th/0502039].

\bibitem{Dilg:1971gqb}
  W.~Dilg, L.~Koester, and W.~Nistler,
  Phys. Lett. {\bf 36B} (1971) 208.

\bibitem{Danilov1961}
G.S. Danilov,
Sov. Phys. JETP {\bf 13} (1961) 349.

\bibitem{Thomas:1935zz}
  L.H. Thomas,
  Phys. Rev. {\bf 47} (1935) 903.

\bibitem{Braaten:2011sz}
  E.~Braaten, D.~Kang, and L.~Platter,
  Phys.\ Rev.\ Lett.\  {\bf 106} (2011) 153005
  [arXiv:1101.2854 [cond-mat.quant-gas]].

\bibitem{Hammer:2001gh}
  H.-W. Hammer and T. Mehen,
  Phys. Lett. B {\bf 516} (2001) 353
  [nucl-th/0105072].

\bibitem{Bedaque:2002yg}
  P.F. Bedaque, G. Rupak, H.W. Grie{\ss}hammer, and H.-W. Hammer,
  Nucl. Phys. A {\bf 714} (2003) 589
  [nucl-th/0207034].

\bibitem{Amado:1992vn}
  R.D. Amado and J.V. Noble,
  Phys. Rev. D {\bf 5} (1972) 1992.

\bibitem{Schoellkopf:1996}
  W. Sch\"ollkopf and J.P. Toennies,
  J. Chem. Phys. {\bf 104} (1996) 1155.

\bibitem{Adhikari:1982zz}
  S.K. Adhikari and L. Tomio,
  Phys. Rev. C {\bf 26} (1982) 83.

\bibitem{Phillips:1968zze}
  A.C. Phillips,
  Nucl. Phys. A {\bf 107} (1968) 209.

\bibitem{Efimov1988}
V. Efimov and E.G. Tkachenko, 
Few-Body Syst. {\bf 4} (1988) 71.

\bibitem{Roudnev2012}
V. Roudnev and M. Cavagnero,
Phys. Rev. Lett. {\bf 108} (2012) 110402
[arXiv:1109.4656 [physics.atm-clus]].

\bibitem{Ji:2011qg}
  C.~Ji, D.R.~Phillips, and L.~Platter,
  Annals Phys. {\bf 327} (2012) 1803
  [arXiv:1106.3837 [nucl-th]].

\bibitem{Ji:2012nj}
  C. Ji and D.R. Phillips,
  Few-Body Syst. {\bf 54} (2013) 2317
  [arXiv:1212.1845 [nucl-th]].

\bibitem{Vanasse:2013sda} 
  J. Vanasse,
  Phys. Rev. C {\bf 88} (2013) 044001
  [arXiv:1305.0283 [nucl-th]].

\bibitem{Ji:2015jgm}
  C.~Ji,
  Int.\ J.\ Mod.\ Phys.\ E {\bf 25} (2016) 
  1641003
  [arXiv:1512.06114 [nucl-th]].

\bibitem{Vanasse:2016jtc}
  J.~Vanasse,
  Int.\ J.\ Mod.\ Phys.\ E {\bf 25} (2016) 
  1641002
  [arXiv:1609.03086 [nucl-th]].

\bibitem{Wilson:1974sk}
  K.G.~Wilson,
  Phys.\ Rev.\ D {\bf 10} (1974) 2445.

\bibitem{Wilson:2004de}
  K.G.~Wilson,
  Nucl.\ Phys.\ Proc.\ Suppl.\  {\bf 140} (2005) 3
  [hep-lat/0412043].

\bibitem{Bulycheva:2014twa}
  K.M.~Bulycheva and A.S.~Gorsky,
  Phys.\ Usp.\  {\bf 57} (2014) 171
  [arXiv:1402.2431 [hep-th]].

\bibitem{Camblong:2003mz}
  H.E.~Camblong and C.R.~Ord\'o\~nez,
  Phys. Rev. D {\bf 68} (2003) 125013
  [hep-th/0303166].

\bibitem{Sornette:1997pb}
  D. Sornette,
  Phys. Rept. {\bf 297} (1998) 239
  [cond-mat/9707012 [cond-mat.stat-mech]].

\bibitem{Braaten:2004rn}
  E. Braaten and H.-W. Hammer,
  Phys. Rept. {\bf 428} (2006) 259
  [cond-mat/0410417].

\bibitem{grimm}
  T. Kraemer {\it et al.},
  Nature {\bf 440} (2006) 315 
  [cond-mat/0512394]. 

\bibitem{grimm2}
  B. Huang {\it et al.},
  Phys. Rev. Lett. {\bf 112} (2014) 190401
  [arXiv:1402.6161 [cond-mat.quant-gas]]. 

\bibitem{Platter:2008cx}
  L.~Platter, C.~Ji, and D.R.~Phillips,
  Phys.\ Rev.\ A {\bf 79} (2009) 022702
  [arXiv:0808.1230 [cond-mat.other]].

\bibitem{Wigner:1936dx}
  E.~Wigner,
  Phys.\ Rev.\  {\bf 51} (1937) 106.

\bibitem{Mehen:1999qs}
  T. Mehen, I.W. Stewart, and M.B. Wise,
  Phys. Rev. Lett.  {\bf 83} (1999) 931
  [hep-ph/9902370].

\bibitem{Vanasse:2016umz}
  J. Vanasse and D.R.~Phillips,
  Few-Body Syst. {\bf 58} (2017) 
  26
  [arXiv:1607.08585 [nucl-th]].

\bibitem{Platter:2004zs}
  L. Platter, H.-W. Hammer, and U.-G. Mei\ss ner,
  Phys. Lett. B {\bf 607} (2005) 254
  [nucl-th/0409040].

\bibitem{Hammer:2006ct}
  H.-W. Hammer and L. Platter,
  Eur. Phys. J. A {\bf 32} (2007) 113
  [nucl-th/0610105].

\bibitem{Kirscher:2009aj}
  J. Kirscher, H.W. Grie{\ss}hammer, D. Shukla, and H.M. Hofmann,
  Eur. Phys. J. A {\bf 44} (2010) 239
  [arXiv:0903.5538 [nucl-th]].

\bibitem{Kirscher:2015yda}
  J. Kirscher, N. Barnea, D. Gazit, F. Pederiva, and U. van Kolck,
  Phys. Rev. C {\bf 92} (2015) 
  054002
  [arXiv:1506.09048 [nucl-th]].

\bibitem{Kolck:2017zzf}
  U.~van Kolck,
  Few-Body Syst. {\bf 58} (2017) 
  112.

\bibitem{Tjon:1975sme}
  J.A. Tjon,
  Phys. Lett. B {\bf 56} (1975) 217.

\bibitem{NakAkaTanLim78} 
  S. Nakaichi, Y. Akaishi, H. Tanaka, and T.K. Lim,
  Phys. Lett. A {\bf 68} (1978) 36.

\bibitem{NakLimAkaTan79} 
  S. Nakaichi, T.K. Lim, Y. Akaishi, and H. Tanaka,
  J. Chem. Phys. {\bf 71} (1979) 4430.

\bibitem{LimNakAkaTan80} 
  T.K. Lim, S. Nakaichi, Y. Akaishi, and H. Tanaka,
  Phys. Rev. A {\bf 22} (1980) 28. 

\bibitem{Lensky:2016djr}
  V.~Lensky, M.C.~Birse, and N.R.~Walet,
  Phys.\ Rev.\ C {\bf 94} (2016) 
  034003
  [arXiv:1605.03898 [nucl-th]].

\bibitem{Bansal:2017pwn}
  A. Bansal, S. Binder, A. Ekstr\"om, G. Hagen, G.R. Jansen, and T. Papenbrock,
  Phys. Rev. C {\bf 98} (2018) 
  054301
  [arXiv:1712.10246 [nucl-th]].

\bibitem{Bazak:2018qnu}
  B. Bazak, J. Kirscher, S. K\"onig, M. Pav\'on Valderrama, N. Barnea,
  and U. van Kolck,
  arXiv:1812.00387 [cond-mat.quant-gas].

\bibitem{Ferlaino:2009zz}
  F. Ferlaino {\it et al.},
  Phys. Rev. Lett. {\bf 102} (2009) 140401
  [arXiv:0903.1276 [cond-mat.other]].

\bibitem{Ste10}  
  J. von Stecher,
  J. Phys. B {\bf 43} (2010) 101002
  [arXiv:0909.4056 [cond-mat.quant-gas]].

\bibitem{Gattobigio:2011ey}
  M. Gattobigio, A. Kievsky, and M. Viviani,
  Phys. Rev. A {\bf 84} (2011) 052503
  [arXiv:1106.3853 [physics.atm-clus]].

\bibitem{vonStecher:2011zz}
  J. von Stecher,
  Phys. Rev. Lett. {\bf 107} (2011) 200402
  [arXiv:1106.2319 [cond-mat.quant-gas]].

\bibitem{Gattobigio:2012tk} 
  M. Gattobigio, A. Kievsky, and M. Viviani,
  Phys. Rev. A {\bf 86} (2012) 042513
  [arXiv:1206.0854 [physics.atm-clus]].

\bibitem{Horinouchi:2016gqg}
  Y. Horinouchi and M. Ueda,
  Phys. Rev. A {\bf 94} (2016) 050702
  [arXiv:1603.05328 [cond-mat.quant-gas]].

\bibitem{Nicholson:2012zp} 
  A.N. Nicholson,
  Phys. Rev. Lett. {\bf 109} (2012) 073003
  [arXiv:1202.4402 [cond-mat.quant-gas]].

\bibitem{Kievsky:2014dua} 
  A. Kievsky, N.K. Timofeyuk, and M. Gattobigio,
  Phys. Rev. A {\bf 90} (2014) 032504
  [arXiv:1405.2371 [cond-mat.quant-gas]].

\bibitem{YanBlu15} 
  Y. Yan and D. Blume,
  Phys. Rev. A {\bf 92} (2015) 033626
  [arXiv:1508.00081 [cond-mat.quant-gas]].

\bibitem{Kievsky:2017mjq}
  A.~Kievsky, A.~Polls, B.~Juli\'a-D\'\i az, and N.K.~Timofeyuk,
  Phys. Rev. A {\bf 96} (2017) 
  040501
  [arXiv:1707.05628 [cond-mat.quant-gas]].

\bibitem{Platter:2005}
  L. Platter,
  Ph.D. dissertation, University of Bonn (2005).

\bibitem{Bedaque:2002mn}
  P.F.~Bedaque and U.~van Kolck,
  Ann.\ Rev.\ Nucl.\ Part.\ Sci.\  {\bf 52} (2002) 339
  [nucl-th/0203055].

\bibitem{Epelbaum:2008ga}
  E.~Epelbaum, H.-W.~Hammer, and U.-G.~Mei{\ss}ner,
  Rev. Mod. Phys. {\bf 81} (2009) 1773
  [arXiv:0811.1338 [nucl-th]].

\bibitem{Machleidt:2011zz}
  R.~Machleidt and D.R.~Entem,
  Phys. Rept. {\bf 503} (2011) 1
  [arXiv:1105.2919 [nucl-th]].

\bibitem{Weinberg:1996kr}
S. Weinberg,
The Quantum Theory of Fields: Modern applications (Vol. 2),
Cambridge University Press (2013).

\bibitem{Ordonez:1992xp}
  C.~Ord\'o\~nez and U.~van Kolck,
  Phys.\ Lett.\ B {\bf 291} (1992) 459.

\bibitem{Fleming:1999ee}
  S.~Fleming, T.~Mehen, and I.W.~Stewart,
  Nucl.\ Phys.\ A {\bf 677} (2000) 313
  [nucl-th/9911001].

\bibitem{Frank:1971xx}
  W.~Frank, D.J.~Land, and R.M.~Spector,
  Rev. Mod. Phys. {\bf 43} (1971) 36.

\bibitem{Beane:2000wh}
  S.R.~Beane, P.F.~Bedaque, L.~Childress, A.~Kryjevski, J.~McGuire,
  and U.~van Kolck,
  Phys. Rev. A {\bf 64} (2001) 042103
  [quant-ph/0010073].

\bibitem{PavonValderrama:2007nu}
  M.~Pav\'on Valderrama and E.~Ruiz Arriola,
  Annals Phys. {\bf 323} (2008) 1037
  [arXiv:0705.2952 [nucl-th]].

\bibitem{Nogga:2005hy}
  A. Nogga, R.G.E. Timmermans, and U. van Kolck,
  Phys. Rev. C {\bf 72} (2005) 054006
  [nucl-th/0506005].

\bibitem{Birse:2005um}
  M.C. Birse,
  Phys. Rev. C {\bf 74} (2006) 014003
  [nucl-th/0507077].

\bibitem{Wu:2018lai}
  S. Wu and B. Long,
  arXiv:1807.04407 [nucl-th].

\bibitem{Long:2007vp}
  B. Long and U. van Kolck,
  Annals Phys. {\bf 323} (2008) 1304
  [arXiv:0707.4325 [quant-ph]].

\bibitem{Frederico:1999ps}
  T.~Frederico, V.S.~Tim\'oteo, and L.~Tomio,
  Nucl.\ Phys.\ A {\bf 653} (1999) 209
  [nucl-th/9902052].

\bibitem{PavonValderrama:2005uj}
  M.~Pav\'on Valderrama and E.~Ruiz Arriola,
  Phys.\ Rev.\ C {\bf 74} (2006) 064004
  [Erratum: Phys.\ Rev.\ C {\bf 75} (2007) 059905]
  [nucl-th/0507075].

\bibitem{Kaplan:1996xu}
  D.B.~Kaplan, M.J.~Savage, and M.B.~Wise,
  Nucl.\ Phys.\ B {\bf 478} (1996) 629
  [nucl-th/9605002].

\bibitem{PavonValderrama:2016lqn}
  M.~Pav\'on Valderrama, M.~S\'anchez S\'anchez, C.-J.~Yang, B.~Long, 
J.~Carbonell, and U.~van Kolck,
  Phys.\ Rev.\ C {\bf 95} (2017) 
  054001
  [arXiv:1611.10175 [nucl-th]].

\bibitem{Long:2012ve}
  B.~Long and C.-J.~Yang,
  Phys.\ Rev.\ C {\bf 86} (2012) 024001
  [arXiv:1202.4053 [nucl-th]].

\bibitem{Song:2016ale}
  Y.-H.~Song, R.~Lazauskas, and U.~van Kolck,
  Phys.\ Rev.\ C {\bf 96} (2017) 
  024002
  [arXiv:1612.09090 [nucl-th]].

\bibitem{Valderrama:2009ei}
  M.~Pav\'on Valderrama,
  Phys.\ Rev.\ C {\bf 83} (2011) 024003
  [arXiv:0912.0699 [nucl-th]].

\bibitem{Valderrama:2011mv}
  M.~Pav\'on Valderrama,
  Phys.\ Rev.\ C {\bf 84} (2011) 064002
  [arXiv:1108.0872 [nucl-th]].

\bibitem{Long:2011qx}
  B.~Long and C.-J.~Yang,
  Phys.\ Rev.\ C {\bf 84} (2011) 057001
  [arXiv:1108.0985 [nucl-th]].

\bibitem{Long:2011xw}
  B.~Long and C.-J.~Yang,
  Phys.\ Rev.\ C {\bf 85} (2012) 034002
  [arXiv:1111.3993 [nucl-th]].

\end{thebibliography}
\end{document}